\documentclass[12pt]{article}
\pdfoutput=1
\usepackage{jheppub}
\usepackage{epsfig}
\usepackage{amsmath}
\usepackage{amssymb}
\usepackage{amsfonts}
\usepackage{amsxtra}
\usepackage{amsthm}
\usepackage{mathrsfs}
\usepackage{makeidx}
\usepackage{graphics}
\usepackage{dsfont}
\usepackage{mathtools}
\usepackage{graphicx}
\usepackage{subcaption}
\usepackage{placeins}
\usepackage{bm}
\usepackage[capitalise]{cleveref}
\usepackage{empheq}
\usepackage{colortbl}
\usepackage{xcolor}
\usepackage{enumerate}
\usepackage{titlesec}
\usepackage{longtable}
\usepackage{float}
\usepackage{color}
\usepackage{tikz}
\usepackage{xfrac}
\usepackage{footnote}
\usepackage{rotating}
\usepackage{lscape}
\usepackage{makecell}
\usepackage{environ}
\usepackage{tabularx}
\usepackage{subfiles}
\usepackage[export]{adjustbox}
\usepackage{ytableau}
\usepackage{tikz-3dplot}
\usepackage{slashed}
\usepackage{pifont}
\usepackage{multirow}
\usepackage{mdframed}
\usepackage{bbm}
\usepackage{ulem}

\usetikzlibrary{positioning,trees,decorations.pathmorphing,decorations.markings,decorations.pathreplacing,calc,shapes,patterns,arrows,chains,arrows.meta,fit,fadings,decorations.markings,graphs,graphs.standard,quotes}

\DeclareMathOperator{\U}{U}

\DeclareMathOperator{\SU}{SU}
\DeclareMathOperator{\SO}{SO}

\DeclareMathOperator{\USp}{USp}
\DeclareMathOperator{\tr}{Tr}

\DeclareMathOperator{\rank}{rank}

\theoremstyle{definition}

\catcode`\@=11   
\newdimen\@rotdimen
\newbox\@rotbox  

\def\@vspec#1{\special{ps:#1}}
\def\@rotstart#1{\@vspec{gsave currentpoint currentpoint translate
   #1 neg exch neg exch translate}}
\def\@rotfinish{\@vspec{currentpoint grestore moveto}}
\def\@rotr#1{\@rotdimen=\ht#1\advance\@rotdimen by\dp#1%
   \hbox to\@rotdimen{\hskip\ht#1\vbox to\wd#1{\@rotstart{90 rotate}%
   \box#1\vss}\hss}\@rotfinish}
\def\@rotl#1{\@rotdimen=\ht#1\advance\@rotdimen by\dp#1%
   \hbox to\@rotdimen{\vbox to\wd#1{\vskip\wd#1\@rotstart{270 rotate}%
   \box#1\vss}\hss}\@rotfinish}%
\def\@rotu#1{\@rotdimen=\ht#1\advance\@rotdimen by\dp#1%
   \hbox to\wd#1{\hskip\wd#1\vbox to\@rotdimen{\vskip\@rotdimen
   \@rotstart{-1 dup scale}\box#1\vss}\hss}\@rotfinish}%
\def\@rotf#1{\hbox to\wd#1{\hskip\wd#1\@rotstart{-1 1 scale}%
   \box#1\hss}\@rotfinish}%
\def\rotate{\@ifnextchar[{\@rotate}{\@rotate[l\right]}}
\def\@rotate[#1]#2{\setbox\@rotbox=\hbox{#2}\@nameuse{@rot#1}\@rotbox}

\catcode`\@=12

\usetikzlibrary{positioning}
\usetikzlibrary{chains}
\usetikzlibrary{arrows, arrows.meta ,fit,decorations.pathreplacing}
\tikzstyle{every picture}+=[remember picture]
\tikzstyle{na} = [baseline]

\usetikzlibrary{arrows, decorations.markings, calc, fadings, decorations.pathreplacing, patterns, decorations.pathmorphing, positioning}
\tikzset{>={Latex[width=1.5mm,length=1.5mm]}}
\usetikzlibrary{graphs,graphs.standard,quotes}

\def\node#1#2{\overset{#1}{\underset{#2}{{\color{gray} \bullet}}}}

\def\node#1#2{\overset{#1}{\underset{#2}{\circ}}}

\tikzstyle{every picture}+=[remember picture]
\tikzstyle{na} = [baseline=-.5ex]

\newcommand{\eg}{e.g.}

\newcommand{\ie}{i.e.}

\numberwithin{equation}{section}
\newcommand{\bes}[1]{\begin{equation} \begin{split} #1\end{split} \end{equation}}

\newcommand{\nn}{\nonumber}

\newcommand{\be}{\begin{equation}} \newcommand{\ee}{\end{equation}}
\newcommand{\bea}{\begin{equation} \begin{aligned}} \newcommand{\eea}{\end{aligned} \end{equation}}

\def\tilde{\widetilde}

\def\hat{\widehat}

\def\bar{\overline}

\def\rt2{\sqrt{2}}

\def\mod{{\rm mod}}

\def\tr{\mathop{\rm tr}}

\def\CA{{\cal A}}

\def\CI{{\cal I}}

\def\CM{{\cal M}}
\def\CN{{\cal N}}

\def\CP{{\cal P}}
\def\CQ{{\cal Q}}

\def\CS{{\cal S}}

\def\CZ{{\cal Z}}

\def\1{{\ds 1}}

\newcommand{\fm}{\mathfrak{m}}
\newcommand{\fn}{\mathfrak{n}}

\def\SO{\mathrm{SO}}

\def\SU{\mathrm{SU}}

\def\Spin{\mathrm{Spin}}

\def\su{\mathfrak{su}}

\def\fp{\mathfrak{p}}
\def\fq{\mathfrak{q}}
\def\fr{\mathfrak{r}}

\def\repa{\raise4pt\hbox{$\square$}\mkern-14mu\raise-4pt\hbox{$\square$}}
\def\repab{\overline{\raise4pt\hbox{$\square$}\mkern-14mu\raise-4pt\hbox{$\square$}\mkern-1mu}}

\def\smileface{\ensuremath{\hbox{\large$\bigcirc$}\mkern-15mu\raise-1pt\hbox{\scriptsize$\smallsmile$}%
\mkern-10mu\raise4pt\hbox{..}\mkern4mu}}
\def\frownface{\ensuremath{\hbox{\large$\bigcirc$}\mkern-15mu\raise-1pt\hbox{\scriptsize$\smallfrown$}%
\mkern-10mu\raise4pt\hbox{..}\mkern4mu}}

\newcommand{\ba}{\begin{array}}
\newcommand{\ea}{\end{array}}
\newcommand{\bi}{\begin{itemize}}
\newcommand{\ei}{\end{itemize}}
\def\vec#1{\bm{#1}}
\def\bea#1\eea{\allowdisplaybreaks \begin{align}#1\end{align}}
 \newcommand{\ben}{\begin{enumerate}}
\newcommand{\een}{\end{enumerate}}
\newcommand{\bean}{\begin{eqnarray*}}
\newcommand{\eean}{\end{eqnarray*}}
\newcommand{\eref}[1]{(\ref{#1})}

\newcommand{\PE}{\mathop{\rm PE}}

\newcommand{\BC}{\mathbb{C}}

\newcommand{\BZ}{\mathbb{Z}}

\definecolor{light-gray}{gray}{0.5}

\def\aup#1 {\overset{#1}{\uparrow} \, \overset{\tilde{#1}}{\downarrow}}

\tikzset{snake it/.style={decorate, decoration={snake, amplitude=.4mm, segment length=2mm,
                       post length=0mm,pre length=0mm}}}

\hypersetup{
	pdftitle={Mixed Anomalies, Two-groups, Non-Invertible Symmetries, and 3d Superconformal Indices},    
	pdfauthor={\textcopyright\ Noppadol Mekareeya, Matteo Sacchi},     
	pdfsubject={hep-th},   
	pdfcreator={pdfLaTex},   
	pdfproducer={LaTex}, 
	pdfkeywords={},
	colorlinks=true,
}

\makeatletter
\newsavebox{\measure@tikzpicture}
\NewEnviron{scaletikzpicturetowidth}[1]{%
  \def\tikz@width{#1}%
  \begin{lrbox}{\measure@tikzpicture}%
  \BODY
  \end{lrbox}%
  \pgfmathparse{#1/\wd\measure@tikzpicture}%
  \BODY
}
\makeatother

\makeatletter
\def\squarecorner#1{
    \pgf@x=\the\wd\pgfnodeparttextbox%
    \pgfmathsetlength\pgf@xc{\pgfkeysvalueof{/pgf/inner xsep}}%
    \advance\pgf@x by 2\pgf@xc%
    \pgfmathsetlength\pgf@xb{\pgfkeysvalueof{/pgf/minimum width}}%
    \ifdim\pgf@x<\pgf@xb%
        \pgf@x=\pgf@xb%
    \fi%
    \pgf@y=\ht\pgfnodeparttextbox%
    \advance\pgf@y by\dp\pgfnodeparttextbox%
    \pgfmathsetlength\pgf@yc{\pgfkeysvalueof{/pgf/inner ysep}}%
    \advance\pgf@y by 2\pgf@yc%
    \pgfmathsetlength\pgf@yb{\pgfkeysvalueof{/pgf/minimum height}}%
    \ifdim\pgf@y<\pgf@yb%
        \pgf@y=\pgf@yb%
    \fi%
    \ifdim\pgf@x<\pgf@y%
        \pgf@x=\pgf@y%
    \else
        \pgf@y=\pgf@x%
    \fi
    \pgf@x=#1.5\pgf@x%
    \advance\pgf@x by.5\wd\pgfnodeparttextbox%
    \pgfmathsetlength\pgf@xa{\pgfkeysvalueof{/pgf/outer xsep}}%
    \advance\pgf@x by#1\pgf@xa%
    \pgf@y=#1.5\pgf@y%
    \advance\pgf@y by-.5\dp\pgfnodeparttextbox%
    \advance\pgf@y by.5\ht\pgfnodeparttextbox%
    \pgfmathsetlength\pgf@ya{\pgfkeysvalueof{/pgf/outer ysep}}%
    \advance\pgf@y by#1\pgf@ya%
}
\makeatother

\pgfdeclareshape{square}{
    \savedanchor\northeast{\squarecorner{}}
    \savedanchor\southwest{\squarecorner{-}}

    \foreach \x in {east,west} \foreach \y in {north,mid,base,south} {
        \inheritanchor[from=rectangle]{\y\space\x}
    }
    \foreach \x in {east,west,north,mid,base,south,center,text} {
        \inheritanchor[from=rectangle]{\x}
    }
    \inheritanchorborder[from=rectangle]
    \inheritbackgroundpath[from=rectangle]
}

\tikzset{stretch/.initial=1}
\newcommand\drawloop[4][]%
   {\draw[shorten <=0pt, shorten >=0pt,#1]
      ($(#2)!\pgfkeysvalueof{/tikz/stretch}!(#2.#3)$)
      let \p1=($(#2.center)!\pgfkeysvalueof{/tikz/stretch}!(#2.north)-(#2)$),
          \n1= {veclen(\x1,\y1)*sin(0.5*(#4-#3))/sin(0.5*(180-#4+#3))}
      in arc [start angle={#3-90}, end angle={#4+90}, radius=\n1]%
   }

\title{Chern-Simons-Trinion Theories: One-form Symmetries and Superconformal Indices}
\author[a,b]{Riccardo Comi,}
\author[a,b]{William Harding,}
\author[b,c]{and Noppadol Mekareeya}
\affiliation[a]{Dipartimento di Fisica, Universit\`a di Milano-Bicocca, \\Piazza della Scienza 3, I-20126 Milano, Italy}
\affiliation[b]{INFN, sezione di Milano-Bicocca, \\Piazza della Scienza 3, I-20126 Milano, Italy}
\affiliation[c]{Department of Physics, Faculty of Science, \\
Chulalongkorn University, Phayathai Road, \\
Pathumwan, Bangkok 10330, Thailand}
\emailAdd{r.comi2@campus.unimib.it}
\emailAdd{w.harding@campus.unimib.it}
\emailAdd{n.mekareeya@gmail.com}
\abstract{We study 3d theories containing $\mathcal{N}=3$ Chern-Simons vector multiplets coupled to the $\mathrm{SU}(N)^3$ flavour symmetry of 3d $T_N$ theories with Chern-Simons levels $k_1$, $k_2$ and $k_3$.  It was formerly pointed out that these theories flow to infrared SCFTs with enhanced $\mathcal{N}=4$ supersymmetry when $1/k_1+1/k_2+1/k_3=0$. We examine superconformal indices of these theories which reveal that supersymmetry of the infrared SCFTs may get enhanced to $4 \leq \mathcal{N} \leq 6$ if such a condition is satisfied. Moreover, even if the Chern-Simons levels do not obey the aforementioned condition, we find that there is still an infinite family of theories that flows to infrared SCFTs with $\mathcal{N}=4$ supersymmetry. The 't Hooft anomalies of the one-form symmetries of these theories are analysed. As a by-product, we observe that there is generally a decoupled topological sector in the infrared. When the infrared SCFTs have $\CN \geq 4$ supersymmetry, we also study the Higgs and Coulomb branch limits of the indices which provide geometric information of the moduli space of the theories in question in terms of the Hilbert series.}
\begin{document}
\maketitle

\section{Introduction}
Three-dimensional (3d) supersymmetric Chern--Simons (CS)--matter theories have rich infrared (IR) behaviours.  For example, the $\CN = 3$ superconformal field theory (SCFT) can be obtained as the IR fixed point of the $\CN = 2$ theory deformed by a certain superpotential \cite{Gaiotto:2007qi}.  Moreover, if the gauge algebra and matter content are chosen appropriately in an $\CN=3$ theory, supersymmetry may get further enhanced up to $\CN=8$ \cite{Gustavsson:2007vu, Bagger:2007jr, Bandres:2008vf, Gaiotto:2008sd, Hosomichi:2008jd, Hosomichi:2008jb, Aharony:2008ug, Schnabl:2008wj, Aharony:2008gk}.  In particular, the $\U(N)_k \times \U(N)_{-k}$ CS theory with two hypermultiplets in the bifundamental representation (in the 3d $\CN=4$ language) describes the system of $N$ M2-branes on $\BC^4/\BZ_k$, where at large $N$ it is dual to M-theory on $\mathrm{AdS}_4 \times S^7/\BZ_k$, and at large $N$ with a fixed ratio $N/k$ is dual to Type IIA string theory on $\mathrm{AdS}_4 \times \mathbf{CP}^3$. This CS--matter theory provides the first explicit realisation of the AdS/CFT correspondence in these dimensions \cite{Aharony:2008ug}.  

In this article, we focus on 3d $\CN=3$ vector multiplets coupled to a certain number of copies of a 3d SCFT, known as the 3d $T_N$ theory, whose flavour symmetry is $\SU(N)^3$.  The 3d $T_N$ theory can be realised by compactifying the 4d $T_N$ theory \cite{Gaiotto:2009we, Gaiotto:2009gz} on a circle, or equivalently by compactifying $N$ M5-branes on a circle times a sphere with three full punctures.  An example of the theories of our interest consists of that constructed from a single copy of the $T_N$ theory such that the $\SU(N)^3$ flavour symmetry is gauged with CS levels $k_1$, $k_2$ and $k_3$. This can be realised by compactifying $N$ M5-branes on a three-manifold which is a Seifert bundle over $S^2$ with three singular fibres, with Seifert parameters $1/k_{1}$, $1/k_2$ and $1/k_3$ \cite{Eckhard:2019jgg, Assel:2022row}.  This example can be generalised further, for example, by involving many copies of the $T_N$ theories where each $\SU(N)$ factor of the $\SU(N)^3$ flavour symmetry of each copy is commonly gauged with CS levels $k_1$, $k_2$ and $k_3$.  The corresponding three-manifolds are then known as graph manifolds \cite{Assel:2022row}.  We remark {\it en passant} that, due to their richness in mathematical and physical properties, 3d theories arising from compactifying M5-branes on three-manifolds have received considerable attention over the recent years; see \eg~ \cite{Terashima:2011qi, Dimofte:2011ju, Gadde:2013sca, Chung:2014qpa, Gukov:2016gkn, Gukov:2017kmk, Gang:2018wek, Assel:2018vtq, Gang:2018huc, Cho:2020ljj, Gang:2021hrd}.


 For the theories of our interest with general $N$, it was pointed out by the authors of \cite{Assel:2022row} that whenever the CS levels satisfy the condition $\sum_{i=1}^3 1/k_i =0$, then the theory flows to an IR SCFT with enhanced $\CN=4$ supersymmetry.\footnote{We remark that for $N=2$ and a single copy of the $T_2$ theory, the enhancement of supersymmetry to $\CN=4$ follows from the result of \cite{Gaiotto:2008sd}.}  For convenience, we shall refer to this condition as the Assel--Tachikawa--Tomasiello (ATT) condition. This statement was supported by a field theoretic argument.  Geometrically, for a single copy of the $T_N$ theory, supersymmetry enhancement is accounted for by the holonomy of the corresponding Seifert manifolds.  However, when the theory contains more than one $T_N$ building blocks, supersymmetry enhancement is left unaccounted for, in general, by the holonomy of graph manifolds.
 
One of the main objectives of this paper is to study supersymmetry enhancement of this family of theories, focusing on $N=2$ and $N=3$, using the superconformal index \cite{Bhattacharya:2008zy,Bhattacharya:2008bja, Kim:2009wb,Imamura:2011su, Kapustin:2011jm, Dimofte:2011py, Aharony:2013dha, Aharony:2013kma}.\footnote{We adopt the same notation as \cite{Aharony:2013kma, Garozzo:2019ejm, Beratto:2021xmn, Mekareeya:2022spm}. Each notation is defined in the main text.}  We find that, when the ATT condition is satisfied, supersymmetry of the IR SCFT generally gets enhanced to $\CN=4$, but there are also a large number of cases with $\CN=5$ and $\CN=6$ supersymmetry.  Surprisingly we also find that, even if the ATT condition is not satisfied, there is still an infinite family of theories whose IR SCFTs have enhanced $\CN=4$ supersymmetry.  In particular, for certain special values of CS levels such as $(k_1, k_2, k_3) = (2,1,1)$, the IR SCFT turns out to be the rank-zero minimal $\CN=4$ SCFT, discussed in \cite{Gang:2018huc, Gang:2021hrd}.  Another main goal of this article is to study the one-form symmetries \cite{Kapustin:2014gua, Gaiotto:2014kfa} of these theories, as well as their (mixed) 't Hooft anomalies along the line of \cite{Hsin:2018vcg, Eckhard:2019jgg}, and the mixed 't Hooft anomalies between the one-form symmetries and zero-form symmetries using the method of \cite{Mekareeya:2022spm} (see also \cite{Sacchi:2023omn}).  As a result, we deduce that there is generally a decoupled topological sector in the IR.  We also identify these topological quantum field theories (TQFTs) using 't Hooft anomalies of the one-form symmetries.  Importantly, we point out that, even if a set of theories flow to the same IR SCFTs (which can be deduced from the fact that the indices are the same or the associated three-manifolds are diffeomorphic to each other), the decoupled topological sectors may be different.  We also gauge the non-anomalous one-form symmetry and study the resulting theories.  Finally, we study the Higgs and Coulomb limits of the superconformal indices \cite{Razamat:2014pta} for theories whose IR SCFTs have $\CN \geq 4$ enhanced supersymmetry.  These provide geometric information of the Higgs and Coulomb branches of the IR SCFTs in question in terms of the Hilbert series \cite{Benvenuti:2006qr, Feng:2007ur}.

The paper is organised as follows.  In Sections \ref{sec:oneT2} and \ref{sec:twoT2}, we study theories with one and two $T_2$ building blocks, respectively.  Their one-form symmetries and 't Hooft anomalies are investigated in Section \ref{sec:oneformanomsingleT2}.  The theories whose CS levels satisfy the ATT condition are then studied in Sections \ref{sec:ATToneT2} and \ref{sec:ATTtwoT2}.  In this class of theories, the Higgs and Coulomb branch limits of the indices are studied in Sections \ref{sec:HBCBoneT2} and \ref{sec:HBCBtwoT2}.  We then move on to explore theories whose CS levels do not satisfy the ATT condition in Sections \ref{sec:nonATToneT2} and \ref{sec:nonATTtwoT2}.  We also study theories coupled to one or many copies of the $T(\SU(2))$ SCFT in Sections \ref{sec:singleT2gluewithS} and \ref{sec:twoT2gluewithS}.  In Section \ref{sec:T3}, we discuss theories with $T_3$ building blocks.  We discuss their indices and 't Hooft anomalies of one-form symmetries.  Due to the technicality of the computations, we focus on the indices of theories whose CS levels satisfy the ATT condition, from which we conclude that the IR SCFT has enhanced $\CN=4$ supersymmetry.  In Appendix \ref{app:fourT2}, we study certain theories with four $T_2$ building blocks as well as their indices.  In Appendix \ref{sec:mixedflvoneT2}, the mixed gauge/zero-form monopole operators in theories with one and two $T_2$ building blocks such that the ATT condition is satisfied are examined.  The potential mixed anomaly between the $\BZ_2$ one-form symmetry and the zero-form flavour symmetry implied by the presence of such monopole operators is discussed.

\paragraph*{Note added.} After this paper had appeared on the arXiv, we found \cite{Li:2023ffx} with some overlapping content. 

\section{Theories with one $T_2$ building block} \label{sec:oneT2}
Let us consider the $T_2$ theory whose three $\SU(2)$ flavour symmetries are gauged with 3d $\CN =3$ Chern-Simons couplings $(k_1, k_2, k_3)$.  We depict this theory diagrammatically by
\bes{ \label{singleT2}
\begin{tikzpicture}[baseline, font=\footnotesize]
\node[draw=none] (T2) at (0,0) {$T_2$};
\node[draw=none] (c1) at (2,1) {};
\node[draw=none] (c2) at (2,0) {};
\node[draw=none] (c3) at (2,-1) {};
\draw[solid, bend left] (T2) to node[above,midway] {$k_1$} (c1);
\draw[solid] (T2) to node[above,midway] {$k_2$} (c2);
\draw[solid, bend right] (T2) to node[above,midway] {$k_3$} (c3);
\end{tikzpicture}
}
where each finite line with label $k_i$ denotes an $\SU(2)_{k_i}$ gauge group.  This theory was studied extensively in \cite[Section 2.2.1]{Assel:2022row}, where it was pointed out that \eref{singleT2} can be realised by compactifying M5-branes on a three-manifold which is  a Seifert bundle over $S^2$ with three singular fibres, with Seifert parameters $1/k_1$, $1/k_2$ and $1/k_3$.

The effective superpotential after integrating out the adjoint scalars is
\bes{
W \propto \sum_{i=1}^3 \frac{1}{k_i} \tr (\mu_i^2)~,
}
where $\mu_i$ are the moment map operators for the $\SU(2)_i$ symmetry of the $T_2$ theory.  In terms of the chiral fields $Q_{\alpha_1 \alpha_2  \alpha_3}$ of the $T_2$ theory, $\mu_i$ can be written as
\bes{ \label{momentmaps}
(\mu_1)_{\alpha_1\alpha'_1} &= \epsilon^{\alpha_2 \alpha'_2} \epsilon^{\alpha_3 \alpha'_3} Q_{\alpha_1 \alpha_2  \alpha_3} Q_{\alpha'_1 \alpha'_2  \alpha'_3}~, \\
(\mu_2)_{\alpha_2\alpha'_2} &= \epsilon^{\alpha_1 \alpha'_1} \epsilon^{\alpha_3 \alpha'_3} Q_{\alpha_1 \alpha_2  \alpha_3} Q_{\alpha'_1 \alpha'_2  \alpha'_3}~, \\
(\mu_3)_{\alpha_3\alpha'_3} &= \epsilon^{\alpha_1 \alpha'_1} \epsilon^{\alpha_2 \alpha'_2} Q_{\alpha_1 \alpha_2  \alpha_3} Q_{\alpha'_1 \alpha'_2  \alpha'_3}~, 
}
where the indices $\alpha_i, \alpha'_i =1,2$ correspond to the $\SU(2)_i$ gauge group (with $i=1,2,3$).  From the above relations, it follows that
\bes{ \label{rel1}
\tr(\mu_1^2) = \tr(\mu_2^2) = \tr(\mu_3^2) \equiv \tr(\mu^2) 
}
and so the effective superpotential can be rewritten as
\bes{ \label{sup1}
W \propto \left( \frac{1}{k_1}+\frac{1}{k_2}+\frac{1}{k_3} \right) \tr (\mu^2)~.
}

If the following condition is satisfied
\bes{\label{ATT}
\frac{1}{k_1}+\frac{1}{k_2}+\frac{1}{k_3}=0~,
} 
then supersymmetry gets enhanced from $\CN=3$ to $\CN=4$.   This follows from the discussion in \cite{Gaiotto:2008sd} and also from \cite{Schnabl:2008wj, Assel:2022row}.  Relation \eref{ATT} is what we referred to as the ATT condition in the introduction.  Subsequently we will show that this is a {\it sufficient}, but {\it not} necessary, condition for supersymmetry enhancement.  In particular, in Section \ref{sec:nonATToneT2}, we will show that even if the ATT condition \eref{ATT} is not satisfied, there are cases in which the IR SCFT has accidental $\CN=4$ supersymmetry.

A main tool that we will use to analyse these theories is the superconformal index.  It is explicitly given by
\bes{ \label{indexsingleT2}
\CI_{\eref{singleT2}}(a, n_a; x) &= \left( \frac{1}{8}\prod_{i=1}^3 \oint \frac{d z_i}{2\pi i z_i} \right) \sum_{(m_1, m_2, m_3) \in \BZ^3}  \left( \prod_{i=1}^3 z_i^{2 k_i m_i}   \CZ^{\SU(2)}_{\text{vec}} (z_i;m_i;x) \right) \times  \\
&\qquad  \prod_{s_1, s_2, s_3 =\pm 1} \CZ^{1/2}_{\chi}(z_1^{s_1} z_2^{s_2} z_3^{s_3} a; s_1 m_1+s_2 m_2+s_3 m_3+n_a;x)~, 
}
where the $\SU(2)$ vector multiplet contribution is
\bes{
\CZ^{\SU(2)}_{\text{vec}} (z;n;x) = x^{-2|n|} \prod_{s=\pm 1}(1- (-1)^{2n} x^{2|n|} z^{2s})~,
}
and the contribution of the chiral multiplet of $R$-charge $R$ is
\bes{
\CZ^R_{\chi}(z; m;x) = \left( x^{1-R} z^{-1} \right)^{|m|/2} \prod_{j=0}^\infty \frac{1-(-1)^m z^{-1} x^{|m|+2-R+2j}}{1-(-1)^m z\,  x^{|m|+R+2j}}~.
}

If the CS levels satisfy the ATT condition \eref{ATT}, it follows from \eref{sup1} that the effective superpotential is zero, and the $\U(1)_a$ flavour symmetry which assigns charge $+1$ to all of the eight chiral multiplets of the $T_2$ theory is a symmetry of the Lagrangian.  We denote by $a$ and $n_a$ the fugacity and background magnetic flux for this flavour symmetry.  Upon computing the series expansion of the index, we will set $n_a = 0$ and drop the $n_a$ dependence from the index, \ie~ we write the latter simply as $\CI_{\eref{singleT2}}(a; x)$.  Note also that, if the CS levels do not satisfy the ATT condition \eref{ATT}, we should set $a=1$ and $n_a=0$ in \eref{indexsingleT2}, since the $\U(1)_a$ flavour symmetry is no longer a symmetry of the effective superpotential \eref{sup1}.

\subsection{One-form symmetries and their 't Hooft anomalies} \label{sec:oneformanomsingleT2}
We now discuss the one-form symmetries of theory \eref{singleT2} and their anomalies.  Let us first consider the $T_2$ theory, whose global form of the manifest flavour symmetry is (see \cite[(4.2)]{Bhardwaj:2021ojs})
\bes{ \label{globalformsymT2}
\frac{\SU(2)_1 \times \SU(2)_2 \times \SU(2)_3}{(\BZ_2)_{13} \times (\BZ_2)_{23}}~,
}
where $(\BZ_2)_{ij}$ denotes the diagonal $\BZ_2$ subgroup of the centre of $\SU(2)_i$ times the centre of $\SU(2)_j$.  In other words, among the three $\BZ_2$ factors that come from the centre of $\prod_{i=1}^3 \SU(2)_i$ in the numerator of \eref{globalformsymT2}, only two combinations present in the denominator act trivially on the four free hypermultiplets of the $T_2$ theory. Gauging each of the $\SU(2)_i$ symmetry therefore leads to the $\BZ_2 \times \BZ_2$ one-form symmetry.  We will see that turning on CS levels $k_i$ for each $\SU(2)_i$ gauge group (with $i=1,2,3$) results in 't Hooft anomalies of a subgroup or the whole $\BZ_2^2$ one-form symmetry.

The 't Hooft anomaly of the one-form symmetry in theory \eref{singleT2} with CS levels $(k_1, k_2, k_3)$ is characterised by the 4d anomaly theory whose action is \cite{Benini:2017dus}
\bes{ \label{anomth1}
\frac{2 \pi}{2} \int_{\CM_4}  \sum_{i=1}^3 k_i \frac{\CP(w_i^{(2)})}{2}~, 
}
where each of $w_i^{(2)} \in H^2(\CM_4, \BZ_2)$ is the two-form background field for each $\BZ_2$ one-form symmetry that arises from the centre of each $\SU(2)_i$ gauge group, $\CP(w^{(2)})$ is the Pontryagin square operation and the integration is performed over a spin manifold $\CM_4$.  Note that $\int_{\CM_4} \CP(w_i^{(2)})$ is even on a spin manifold $\CM_4$.

Since among the $\BZ_2^3$ centres of $\SU(2)^3$ only $\BZ_2^2$ acts non-trivially on the trifundamental matter of the $T_2$ theory, we have the condition
\bes{ \label{constrT2}
\sum_{i=1}^3 w_i^{(2)} =0~.
}  
Using the identity
\bes{
\int_{\CM_4} \CP(A+B) &= \int_{\CM_4} \CP(A) + \int_{\CM_4} \CP(B)  + 2 \int_{\CM_4} A \cup B~, \\
}
and the fact that $\int_{\CM_4} \CP(w_i^{(2)})$ is even on spin manifold $\CM_4$, we rewrite \eref{anomth1} as
\bes{ \label{Sanom1}
S_{\text{anom}} = \frac{2 \pi}{2} \int_{\CM_4} \left[ (k_1+k_3) \frac{\CP(w_1^{(2)})}{2}+ (k_2+k_3) \frac{\CP(w_2^{(2)})}{2} - k_3 (w_1^{(2)} \cup  w_2^{(2)})\right]
}
after dropping terms that are integer multiples of $2\pi$. Using the notation of \cite[(F.8)]{Hsin:2018vcg} 
\bes{ \label{anomtheory}
S_{\text{anom}} = \frac{2\pi}{2} \left[ \sum_{I=1}^2 p_{II} \int_{\CM_4} \frac{\CP(w_I^{(2)})}{2} + p_{12} \int_{\CM_4} w_1^{(2)} \cup  w_2^{(2)} \right]~, 
}
we see that the above anomalies can be summarised in following symmetric matrix $p$:
\bes{ \label{anompmatrix}
p = \begin{pmatrix} 
k_1 +k_3 \quad & \quad- k_3 \\
-k_3 \quad & \quad k_2+k_3
\end{pmatrix} \quad \mod ~2~.
}
For $k_3=1$, this is in agreement with the anomaly matrix given by \cite[(4.58)]{Eckhard:2019jgg} with $N=n=2$ and $k_{1,2} \rightarrow k_{1,2}+1$.   

We can decompose the $\BZ_2^2$ one-form symmetry of \eref{singleT2} into two parts: a subgroup $\Gamma_A$ of $\BZ_2^2$ that has an 't Hooft anomaly and the 't Hooft anomaly free part $\BZ_2^2/\Gamma_A$. The subgroup $\Gamma_A$ is given by $\Gamma_A = \BZ_2^{\mathrm{rank}(p)}$, where $\mathrm{rank}(p)$ is the rank of matrix $p$. The 't Hooft anomaly free part is then $\BZ_2^{\dim(\mathrm{ker}\, p)}$, where $\mathrm{ker}\, p$ denotes the kernel (nullspace) of matrix $p$.

According to \cite[(4.44)]{Eckhard:2019jgg}, it was proposed that theory \eref{singleT2} flows to an IR theory that splits into two subsectors:
\ben
\item the minimal abelian TQFT \cite{Hsin:2018vcg}:
\bes{
\CA^{\Gamma_A, p} = \begin{cases} 
\CA^{2,1} \cong \SU(2)_1 \cong \mathrm{U}(1)_2 \quad & \text{if $\mathrm{rank}(p)=1$}\\
\text{$\CA^{\{2,2\},p}$ of \cite[Appendix F]{Hsin:2018vcg}} \quad & \text{if $\mathrm{rank}(p)=2$}
\end{cases}~;}
\item the subsector $\eref{singleT2}_\text{AF}$, which can be an interacting SCFT or another TQFT, whose one-form symmetry is 't Hooft anomaly free.  
\een
The above statement can be summarised as\footnote{A piece of evidence that supports this statement is as follows. We observe that there exist theories associated with diffeomorphic three-manifolds (their superconformal indices are equal) whose anomaly matrices $p$ are inequivalent; see Footnote \ref{foot:diffTQFT}. For such theories, we interpret this phenomenon in the following way. The subsectors $\eref{singleT2}_\text{AF}$, characterised by the index, are the same. However, the topological sectors, characterised by the anomaly matrices, are different.}
\bes{
\eref{singleT2}  = \eref{singleT2}_\text{AF} \otimes  \CA^{\Gamma_A, p}~.
}
This relation can be inverted as in \cite[(1.13)]{Hsin:2018vcg} and \cite[(4.45)]{Eckhard:2019jgg}, namely
\bes{
\eref{singleT2}_\text{AF} = \frac{ \eref{singleT2} \otimes \CA^{\Gamma_A, -p} }{\Gamma_A}~.
}

\subsection{Cases that satisfy the ATT condition} \label{sec:ATToneT2}
We consider theories with CS levels satisfying \eref{ATT}.

\subsubsection{Special case of $(k_1, k_2, k_3) = (-k,2k,2k)$}
These CS levels satisfy the ATT condition.  Theory \eref{singleT2} with these CS levels, namely 
\bes{ \label{singleT2mk2k2k}
\begin{tikzpicture}[baseline, font=\footnotesize]
\node[draw=none] (T2) at (0,0) {$T_2$};
\node[draw=none] (c1) at (2,1) {};
\node[draw=none] (c2) at (2,0) {};
\node[draw=none] (c3) at (2,-1) {};
\draw[solid, bend left] (T2) to node[above,midway] {$k_1=-k$} (c1);
\draw[solid] (T2) to node[above,midway] {$k_2=2k$} (c2);
\draw[solid, bend right] (T2) to node[below,midway] {$k_3=2k$} (c3);
\end{tikzpicture}
}
can also be regarded as the $\USp(2)_{-k} \times \Spin(4)_{2k}$ gauge theory with a bifundamental half-hypermultiplet in the representation $[\mathbf{2}; \mathbf{4}]$ whose quiver diagram is
\bes{ \label{USp2Spin4}
\begin{tikzpicture}[baseline, font=\footnotesize]
\node[draw, circle, minimum size=1.5cm] (USp) at (-1.5,0) {\scalebox{0.75}{$\USp(2)_{-k}$}};
\node[draw, circle, minimum size=1.5cm] (Spin) at (1.5,0) {\scalebox{0.75}{$\Spin(4)_{2k}$}};
\draw[solid] (USp) to (Spin);
\end{tikzpicture}
}
The equivalence of these two theories is due to the fact that $\Spin(4) \cong \SU(2) \times \SU(2)$ and that the vector representation $[\mathbf{4}]$ of $\Spin(4)$ is equivalent to the representation $[\mathbf{2}; \mathbf{2}]$ of $\SU(2) \times \SU(2)$.  The index of this theory can be derived as in \cite{Aharony:2013kma, Beratto:2021xmn}. We first compute the index for the $\USp(2)_{-k} \times \SO(4)_{2k}$ gauge theory with the same matter content, namely
\bes{ \label{USp2SO4}
\begin{tikzpicture}[baseline, font=\footnotesize]
\node[draw, circle, minimum size=1.5cm] (USp) at (-1.5,0) {\scalebox{0.75}{$\USp(2)_{-k}$}};
\node[draw, circle, minimum size=1.5cm] (Spin) at (1.5,0) {\scalebox{0.75}{$\SO(4)_{2k}$}};
\draw[solid] (USp) to (Spin);
\end{tikzpicture}
}
The index of this theory is
\bes{ \label{indUSp2SO4}
&\CI_{\eref{USp2SO4}}(\zeta, a;x) \\
&= \frac{1}{8} \sum_{(\fm_1, \fm_2) \in \BZ^2} \, \sum_{\fn \in \BZ} \left( \prod_{j=1}^2 \oint \frac{d v_j}{2 \pi i v_j}  \, v_j^{2k \fm_j}  \right)   \zeta^{\fm_1+\fm_2} \oint \frac{d u}{2 \pi i u}\, u^{-2k n} \times \\
& \quad  \CZ^{\SO(4)}_{\text{vec}}(v_1,v_2; \fm_1, \fm_2; x)  \CZ^{\USp(2)}_{\text{vec}}(u; \fn; x)  \times \\
& \quad \prod_{i=1}^2 \,  \prod_{s_1, s_2 = \pm 1}  \CZ^{1/2}_{\chi} \left( v_i^{s_1} u^{s_2} a; s_1 \fm_i+{s_2} \fn; x\right) ~,
}
where $\zeta$ is the fugacity for the zero-form magnetic symmetry such that $\zeta^2=1$, and the contribution of the $\SO(4)$ vector multiplet is
\bes{
\CZ^{\SO(4)}_{\text{vec}}(v_1,v_2; \fm_1, \fm_2; x)  &= x^{-|\fm_1-\fm_2|-|\fm_1+\fm_2|} \times \\
& \qquad  \prod_{s_1, s_2 = \pm 1} \left( 1- (-1)^{s_1 \fm_1 +s_2 \fm_2} x^{|s_1 \fm_1+s_2 \fm_2|} v_1^{s_1} v_2^{s_2}\right)~.
} 
For simplicity, we have set the fugacity $\chi$ for the charge conjugation symmetry to unity. Theory \eref{USp2Spin4} can be obtained by gauging the magnetic symmetry; its index reads
\bes{ \label{indUSp2Spin4}
\CI_{\eref{USp2Spin4}}(a; x)  = \frac{1}{2} \left[\CI_{\eref{USp2SO4}}(\zeta=1, a;x) + \CI_{\eref{USp2SO4}}(\zeta=-1, a;x) \right]~.
}
It can be checked that \eref{indexsingleT2} and \eref{indUSp2Spin4} yield the same result for $(k_1, k_2, k_3) = (-k,2k,2k)$.  In fact, the gauge fugacities and magnetic fluxes in \eref{indexsingleT2} and \eref{indUSp2SO4} can be mapped to each other as follows:
\bes{
\begin{array}{lll}
{z_1}={u}~, &\qquad {z_2^{2}}={v_1}{v_2}~, & \qquad {z_3^{2}}={v_1}{v_2^{-1}}~, \\
{m_1}={\mathfrak{n}}~, &  \qquad {2}{m_2}={\mathfrak{m}_1}+{\mathfrak{m}_2}~, &\qquad  {2}{m_3}={\mathfrak{m}_1}-{\mathfrak{m}_2}~.
\end{array}
} 
The indices up to order $x^4$ are as follows.
\bes{ \label{indmk2k2k}
\begin{tabular}{|c|c|}
\hline
$k$ & Index \\
\hline
$1$, $2$ & diverges \\
$3$& $1 +0x + (2a^4 -1) x^2 + \left(a^6-a^2+a^{-2}\right) x^3+\left(3 a^8-a^4-2\right) x^4 + \ldots$ \\
$4$ & $1+0x + (a^4 -1) x^2 + a^{-2} x^3 + (2a^8-2) x^4 + \ldots$ \\
$\geq\, 5$ & $1+0x + (a^4 -1) x^2 + a^{-2} x^3 + (a^8-2) x^4 + \ldots$ \\
\hline
\end{tabular}
}
\noindent Note that, for $k \geq 5$, the indices for these cases differ from each other at a higher order than $x^4$.  For $k=1, \,2$, the index diverges and the theories are {\it bad} in the sense of Gaiotto and Witten \cite{Gaiotto:2008ak}.

Let us now analyse the superconformal multiplets and enhancement of supersymmetry for the cases of $k \geq 3$ using information from \cite{Cordova:2016emh, Razamat:2016gzx, Evtikhiev:2017heo} (see also the argument in \cite{Garozzo:2019ejm, Beratto:2020qyk}). The vanishing coefficient of $x$ implies that there is no 3d $\CN=3$ flavour current multiplet $B_1[0]_1^{(2)}$. In general, the negative terms at order $x^2$ receive the contribution from the $\CN=3$ flavour current multiplet $B_1[0]_1^{(2)}$ and $\CN=3$ extra-SUSY current multiplet $A_2[0]_1^{(0)}$.  Since the former is absent and the only negative term at order $x^2$ is $-1$, we conclude that there is precisely one $\CN=3$ extra SUSY-current multiplet that leads to the enhanced $\CN=4$ supersymmetry in the IR.

As can be read off from the index, the bare monopole operator with magnetic fluxes $(m_1, m_2, m_3)$ has dimension 
\bes{ \label{foot:chargesbare}
\Delta(m_1, m_2, m_3)=  \frac{1}{4} \sum_{s_1, s_2, s_3 =\pm 1} \left|  \sum_{i=1}^3 s_i m_i \right| -\sum_{i=1}^3 2|m_i| ~,
}
its charge under the $\U(1)_a$ flavour symmetry is 
\bes{
a(m_1, m_2, m_3) = -\frac{1}{2} \sum_{s_1, s_2, s_3 =\pm 1} \left|  \sum_{i=1}^3 s_i m_i \right|}
and it carries charge $2 k_i m_i$ under the Cartan subalgebra of each $\SU(2)_i$ gauge factor. 

Viewing the theory in question from the 3d $\CN=2$ perspective, the $\U(1)_a$ flavour current multiplet also contributes $-1$ at order $x^2$ in the index.  As a consequence the aforementioned $\CN=3$ extra SUSY-current should be identified with that of the flavour symmetry. Furthermore, we see that there is no relevant operator due to the vanishing coefficient of $x$.  For $k \geq 4$, there is one marginal operator corresponding to $\tr(\mu_1^2)=\tr(\mu_2^2)= \tr(\mu_3^2)$, where $\mu_i$ is the moment map of the $\SU(2)_i$ flavour symmetry of the $T_2$ theory.  For $k=3$, we have an extra marginal operator which, according to \eref{indexsingleT2}, receives the contribution from the eight gauge magnetic fluxes $(m_1, m_2, m_3) = (\pm 2, \pm 1, \pm 1)$, where $\pm$ here denotes all possible 8 sign combinations that can appear.  Since the bare (non-gauge-invariant) monopole operators with such fluxes contribute $z_1^{\mp 12} z_2^{\pm 12} z_3^{\pm 12}$, $x^{-4}$ and $a^{-8}$ to the index, we interpret the aforementioned marginal operator as a dressed monopole operator, where such a bare monopole operator is dressed in a gauge invariant way with a combination of $12$ chiral fields of the $T_2$ theory.

\subsubsection*{One-form symmetries and gauging thereof}
We now examine the one-form symmetry of \eref{singleT2} with $(k_1, k_2, k_3) = (-k, 2k, 2k)$. From the discussion in Section \ref{sec:oneformanomsingleT2}, we see that
\bes{ \label{anommk2k2k}
\scalebox{0.8}{
\begin{tabular}{|c|c|c|c|c|}
\hline
CS levels & Anomaly  & Non-anomalous  & Anomalous  & TQFT with \\
$(-k, 2k, 2k)$ & matrix $p$ & 1-form symmetry & 1-form symmetry & anom. symmetry  \\
\hline
$k$ even &  $\begin{pmatrix} 0 & 0 \\ 0 & 0 \end{pmatrix}$ & $\BZ_2^2$ & $\mathbf{1}$ &   $-$ \\             
\hline
$k$ odd & $\begin{pmatrix} 1 & 0 \\ 0 & 0 \end{pmatrix}$ & $\BZ_2$ & $\BZ_2$ & $\CA^{2, 1}  \cong \SU(2)_1\cong \U(1)_2 $\\
\hline  
\end{tabular}}
}   
For $k$ even, theory \eref{singleT2mk2k2k} flows to an $\CN=4$ SCFT with a non-anomalous $\BZ_2^2$ one-form symmetry.  However, for $k$ odd, theory \eref{singleT2mk2k2k} flows to an $\CN=4$ SCFT with a non-anomalous $\BZ_2$ one-form symmetry and a decoupled topological sector $\CA^{2, 1}$.


We can understand the above statement from another point of view. A slight modification of \cite[(3.27)]{Tachikawa:2019dvq} (see also \cite[(3.18), (3.19)]{Tachikawa:2019dvq}) states that the $\USp(2N)_{-k} \times \SO(2M)_{2k}$ gauge theory with bifundamental matter admits a quotient by a $\BZ_2$ symmetry, whose generator is a combination of the $\BZ_2$ centres of $\SO(2M)$ and $\USp(2N)$, if\footnote{More generally, for the $\USp(2N)_{k_2} \times \SO(2M)_{k_1}$ gauge theory with bifundamental matter, this condition reads $\frac{1}{4}k_1 M + \frac{1}{2} k_2 N \in \BZ$. \label{condnonanomZ2}}
\bes{ \label{condMN}
\frac{1}{2}k(M-N) \in \BZ~.
}
In other words, the $\BZ_2$ one-form symmetry of the $\USp(2N)_{-k} \times \SO(2M)_{2k}$ gauge theory is non-anomalous if \eref{condMN} is satisfied. Applying this to \eref{USp2SO4}, namely $M=2$ and $N=1$, we see that its non-anomalous one-form symmetry is
\bes{ \label{oneformUSp2SO4}
\text{one-form symmetry of}~ \eref{USp2SO4} = \begin{cases}  \BZ_2 &\quad \text{$k$ even} \\  \text{trivial} &\quad \text{$k$ odd} \end{cases}~.
}
Recall that theory \eref{USp2Spin4}, or equivalently theory \eref{singleT2mk2k2k}, arises from gauging the $\BZ_2$ zero-form magnetic symmetry of \eref{USp2SO4}.  Since gauging a discrete $\BZ_2$ zero-form symmetry in 3d leads to a dual $\BZ_2$ one-form symmetry, we conclude that for $k$ odd theory \eref{USp2Spin4} = \eref{singleT2mk2k2k} has a $\BZ_2$ one-form symmetry.  For $k$ even, the one-form symmetry of theory \eref{USp2Spin4} = \eref{singleT2mk2k2k} can be either $\BZ_2 \times \BZ_2$ or its extension $\BZ_4$.  Note that the extension is formed if theory \eref{USp2SO4}, with $k$ even, has a mixed anomaly between the $\BZ_2$ zero-form magnetic symmetry and the $\BZ_2$ one-form symmetry \cite{Tachikawa:2017gyf}.  Subsequently, we will explicitly show that there is no extension of the symmetry to $\BZ_4$.

For this purpose, let us gauge the {\it whole} one-form symmetry of theory \eref{USp2Spin4}, \ie~ we turn it into a dual zero-form symmetry.  This is equivalent to gauging the $\BZ_2$ one-form symmetry in \eref{USp2SO4}, namely considering the $[\USp(2)_{-k} \times \SO(4)_{2k}]/\BZ_2$ theory.  The index of the latter can be computed as in \eref{indUSp2SO4} but with the inclusion of the summation over half-odd-integral fluxes; in particular, we modify the summation in \eref{indUSp2SO4} as follows:
\bes{
\sum_{(\fm_1, \fm_2) \in \BZ^2} \, \sum_{\fn \in \BZ}  \quad \longrightarrow \quad \sum_{p'=0}^1 s^{p'} \sum_{(\fm_1, \fm_2) \in \left(\BZ+\frac{p'}{2} \right)^2} \, \sum_{\fn \in \BZ+ \frac{p'}{2} }~, 
}
where $s$ is a fugacity for the $\BZ_2$ zero-form symmetry arising from gauging the one-form symmetry such that $s^2=1$.  For $k$ odd, we see that the half-odd-integral fluxes (\ie those correspond to $p'=1$) do not contribute to the index; in other words, the index of the $[\USp(2)_{-k} \times \SO(4)_{2k}]/\BZ_2$ theory is equal to that of the $\USp(2)_{-k} \times \SO(4)_{2k}$ theory. This is in agreement with the proposal that the one-form symmetry of \eref{USp2SO4} for $k$ odd has a non-trivial 't Hooft anomaly.\footnote{One can also check, in the same way as in \cite{Mekareeya:2022spm, Sacchi:2023omn}, that for $k$ odd the integrand of the index \eref{indUSp2SO4} contains fractional powers of $\SO(4)$ gauge fugacities for the magnetic fluxes $(\fn, \fm_1, \fm_2) = (1/2, 1/2,1/2)$.}  Let us focus on $k$ even.  We see that for $p'=0$ the contribution from $\zeta^{\fm_1+\fm_2}$ is either 1 or $\zeta$, whereas for $p'=1$ we have either $s$ or $s\zeta$.  The elements of $\{1, \zeta, s, s \zeta\}$ form the $\BZ_2 \times \BZ_2$ zero-from symmetry.  Observe that the $\BZ_2$ zero-form symmetry associated with $s$ and the $\BZ_2$ zero-form magnetic symmetry associated with $\zeta$ do not form an extension to $\BZ_4$, since there is no element of order $4$.

We can also see this from the perspective of theory \eref{singleT2mk2k2k}.  In order to gauge the whole one-form symmetry, we can do it in two steps. First, gauge the diagonal $\BZ_2$ one-form symmetry, whose generator is a combination of the $\BZ_2$ centres of $\SU(2)_{2k}$ and $\SU(2)_{2k}$; in other words, we consider the quotient $\SU(2)_{-k} \times [\SU(2)_{2k} \times \SU(2)_{2k}]/\BZ_2$.  The latter is equivalent to \eref{USp2SO4}, namely the $\USp(2)_{-k}  \times \SO(4)_{2k}$ gauge theory, since $\SO(4) \cong [\SU(2)\times \SU(2)]/\BZ_2$. We report the indices up to order $x^4$ below.
\bes{ \label{indmk2k2kmodZ2zeta}
\scalebox{0.95}{
\begin{tabular}{|c|c|}
\hline
$k$ & Index \\
\hline
$1$, $2$ & diverges \\
\hline
$3$& $1 +\zeta a^2 x + \left[(2+\zeta)a^4 -(1+\zeta)\right] x^2 +\left[(1+2 \zeta)a^6-(1+\zeta)a^2+a^{-2}\right] x^3+$\\ & $\left[(3+2 \zeta) a^8-(1+\zeta)a^4-2\right] x^4 + \ldots$ \\
\hline
$4$ & $1 +0x + \left[(1+\zeta)a^4 -1\right] x^2 +\left[\zeta (a^6-a^2)+a^{-2}\right] x^3+$\\ & $\left[(2+\zeta) a^8-\zeta a^4-2\right] x^4 + \ldots$ \\
\hline
$5$ & $1 +0x + \left(a^4 -1\right) x^2 +\left(\zeta a^6+a^{-2}\right) x^3+\left[(1+\zeta) a^8-\zeta a^4-2\right] x^4 + \ldots$ \\
\hline
$\geq\, 6$ & $1+0x + (a^4 -1) x^2 + a^{-2} x^3 + \left[(1+\zeta) a^8-2\right] x^4 + \ldots$ \\
\hline
\end{tabular}}
}
For $k\geq 6$, the terms with fugacity $\zeta$ appear at a higher order than $x^4$. 

The second step is to gauge the remaining of $\BZ_2$ one-form symmetry, \ie~ we consider the further quotient $[\SU(2)_{-k} \times [\SU(2)_{2k} \times \SU(2)_{2k}]/\BZ_2]/\BZ_2$.  The index of the latter can be obtained from \eref{indexsingleT2} by replacing the summation as
\bes{ \label{replacesumT2}
\sum_{(m_1, m_2, m_3) \in \BZ^3} \quad \longrightarrow \quad    \sum_{p'=0}^1 s^{p'} \sum_{m_1 \in \left( \BZ+\frac{p'}{2} \right) } \,\,\, \sum_{p=0}^1 \zeta^p \sum_{(m_2,m_3) \in \left(\BZ+\frac{p}{2}  \right) \times  \left( \BZ+\frac{p}{2} +\frac{p'}{2} \right)}~,
}
where $\zeta$ is the fugacity of the $\BZ_2$ zero-form symmetry arising from the first step and $s$ is that arising from the second step.\footnote{Observe that we have four mutually exclusive cases, namely (1) $p=0$ and $p'=0$: $m_1, m_2, m_3$ are integral and we have $\zeta^0 s^0=1$; (2) $p=1$ and $p'=0$: $m_1$ is integral, $m_2, m_3$ are half-odd-integral and we have $\zeta^1 s^0=\zeta$; (3) $p=0$ and $p'=1$: $m_1$ is half-odd-integral, $m_2$ is integral, $m_3$ is half-odd-integral and we have $\zeta^0 s^1=s$;  (4) $p=1$ and $p'=1$: $m_1$ is half-odd-integral, $m_2$ is half-odd-integral, $m_3$ is integral and we have $\zeta s$.}  We deliberately used the same notation as those for the $[\USp(2)_{-k} \times \SO(4)_{2k}]/\BZ_2$ theory since they can be identified with each other.  As before, the elements of $\{1, \zeta, s, s \zeta\}$ form the $\BZ_2 \times \BZ_2$ zero-from symmetry.  Moreover, it is clear from \eref{replacesumT2} that the order of gauging of the $\BZ_2$ one-form symmetry in each of the two steps is immaterial; this confirms that the one-form symmetry of \eref{singleT2mk2k2k} is indeed $\BZ_2 \times \BZ_2$ for $k$ even.  

For reference, we report the index for the $[\USp(2)_{-k} \times \SO(4)_{2k}]/\BZ_2$ theory or the $[\SU(2)_{-k} \times [\SU(2)_{2k} \times \SU(2)_{2k}]/\BZ_2]/\BZ_2$ theory, with $k=4$, up to order $x^9$ as follows:
\bes{ \label{gaugefurtherZ2}
1 +& (-1 + a^4 + a^4 \zeta) x^2 + (a^{-2} - a^2 \zeta + a^6 \zeta) x^3  +  (-2 + 2 a^8 - a^4 \zeta + a^8 \zeta) x^4 +\\
& (-a^6 + a^{10} + a^{10} \zeta) x^5 + (1 + a^{-4} - a^4 - a^8 + 2 a^{12} - 2 a^4 \zeta +  2 a^{12} \zeta) x^6 +\\
& (-2 a^{-2} - a^6 + a^{14} + 2 a^2 \zeta - 2 a^6 \zeta - a^{10} \zeta + 2 a^{14} \zeta) x^7 +\\
&  (-a^4 - 2 a^8 + 3 a^{16} - s + a^4 s - \zeta + a^4 \zeta - 
    a^{12} \zeta + 2 a^{16} \zeta - s \zeta + 
    a^4 s \zeta) x^8 +\\
& (a^{-6} + 4 a^2 + 2 a^6 - a^{10} - a^{14} + 2 a^{18} + \\
& \   2 a^{-2} s - 2 a^2 s - 5 a^6 \zeta - a^{10} \zeta + 
    2 a^{18} \zeta + 2 a^{-2} s \zeta - 2 a^2 s \zeta) x^9 + \ldots~.
}    
On the other hand, for $k$ odd, we find that the magnetic fluxes in the sector $p'=1$ contribute zero to the index, and so the fugacity $s$ does not appear.

\subsubsection*{Higgs and Coulomb branches}
Let us now explore the Higgs and Coulomb branches of the IR 3d $\CN=4$ SCFT associated with this class of theories.  We can take the Higgs and Coulomb branch limit of the index in a similar way as in \cite{Razamat:2014pta} to obtain the Higgs and Coulomb branch Hilbert series as follows.  We define
\bes{ \label{CBHBlimits}
&h = x a^2~, ~ c = x a^{-2}~, \\ 
&\text{or equivalently} \quad  x = (h c)^{1/2} ~,~ a = (h/c)^{1/4}
}
and substitute them in the index \eref{indexsingleT2}.  In the Higgs branch limit we send $c \rightarrow 0$ and keep $h$ fixed, whereas in the Coulomb branch limit we send $h \rightarrow 0$ and keep $c$ fixed.

Let us now apply this to \eref{singleT2} with CS levels $(-k,2k,2k)$.  We obtain the Higgs and Coulomb branch limits of \eref{indexsingleT2} to be
\bes{
\text{Higgs limit $\eref{singleT2}_{(-k,2k,2k)}$:} &\qquad  \PE \left[ h^2 + h^{2k-4} + h^{2k-3} - h^{4k-6} \right]~, \\
\text{Coulomb limit $\eref{singleT2}_{(-k,2k,2k)}$:} &\qquad  1~.
}
The former is indeed the Hilbert series of $\BC^2/\hat{D}_{2k-2}$ \cite[(5.31)]{Benvenuti:2006qr}.  These indicate that Higgs and Coulomb branches of the IR $\CN=4$ interacting SCFT associated with \eref{singleT2} with CS levels ${(-k,2k,2k)}$ are $\BC^2/\hat{D}_{2k-2}$ and trivial, respectively.  The generators of the Higgs branch, corresponding to $h^2$, $h^{2k-4}$ and $h^{2k-3}$, are respectively
\bes{
w= \tr (\mu^2)~, \quad v= X_{(2,1,1)} Q^{4k}~, \quad u = X_{(2,1,1)} Q^{4k-4} \mu_1 \mu_2  \mu_3~,
}
satisfying the relation
\bes{ \label{relDsing1}
u^2 + v^2 w = w^{2k-3}~. 
}
Here $X_{(2,1,1)}$ denotes the bare monopole operator of flux $(2, 1, 1)$, where it has dimension $-4$ and carries the flavour charge $-8$ as well as gauge charges $(-4k, 4k,  4k)$ under the Cartan subalgebras of each $SU(2)_i$; see around \eref{foot:chargesbare}.  In the above $v= X_{(2,1,1)} Q^{4k}$ denotes the gauge invariant dressed monopole operator, where the bare monopoles with fluxes $(\pm 2, \pm 1, \pm 1)$ are dressed with appropriate combinations of $4k$ chiral multiplets $Q$ of the $T_2$ theory.  Note that $X_{(2,1,1)}$ contains $4k$ gauge indices of each $\SU(2)_i$ and these are contracted with the gauge indices in $Q^{4k}$ to form $v$.  Similarly, for $u$, the gauge indices of $X_{(2,1,1)}$ are contracted with those of $Q^{4k-4}$ as well as $\mu_1 \mu_2  \mu_3$, and the remaining indices are contracted with epsilon tensors.

Let us revisit the special case of $k=2$, \ie~ the CS levels $(-2,4,4)$. Although the index diverges, the above computation shows that the Higgs branch is $\BC^2/\hat{D}_{2}$.  This is reminiscent of the Higgs branch of the 3d $\CN=4$ $\SU(2)$ gauge theory with $2$ hypermultiplets in the fundamental representation.  The latter is the union of two isomorphic hyperK\"ahler cones, each described by $\BC^2/\BZ_2$ \cite{Seiberg:1996nz, Gaiotto:2008ak, Ferlito:2016grh, Assel:2018exy}.  We believe that the Higgs branch of the case of $k=2$ has the same structure.

We can also study of the Higgs and Coulomb branch limits of the $\USp(2)_{-k} \times \SO(4)_{2k} $ gauge theory, or equivalently $\SU(2)_{-k} \times  [\SU(2)_{2k} \times \SU(2)_{2k}]/\BZ_2$, which comes from gauging a non-anomalous $\BZ_2$ subgroup of $\BZ_2^2$ one-form symmetry of the aforementioned theory.   We find that the Higgs and Coulomb branch limits are\footnote{Throughout the paper, we put superscript $[1]$ whenever we would like to emphasise that the corresponding symmetry is one-form.}
\bes{
\text{Higgs limit $\eref{singleT2}_{(-k,2k,2k)}/\BZ_2^{[1]}$:} &\qquad  \PE \left[ h^2 + h^{k-2} + h^{k-1} - h^{2k-2} \right]~, \\
\text{Coulomb limit $\eref{singleT2}_{(-k,2k,2k)}/\BZ_2^{[1]}$:} &\qquad  1~.
}
The former is indeed the Hilbert series of $\BC^2/\hat{D}_{k}$ \cite[(5.31)]{Benvenuti:2006qr}.   This means that the Higgs branch is $\BC^2/\hat{D}_{k}$, and the Coulomb branch is trivial. The generators of the Higgs branch, corresponding to $h^2$, $h^{k-2}$ and $h^{k-1}$, are respectively
\bes{
w= \tr (\mu^2)~, \quad v= X_{\left(1,\frac{1}{2},\frac{1}{2} \right)} Q^{2k}~, \quad u = X_{\left(1,\frac{1}{2},\frac{1}{2}\right)} Q^{2k-4} \mu_1 \mu_2  \mu_3~,
}
satisfying the relation
\bes{
u^2 + v^2 w = w^{k-1}~. 
}
Here $X_{\left(1,\frac{1}{2},\frac{1}{2} \right)}$ denotes the bare monopole operator of flux $\left(1,\frac{1}{2},\frac{1}{2} \right)$, where it has dimension $-2$ and carries the flavour charge $-4$ as well as gauge charges $(-2k, 2k,  2k)$ under the Cartan subalgebras of each $SU(2)_i$.  The notations and contractions of the gauge indices are as described above.

\subsubsection{General results for  $(k_1, k_2, k_3) = k(\fp \fq, -\fp \fr, -\fq \fr )$ with $\fr = \fp+\fq$}
As pointed out in \cite[Footnote 7]{Assel:2022row}, the ATT condition \eref{ATT} admits the general solution of the form
\bes{ \label{genCS}
&k_1 = \fp \fq k~, \quad k_2 = - \fp \fr k~, \quad k_3 = - \fq \fr k~, \\
& \text{with $\fr = \fp+\fq$ and $\fp, \, \fq, \,  \fr, \, k \in \BZ_{\neq 0}$}~.
}
For simplicity, we will consider the cases of $\fp>0$, $\fq >0$ and $k>0$.  The index \eref{indexsingleT2} receives non-trivial contributions from gauge fluxes $(0,0,0)$ and $(\pm \fr, \pm \fq, \pm \fp) n$ with $n \in \BZ_{\geq 1}$.  The contribution from flux $(0,0,0)$, up to order $x^4$, reads 
\bes{ \label{index000singleT2}
1+0x + (a^4 -1) x^2 + a^{-2} x^3 + (a^8-2) x^4 + \ldots~.
}
The contributions from the eight fluxes $(\pm \fr, \pm \fq, \pm \fp) n$ correspond to the gauge-invariant dressed monopole operators. According to the discussion around \eref{foot:chargesbare}, it follows that the bare monopole operators associated with these fluxes have dimension $-2 \fr n$. The charge under the flavour symmetry is $-4 \fr n$, and the charges under the Cartan subalgebra of each $\SU(2)_i$ are $2k \fp \fq \fr n(\pm 1, \mp 1, \mp 1 )$.  This bare monopole can be dressed with a combination of $2k \fp \fq \fr n$ chiral fields of the $T_2$ theory to form gauge invariant quantities.  As a consequence, such gauge-invariant dressed monopole operators have dimension $(-2  + k \fp \fq ) \fr n$ and the charge under the flavour symmetry $(-4 + 2 k \fp \fq) \fr n$.  Indeed, if $(-2  + k \fp \fq ) \fr n$ is sufficiently large, the index \eref{index000singleT2} at sufficiently low order in $x$ does not get affected by these dressed monopole operators.  In any case, using the same argument as above, we see from \eref{index000singleT2} that the SCFT in the IR has enhanced $\CN=4$ supersymmetry, regardless of the contribution of the dressed monopole operators.

In the previous example of $(-k, 2k, 2k)$, corresponding to $\fp = 1$, $\fq=1$ and $\fr=2$, we see that the dimension of such gauge-invariant dressed monopole operators is $(2k-4)n$ and the charge under flavour symmetry is $(4k-8)n$.  Indeed, for $k=3$, in addition to \eref{index000singleT2}, we have the terms $a^4 x^2$ and $a^8 x^4$ coming from $n=1$ and $n=2$ (up to order $x^4$); as reported in \eref{indmk2k2k}.  Similarly for $k=4$, we have an additional term $a^8 x^4$ coming from $n=1$; see \eref{indmk2k2k}.  For $k=1$ and $k=2$, the dimensions are negative (or zero) and this is why the indices diverge; as reported in \eref{indmk2k2k}.


\subsubsection*{One-form symmetries}

Let us now discuss the one-form symmetries as well as the topological sector in the IR.  From the discussion in Section \ref{sec:oneformanomsingleT2}, we see that if $k$ is even, then all of the CS levels $(k_1, k_2, k_3)$ are even and it follows that the anomaly action \eref{Sanom1} is an integral multiple of $2 \pi$, which means that the corresponding anomaly theory is trivial.  We thus turn to the case in which $k$ is odd, in which case the anomaly theory is the same as that for $k=1$.  We summarise the information in the table below.
\bes{ \label{oneformpqrt}
\scalebox{0.77}{
\begin{tabular}{|c|c|c|c|c|}
\hline
CS levels & Anomaly  & Non-anomalous  & Anomalous  & TQFT with \\
$(\fp \fq, -\fp \fr, -\fq \fr )$ & matrix $p$ \eref{anompmatrix} & 1-form symmetry & 1-form symmetry & anom. symmetry  \\
\hline
Both $\fp, \, \fq$ even &  $\begin{pmatrix} 0 & 0 \\ 0 & 0\end{pmatrix}$ & $\BZ^2_2$ & $\mathbf{1}$ & $-$ \\             
\hline
$\fp$ even, $\fq$ odd & $\begin{pmatrix} 1 & 1 \\ 1 & 1 \end{pmatrix}$ & $\BZ_2$ & $\BZ_2$ & $\CA^{2, 1}  \cong \SU(2)_1\cong \U(1)_2 $\\
\hline  
$\fp$ odd, $\fq$ even & $\begin{pmatrix} 0 & 0 \\ 0 & 1 \end{pmatrix}$ & $\BZ_2$ & $\BZ_2$ & $\CA^{2, 1}  \cong \SU(2)_1\cong \U(1)_2 $\\
\hline
Both $\fp, \, \fq$ odd & $\begin{pmatrix} 1 & 0 \\ 0 & 0 \end{pmatrix}$ & $\BZ_2$ & $\BZ_2$ & $\CA^{2, 1}  \cong \SU(2)_1\cong \U(1)_2 $\\
\hline
\end{tabular}}
}   
where the third and fourth columns are given by $\BZ_2^{\dim(\ker\, p)}$ and $\BZ_2^{\rank(p)}$ respectively.

We thus conclude that if both $\fp$ and $\fq$ are even, the theory flows to an IR $\CN=4$ SCFT with a non-anomalous $\BZ^2_2$ one-form symmetry with no decoupled topological sector; otherwise, it flows to an IR $\CN=4$ SCFT with a non-anomalous $\BZ_2$ one-form symmetry with the decoupled $\CA^{2, 1}$ TQFT.

\subsubsection{Higgs and Coulomb branches} \label{sec:HBCBoneT2}
We can compute the Coulomb and Higgs branch limits of \eref{indexsingleT2} as in \eref{CBHBlimits}.  The result is as follows:
\bes{
\text{Higgs limit:} &\qquad  \PE \left[ h^2 + h^{K} + h^{K+1} - h^{2K+2} \right]~, \quad K= \fp \fq \fr k-2 \fr~, \\
\text{Coulomb limit:} &\qquad  1~.
}
These indicate that the Higgs branch is $\BC^2/\hat{D}_{K+2}$, and the Coulomb branch is a point.  The generators of the Higgs branch, corresponding to $h^2$, $h^{K}$ and $h^{K+1}$, are respectively
\bes{
w= \tr (\mu^2)~, \quad v= X_{(\fr, \fq, \fp) } Q^{2\fp \fq \fr k}~, \quad u = X_{(\fr, \fq, \fp) } Q^{2\fp \fq \fr k-4} \mu_1 \mu_2  \mu_3~,
}
satisfying the relation
\bes{
u^2 + v^2 w = w^{K+1}~. 
}
In the above, $X_{(\fr, \fq, \fp) }$ denotes the bare monopole operator of flux $(\fr, \fq, \fp) $, where it has dimension $-2 \fr$ and carries the flavour charge $-4 \fr$ as well as the gauge charges $2\fp \fq \fr k(1,-1,-1)$ under the Cartan subalgebras of each $\SU(2)_i$.  The notations and contractions of the gauge indices are as described below \eref{relDsing1}.

\subsection{Cases that do not satisfy the ATT condition} \label{sec:nonATToneT2}
In this subsection, we focus on \eref{singleT2} whose CS levels do not satisfy the ATT condition \eref{ATT}.  As can be seen from the effective superpotential \eref{sup1}, the flavour symmetry associated with the fugacity $a$ is explicitly broken, and the marginal operator $\tr(\mu^2)$ is set to zero in the chiral ring. 

We will consider three interesting families of theories arising from M5-branes compactified on the quotient of the three-sphere $S^3/\Gamma$, where $\Gamma$ is a finite subgroup of $\SU(2)$.  The CS levels for each of these families are as follows \cite[Section 5.4.3]{Eckhard:2019jgg}:
\bi
\item  Lens space\footnote{Recall that the Lens space $L(p,q)$ can be viewed as the quotient space $S^3/\BZ_p$ with the identification $(z_1, z_2) \sim (e^{2 \pi i/p} z_1, e^{2 \pi i q/p} z_2)$.} $L(p,q)$ with\footnote{According to \cite[Theorem 2.5]{Hatcher:3manifolds}, the Lens spaces $L(p,q)$
 and $L(p',q')$ are diffeomorphic if and only if $p' = p$ and $q' = \pm q^{\pm 1} \, (\mod \, p)$.  \label{homeolens}} $\frac{p}{q} = (k_1+1)-\frac{1}{k_2+1}$ or $(k_2+1)-\frac{1}{k_1+1}$.  Explicitly, we take
\bes{ \label{pqdef}
&\text{$p= |k_1k_2+k_1+k_2|$ and $q= \pm (k_1+1)$ or $ \pm(k_2+1)$~.}
}  
This corresponds to the CS levels $(k_1, k_2,1)$.
\item $S^3/D_n$: This corresponds to the CS levels $(-2,2, n-2)$.
\item $S^3/E_m$: This corresponds to the CS levels $(-2,3,m-3)$.
\ei
We summarise the results in each case below.
\ben
\item \label{caseLens} Let us consider the CS levels $(k_1, k_2, 1)$. We have three cases as follows:
\bi
\item If $p = 1$, \ie~ the Lens space is diffeomorphic to the three-sphere, namely $L(p = 1, q) \cong S^3$, then the index vanishes and IR theory is trivial. 
\item If $p \neq 1$ and both choices of $q$ in \eref{pqdef} satisfy either of the following conditions:\footnote{These conditions lead to the following identifications: $(z_1, z_2) \sim (e^{2 \pi i/p} z_1, e^{ \pm 2 \pi i /p} z_2)$ or $(z_1, z_2) \sim (e^{2 \pi i/p} z_1, z_2)$, with $p\neq 1$.}
\bes{ \label{poverq}
&\text{One of the $q$ is $\pm 1$ $(\mod \, p)$ and the other is divisible by $p$,} \\ 
&\text{or both choices of $q$ are $\pm 1$ $(\mod \, p)$}~,
}
then the index $\eref{indexsingleT2}_{a=1}$ is equal to unity, and so theory \eref{singleT2} flows to a TQFT.  
\item Otherwise, the index $\eref{indexsingleT2}_{a=1}$ takes the form
\bes{
1+0 x - x^2+2 x^3+\ldots~,
}
where $0 x$ indicates that there is no $\CN=3$ flavour current and $-x^2$ indicates that there is one $\CN=3$ extra SUSY-current; therefore, each theory in this subclass flows to a {\bf 3d $\CN=4$ interacting SCFT}, where supersymmetry gets {\bf enhanced} in the IR, with a decoupled TQFT.
\ei

Using the information in Section \ref{sec:oneformanomsingleT2}, we summarise the information about the anomalies and TQFTs below.\footnote{As before, the third and fourth columns are given by $\BZ_2^{\dim(\ker\, p)}$ and $\BZ_2^{\rank(p)}$ respectively.  These deserve some explanations. Let us consider, for example, the second row, \ie~ that with $(\CZ_2)_2$ TQFT. The matrix element $1$ in the first row and first column means the first $\BZ_2$ one-form symmetry is individually anomalous, in the sense that $\pi (k_1+1) \int_{\CM_4} \CP(w_1^{(2)})/2 \equiv \pi \int_{\CM_4} \CP(w_1^{(2)})/2$ in \eref{Sanom1} is non-trivial.  The matrix element $0$ means that the second $\BZ_2$ one-form symmetry is individually non-anomalous. The off-diagonal element 1 means that there is a mixed anomaly between the first and second $\BZ_2$ symmetries. If one takes the diagonal subgroup of these two $\BZ_2$, namely setting $w_1^{(2)} = w_2^{(2)} \equiv B^{(2)}$, it is anomalous due to the non-trivial term $\pi (k_1+1)  \int_{\CM_4}  \CP(B^{(2)})/2 - 2 \pi \int_{\CM_4}  \CP(B^{(2)})/2 = \pi (k_1-1)  \int_{\CM_4}  \CP(B^{(2)})/2$ in \eref{Sanom1}. We will also discuss this below \eref{condZ211k}.  Here, $\ker(p) = \emptyset$ means that any two $\BZ_2$ symmetries associated with linearly independent combinations of $w_1^{(2)}$ and $w_2^{(2)}$ (with the coefficients in $\BZ_2$) will be involved in an individual and/or a mixed 't Hooft anomaly. This should be contrasted with the last row: the vector $(-1, 1)^T \equiv (1,1)^T \in \ker(p)$ means that the $\BZ_2$ one-form symmetry associated with $w_1^{(2)}+w_2^{(2)}$ is free from both individual and mixed 't Hooft anomalies.}
\bes{ \label{tab:k1k21}
\scalebox{0.76}{
\begin{tabular}{|c|c|c|c|c|}
\hline
CS levels & Anomaly  & Non-anomalous  & Anomalous  & TQFT with  \\
$(k_1, k_2,1)$ & matrix $p$ & 1-form symmetry & 1-form symmetry &   anom. symmetry \\
\hline              
Both $k_1$ and   & $\begin{pmatrix} 0 & 1 \\ 1 & 0 \end{pmatrix}$ & $\mathbf{1}$ & $\BZ_2^2$ & $\CA^{\{2,2\}, p} \equiv (\CZ_2)_0$ \\
$k_2$ are odd & & & & in the notation of \cite{Hsin:2018vcg} \\
\hline
$k_1$ is even  & $\begin{pmatrix} 1 & 1 \\ 1 & 0 \end{pmatrix}$ & $\mathbf{1}$ & $\BZ_2^2$ & $\CA^{\{2,2\}, p}\equiv (\CZ_2)_2$ \\
and $k_2$ is odd & & & & \\
\hline
Both $k_1$ and   & $\begin{pmatrix} 1 & 1 \\ 1 & 1 \end{pmatrix}$ & $\BZ_2$ & $\BZ_2$ & $\CA^{2, 1}  \cong \SU(2)_1$ \\
$k_2$ are even & & & & $\cong \U(1)_2$ \\
\hline
\end{tabular}}
}
Note that, in the case of $k_1$ and $k_2$ even, theory \eref{singleT2} flows to either
\bi
\item an $\CN=4$ SCFT with a non-anomalous $\BZ_2$ one-form symmetry $\otimes$ $\CA^{2, 1}$, or
\item a TQFT with non-anomalous $\BZ_2$ one-form symmetry $\otimes$ $\CA^{2, 1}$
\ei  
depending on the value of $p/q$ as discussed above.
\item Let us consider the CS levels $(-2,2, n-2)$ with $n\geq 4$.\footnote{The case of $n=3$ was discussed in Case \ref{caseLens}.} The index $\eref{indexsingleT2}_{a=1}$ is unity. This indicates that the theory flows to a TQFT in the IR. 
\bes{
\scalebox{0.8}{
\begin{tabular}{|c|c|c|c|c|}
\hline
CS levels & Anomaly  & Non-anomalous  & Anomalous  & TQFT with \\
                & matrix $p$ & 1-form symmetry & 1-form symmetry & anom. symmetry  \\
\hline
$n$ even &  $\begin{pmatrix} 0 & 0 \\ 0 & 0 \end{pmatrix}$ & $\BZ_2^2$ & $\mathbf{1}$ &   $-$ \\             
\hline
$n$ odd & $\begin{pmatrix} 1 & 1 \\ 1 & 1 \end{pmatrix}$ & $\BZ_2$ & $\BZ_2$ & $\CA^{2, 1}  \cong \SU(2)_1\cong \U(1)_2 $\\
\hline  
\end{tabular}}
}                     
For $n$ even, theory \eref{singleT2} flows to a TQFT that has a non-anomalous $\BZ_2^2$ one-form symmetry, whereas for $n$ odd, it flows to a TQFT that has a non-anomalous $\BZ_2$ one-form symmetry $\otimes$ $\CA^{2, 1}$.
 
\item  Let us consider the CS levels $(-2,3, m-3)$ with $m = 6,\, 7, \, 8$. The index $\eref{indexsingleT2}_{a=1}$ is unity, and so theory \eref{singleT2} flows to a TQFT in the IR.
\bes{
\scalebox{0.8}{
\begin{tabular}{|c|c|c|c|c|}
\hline
CS levels & Anomaly  & Non-anomalous  & Anomalous  & TQFT with \\
                & matrix $p$ & 1-form symmetry & 1-form symmetry & anom. symmetry  \\
\hline
$m$ even &  $\begin{pmatrix} 1 & 1 \\ 1 & 0 \end{pmatrix}$ & $\mathbf{1}$ & $\BZ_2^2$ & $\CA^{\{2,2\}, p}\equiv (\CZ_2)_2$ \\            
\hline
$m$ odd & $\begin{pmatrix} 0 & 0 \\ 0 & 1 \end{pmatrix}$ & $\BZ_2$ & $\BZ_2$ & $\CA^{2, 1}  \cong \SU(2)_1\cong \U(1)_2 $\\
\hline  
\end{tabular}}
} 
For $m$ even, theory \eref{singleT2} flows to the TQFT $\CA^{\{2,2\}, p}\equiv (\CZ_2)_2$, whereas for $m$ odd, it flows to a TQFT that has a non-anomalous $\BZ_2$ one-form symmetry $\otimes$ $\CA^{2, 1}$.     
\een

\subsubsection{Special case of $(k_1, k_2, k_3) = (k,1,1)$}
For any integer $k$, these CS levels do {\bf not} satisfy the ATT condition.\footnote{Other CS levels that leads to the same IR SCFTs in accordance with Footnote \ref{homeolens} are, for example, $(k_1, k_2, k_3) = (-k-1,1,1)$ and $(k-1,-3,1)$.  We have checked that the indices of the corresponding theories are equal. However, if $k$ is even, the topological sector for $(k,1,1)$ is $(\CZ_2)_2$, whereas that for $(-k-1,1,1)$ and $(k-1,-3,1)$ is $(\CZ_2)_0$.  On the other hand, if $k$ is odd, the situation is reverse: the topological sector for $(k,1,1)$ is $(\CZ_2)_0$ and that for $(-k-1,1,1)$ and $(k-1,-3,1)$ is $(\CZ_2)_2$.  We expect that different topological sectors (despite the diffeomorphism of the three-manifolds) arise from different choices of polarisation of the 6d $\CN=(2,0)$ theory; see also \cite{Eckhard:2019jgg}.  \label{foot:diffTQFT}} 
Let us report the indices, up to order $x^{10}$, for the cases of $k\geq 1$ below:
\bes{ \label{indicesk11}
\begin{tabular}{|c|c|}
\hline
$k$ & Index \\
\hline
$1$ & $1$ \\
$2$ & $1 - x^2 + 2 x^3 - 2 x^4 + 2 x^5 - 2 x^6 + 2 x^7 - 2 x^8 + 2 x^{10}+\ldots$ \\
$3$ & $1 - x^2 + 2 x^3 - 2 x^4 + x^5 + x^8 - 4 x^9 + 7 x^{10} +\ldots$ \\
$\geq 4$ & $1 - x^2 + 2 x^3 - 2 x^4 + x^5 - 2 x^9 + 5 x^{10} +\ldots$ \\
\hline
\end{tabular}
}
where for $k \geq 4$ the indices differ from each other at higher order than $x^{10}$.  

For $k=1$, the theory flows to the TQFT given by the first row of \eref{tab:k1k21}.  For $k \geq 2$, each of these theories flows to an {\it interacting SCFT with enhanced $\CN=4$ supersymmetry}, along with a decoupled TQFT, in the IR.\footnote{After this paper had been published in the JHEP, Ref. \cite{Gang:2023rei} appeared on the arXiv. A class of 3d $\CN=4$ rank-zero SCFTs was proposed.  Each of them (labelled by $r \in \BZ_{\geq 1}$) admits a Lagrangian description in terms of a 3d $\CN=2$ CS-matter theory with $r$ $\U(1)$  gauge groups, CS couplings between them, chiral multiplets carrying charge $+1$ under each gauge group, and a monopole superpotential. Supersymmetry of such a theory gets enhanced to $\CN=4$ in the IR.  We find that the index of the theory labelled by $r=k-1$ is equal to that of our CS-$T_2$ theory with CS levels $(k,1,1)$ listed in \eref{indicesk11}, up to a very high order in $x$. As an example, for $r=2$, the index of the theory in \cite{Gang:2023rei} is given by
\bes{ \nn
&\sum_{m_1, m_2 \in \BZ} \oint \frac{dz_1}{ 2 \pi i z_1} \oint \frac{dz_2}{ 2 \pi i z_2} z_1^{(2-\frac{1}{2}) m_1} z_2^{(4-\frac{1}{2}) m_2} z_1^{2m_2} z_2^{2m_1} v^{m_1} v^{2 m_2} \times \\
& \qquad \qquad \CZ^{R=1}_\chi (x^{\fr_1-1} z_1, m_1;x) \CZ^{R=1}_\chi (x^{\fr_2-1} z_2, m_2;x)~, \qquad \fr_1= -1, \, \fr_2=1 \\
&= 1 - x^2 + (v+ v^{-1}) x^3 - 2 x^4 + v x^5 + ( v^{-2}-1) x^6 + (v-v^{-1}) x^7 + \\
& \qquad \quad v^{-2} x^8 - (v+ 3 v^{-1}) x^9 + (4  + v^2+2v^{-2}) x^{10}+\ldots~.
}
Upon setting $v=1$, we obtain the index \eref{indicesk11} with $k=3$.  An advantage of such a Lagrangian description is that one can turn on the fugacity $v$ associated with the manifest $\U(1)$ global symmetry in the $\CN=2$ theory.  The conserved current of this symmetry plays a role as the $\CN=3$ extra SUSY-current for our CS-$T_2$ theory.  We thank Dongmin Gang for a useful discussion and for comparing the result in \cite{Gang:2023rei} with ours during the workshop ``Recent Trends in Supersymmetric Field Theories 2023''.}  The enhancement of supersymmetry from $\CN=3$ to $\CN=4$ can be deduced using the same argument as in the precedent subsection: due to the absence of the $\CN=2$ preserving marginal operator, the term $-x^2$ indicates that there is one $\CN=3$ extra SUSY-current, rendering the enhanced supersymmetry.  From \eref{tab:k1k21}, we see that the interacting SCFT does not have a non-anomalous one-form symmetry, and the decoupled TQFT has an anomalous $\BZ_2^2$ one-form symmetry whose anomaly is given by the first or second row of \eref{tab:k1k21} depending whether $k$ is odd or even.

\subsubsection*{The case of $k=2$} 
The case of $k=2$ is of particular importance: the SCFT in question turns out to be the (rank-zero) minimal 3d $\CN = 4$ SCFT, discussed in \cite{Gang:2018huc, Gang:2021hrd}.
From \eref{indexsingleT2}, the index of this theory, up to order $x^{12}$, is
\bes{
1 - x^2 + 2 x^3 - 2 x^4 + 2 x^5 - 2 x^6 + 2 x^7 - 2 x^8 + 2 x^{10}  - 
 2 x^{11} + 3 x^{12}
 +\ldots~.
}
This turns out to be equal to the index of the (rank-zero) minimal 3d $\CN = 4$ SCFT, described by 3d $\CN=2$ $\U(1)_{-3/2}$ gauge theory with one chiral multiplet of charge 1 \cite{Gang:2018huc, Gang:2021hrd}, namely
\bes{ \label{index3dN4min}
& \oint \frac{d z}{2\pi i z} \sum_{m \in \BZ} w^m z^{-\frac{3}{2} m} \CZ^{1/3}_{\chi}(z, m;x) \\
& = 1 - x^2 + (w + w^{-1}) x^3 - 2 x^4 + (w + w^{-1}) x^5 - 2 x^6 +\\
& \qquad (w + w^{-1}) x^7 - 2 x^8 + (w^2 + w^{-2}) x^{10} -  (w + w^{-1}) x^{11} +\\
& \qquad (w^2 + 1 + w^{-2}) x^{12}+ \ldots~,
}
upon setting the fugacity $w$ for the topological symmetry to $1$.  It was pointed out in \cite{Gang:2018huc} that, in the $\U(1)_{-3/2}$ CS theory, supersymmetry gets enhanced from $\CN=2$ to $\CN=4$ in the IR.  Viewing these as $\CN=2$ indices, the absence of the term at order $x$ and the positive term at order order $x^2$ implies that there is no $\CN=2$ preserving relevant and marginal deformation. As pointed out by \cite{Razamat:2016gzx, Evtikhiev:2017heo} (see also \cite{Cordova:2016emh}), at order $x^3$, there are two $\CN=2$ extra SUSY-current multiplets $A_1 \bar{A}_1[1]_{3/2}^{(0)}$ that render supersymmetry enhancement.  
Here we find a {\it new description}, in terms of the  $\CN=3$ gauge theory \eref{singleT2} with the CS levels $(2,1,1)$, of the minimal 3d $\CN = 4$ SCFT, along with the decoupled TQFT described in the second row of \eref{tab:k1k21}.  The $\CN=3$ extra SUSY-current of this theory should actually be identified with the current of the topological symmetry that is manifest in the 3d $\CN=2$ $\U(1)_{-3/2}$ CS theory.

In fact, there is another theory that has a similar behaviour: the 3d $\CN=2$ $\U(1)_0$ gauge theory with one chiral multiplet of charge $2$.  This was, in fact, mentioned in \cite[(36)]{Gang:2018huc}.  The three-sphere partition function of this theory is $Z = \int_{-\infty}^\infty \mathrm{d} s \, \Gamma_h(i r + 2s) $,\footnote{Here we use the same convention as \cite[Section 5.1]{Aharony:2013dha} and turn off the FI and mass parameters.} where $r$ is the R-charge of the chiral multiplet. It turns out that the value of $|Z|$ is independent of $r$ such that $0\leq r <1$, and we find that the free energy is $F \equiv -\log |Z| = -\log  \sqrt{\frac{5-\sqrt{5}}{10}} +\log \sqrt{2}$.  According to \cite[(3.6)]{Gang:2021hrd}, the free energy of the minimal 3d $\CN=4$ SCFT is $-\log  \sqrt{\frac{5-\sqrt{5}}{10}}$.  The other contribution comes from the TQFT $\CA^{2,1} \cong \U(1)_2$, whose free energy is given by $-\log \left | \int_{-\infty}^\infty \mathrm{d}s \, e^{2 \pi i s^2} \right | =\log \sqrt{2} $.  Hence, we conclude that the 3d $\CN=2$ $\U(1)_0$ gauge theory with one chiral multiplet of charge $2$ flows to the minimal 3d $\CN=4$ SCFT along with the $\CA^{2,1}$ TQFT.  Indeed, the index of the $\U(1)_0$ gauge theory is given by the integral $\oint \frac{d z}{2\pi i z} \sum_{m \in \BZ} w^m  \CZ^{r}_{\chi}(z^2, 2m;x)$ and the result turns out to be equal to \eref{index3dN4min}.

\subsubsection*{Description in terms of the $\USp(2)_{k} \times \Spin(4)_1$ gauge theory} 
Theory \eref{singleT2} with the CS levels $(k,1,1)$ can also be described by the 3d $\CN=3$ $\USp(2)_{k} \times \Spin(4)_1$ gauge theory with a half-hypermultiplet in the representation $[\mathbf{4}; \mathbf{2}]$.  In fact, it can be checked that the index of this theory is equal to that of the $\USp(2)_{k} \times \SO(4)_1$ gauge theory with the same matter content.  This is because in the latter the bare monopole operators cannot be dressed by the half-hypermultiplet to form a gauge invariant operator.   Since the fugacity $\zeta$ of the $\BZ_2$ zero-form magnetic symmetry does not appear in the index of the $\USp(2)_{k} \times \SO(4)_1$ gauge theory, this means it acts trivially on the local operators.  This indeed signalises that the corresponding dual one-form symmetry in the $\USp(2)_{k} \times \Spin(4)_1$ gauge theory acts on the line operators of the decoupled topological sector, in accordance with the above statement that the interacting SCFT does not have a non-anomalous $\BZ_2$ one-form symmetry.  Note that the topological sector is invisible to the index computation \eref{indexsingleT2}.

The $\USp(2)_{k} \times \SO(4)_1$ gauge theory may have a non-anomalous $\BZ_2$ one-form symmetry, depending on $k$.  As before, the condition for the existence of a  $\BZ_2$ one-form symmetry of this theory can be determined by a simple generalisation of \cite[(3.27)]{Tachikawa:2019dvq} (see Footnote \ref{condnonanomZ2} with $k_1=1$, $M=2$, $k_2=k$ and $N=1$):
\bes{ \label{condZ211k}
\frac{1}{4} \times 1 \times 2 + \frac{1}{2} k  =\frac{1}{2} (k+1) \in \BZ \qquad \Leftrightarrow \qquad \text{$k$ is odd}~. 
}
We interpret this results as follows.  Although theory \eref{singleT2} with the CS levels $(k,1,1)$, or equivalently the $\USp(2)_{k} \times \Spin(4)_1$ gauge theory, has an anomalous $\BZ_2^2$ one-form symmetry, its $\BZ_2$ diagonal subgroup is non-anomalous for $k$ odd. This can be seen directly from the action \eref{Sanom1} of the anomaly theory: $S_{\text{anom}} = -\frac{2 \pi}{2} \int_{\CM_4}  (w_1^{(2)} \cup  w_2^{(2)})$, where the first two terms of \eref{Sanom1} can be dropped.  Upon taking $w_1^{(2)} =  w_2^{(2)}  \equiv B^{(2)}$, we have $S_{\text{anom}} = - 2 \pi \int_{\CM_4} \CP(B^{(2)})/2$.  Since $\int_{\CM_4} \CP(B^{(2)})$ is even on a spin manifold ${\CM_4}$, this action is an integer multiple of $2 \pi$; this indicates that the $\BZ_2$ diagonal subgroup of $\BZ_2^2$ one-form symmetry is non-anomalous for $k$ odd.  On the other hand, for $k$ even, we also have a non-trivial contribution from the first term of \eref{Sanom1}, namely $\pi (k-1)  \int_{\CM_4}  \CP(w_1^{(2)})/2$; this renders the anomaly of the $\BZ_2$ diagonal subgroup non-trivial.

For $k$ odd, the above interpretation can be supported by an explicit realisation of the topological sector $(\CZ_2)_0$; see the first row of \eref{tab:k1k21}. As discussed around \cite[(2.7)]{Hsin:2018vcg}, the $\BZ_2^2$ one-form symmetry of the $(\CZ_2)_0$ TQFT is generated by the basic electric and magnetic lines $V_E$, $V_M$ of integer spins, where each of such lines generates a $\BZ_2$ non-anomalous one-form symmetry labelled by $p=0$.  Due to a non-trivial mutual braiding phase $e^{-i \pi}$ of $V_E$ and $V_M$, we can always find a line $b$ that generates a $\BZ_2$ subgroup of the $\BZ_2^2$ one-form symmetry with anomaly characterised by $p$ (with $p =0,\, 1 \,\, \mod \,2$), namely $b = V_E^{p/2} V_M$.  We see that the $\BZ_2$ diagonal subgroup, generated by the line $V_E V_M$, of the $\BZ_2^2$ one-form symmetry is indeed anomaly free.

\subsection{Gluing with $T(\SU(2))$ theories} \label{sec:singleT2gluewithS}
Let us now include the $T(\SU(2))$ SCFT \cite{Gaiotto:2008ak} into the discussion.  This theory can be realised as an IR SCFT of the 3d $\CN=4$ $\U(1)$ gauge theory with two hypermultiplets of charge $1$ and it has an $\SU(2)_H \times \SU(2)_C$ flavour symmetry with the mixed anomaly given by the following anomaly theory (see \cite{Gang:2018wek,Genolini:2022mpi, Bhardwaj:2022dyt} and also \cite{Komargodski:2017dmc})
\bes{ \label{mixedanomTSU2}
\pi \int_{\CM_4} w_2^H \cup w_2^C ~,
}
where $w_2^H$ and $w_2^C$ are, respectively, the second Stiefel-Whitney classes associated with the $\SO(3)_H$ and $\SO(3)_C$ bundles that obstruct the lift to the $\SU(2)_H$ and $\SU(2)_C$ bundles.

As discussed in \cite{Assel:2022row}, an interesting generalisation of \eref{singleT2} is to gauge the diagonal subgroup of the $\SU(2)_i$ symmetries (with $i=1,2,3$) of the $T_2$ theory and the $\SU(2)_C$ global symmetry of the $i$-th copy of the $T(SU(2))$ theory with CS level $k^{(1)}_i$, and then gauge the $\SU(2)_H$ global symmetry of the $i$-th copy of the $T(\SU(2))$ theory with CS level $k^{(2)}_i$.  The resulting theory can be represented as
\bes{ \label{singleT2withTSU2}
\begin{tikzpicture}[baseline, font=\footnotesize]
\node[draw=none] (T2) at (0,0) {$T_2$};
\node[draw=none] (s1) at (2,1) {$S$~};
\node[draw=none] (s2) at (2,0) {$S$~};
\node[draw=none] (s3) at (2,-1) {$S$~};
\node[draw=none] (c1) at (4,1) {};
\node[draw=none] (c2) at (4,0) {};
\node[draw=none] (c3) at (4,-1) {};
\draw[solid, bend left] (T2) to node[above,midway] {$k^{(1)}_1$} (s1);
\draw[solid] (s1) to node[above,midway] {$k^{(2)}_1$} (c1);
\draw[solid] (T2) to node[above,midway] {$k^{(1)}_2$} (s2);
\draw[solid] (s2) to node[above,midway] {$k^{(2)}_2$} (c2);
\draw[solid, bend right] (T2) to node[above,midway] {$k^{(1)}_3$} (s3);
\draw[solid] (s3) to node[above,midway] {$k^{(2)}_3$} (c3);
\end{tikzpicture}
}
where $S$ stands for the $T(SU(2))$ theory.  More generally, we could consider the following longer `tail':
\bes{
\begin{tikzpicture}[baseline, font=\footnotesize]
\node[draw=none] (s0) at (-1.5,0) {~};
\node[draw=none] (s1) at (0,0) {$S$~};
\node[draw=none] (s2) at (1.5,0) {$S$~};
\node[draw=none] (s3) at (3,0) {$\cdots$~};
\node[draw=none] (s4) at (4.5,0) {$S$~};
\node[draw=none] (s5) at (6,0) {~};
\draw[solid] (s0) to node[above,midway] {$k^{(1)}_i$} (s1) to node[above,midway] {$k^{(2)}_i$} (s2) to node[above,midway] {$k^{(3)}_i$} (s3) to node[above,midway] {$k^{(a_i-1)}_i$} (s4) to node[above,midway] {$k^{(a_i)}_i$} (s5);
\end{tikzpicture}
}
For simplicity, we focus on the configuration \eref{singleT2withTSU2}.  As pointed out in \cite[Section 3.3]{Assel:2022row}, this model can be realised by compactifying M5-branes on a three-manifold given by a Seifert bundle over $S^2$ with three singular fibers, with Seifert parameters $q_1/p_1$, $q_2/p_2$ and $q_3/p_3$, where
\bes{ \label{defratios}
\frac{p_i}{q_i} = k^{(1)}_i - \frac{1}{k^{(2)}_i}~, \qquad \text{$i=1,2,3$}~.
}
The effective superpotential after integrating out the adjoint scalars is
\bes{ \label{supgluesingleT2}
W = \frac{1}{2} \left( \frac{q_1}{p_1}+ \frac{q_2}{p_2} + \frac{q_3}{p_3} \right) \tr(\mu^2) +\sum_{i=1}^3 \frac{q_i}{p_i} \tr( \mu_i \mu_{i,C}) ~,
}
where $\mu_i^C$ denotes the moment map of the $\SU(2)_C$ global symmetry of the $i$-th copy of the $T(\SU(2))$ theory, and $\mu_i$ is the moment map of the $\SU(2)_i$ global symmetry of the $T_2$ theory satisfying \eref{rel1}. The ATT condition \eref{ATT} is then generalised to \cite{Assel:2022row}
\bes{ \label{ATT2}
\frac{q_1}{p_1}+ \frac{q_2}{p_2} + \frac{q_3}{p_3} = 0~.
}
Similarly to \eref{singleT2}, we will again show that this is a {\it sufficient} condition for supersymmetry enhancement in the IR.  However, even if \eref{ATT2} is not satisfied, there are cases in which the IR SCFT has accidental $\CN=4$ supersymmetry.  

The index of the $T(\SU(2))$ SCFT can be written as
\bes{
&\CI_{T(\SU(2))} (w, n| f, m|a, n_a; x) \\
&= \sum_{l \in \BZ} (w^2)^l \oint \frac{dz}{2\pi i z} z^n \CZ_{\chi}^{1}(a^{-2};-2 n_a; x) \times \\
& \qquad \quad \prod_{s = \pm 1} \CZ_{\chi}^{1/2}( (z f)^s a; s(l+m) + n_a; x)  \CZ_{\chi}^{1/2}( (z^{-1} f)^s a; s(-l+m) + n_a; x)~,
}
where $(w,n)$ are the (fugacity, background magnetic flux) for the topological symmetry, $(f, m)$ are those for the flavour symmetry, and $(a, n_a)$ are those for the axial symmetry.   Here we normalise the power of the fugacity $w$ in such a way that the elementary monopole operators $V_\pm$ carry the fugacity $a^{-2} w^{\pm 2}$.  In this way, the Coulomb branch moment maps correspond to the term $a^{-2} \chi^{\su(2)_C}_{[2]} (w) x $ in the index, and the Higgs branch moment maps correspond to the term $a^{2} \chi^{\su(2)_H}_{[2]} (f) x$. The index for theory \eref{singleT2withTSU2} is therefore given by
\bes{ \label{indsingleT2withTSU2}
&\CI_{\eref{singleT2withTSU2}}(a, n_a; x) \\ &= \left( \frac{1}{8}\prod_{i=1}^3 \oint \frac{d z_i}{2\pi i z_i} \right) \sum_{(m_1, \cdots, m_3) \in \BZ^3}  \left( \frac{1}{8}\prod_{i=1}^3 \oint \frac{d f_i}{2\pi i f_i} \right) \sum_{(\hat{m}_1, \cdots, \hat{m}_3) \in \BZ^3} \times  \\
& \quad \ \left( \prod_{i=1}^3 z_i^{2 k^{(1)}_i m_i}   \CZ^{\SU(2)}_{\text{vec}} (z_i; m_i; x) \right) \left( \prod_{i=1}^3 f_i^{2 k^{(2)}_i \hat{m}_i}   \CZ^{\SU(2)}_{\text{vec}} (f_i;\hat{m}_i;x) \right) \times \\
&\quad \ \prod_{s_1, s_2, s_3 =\pm 1} \CZ^{1/2}_{\chi}(z_1^{s_1} z_2^{s_2} z_3^{s_3} a; s_1 m_1+s_2 m_2+s_3 m_3+n_a;x) \times \\
& \quad \  \prod_{i=1}^3 \CI_{T(\SU(2))} (z_i, m_i| f_i, \hat{m}_i |a, n_a; x) ~.
}

When the ATT condition \eref{ATT2} is satisfied, the first term in \eref{supgluesingleT2} vanishes and the $\U(1)_a$ symmetry associated with the fugacity $a$ assigned as above is a symmetry of the theory, since $\mu_i$ carries charge $+2$ and $\mu_{i, C}$ carries charge $-2$. However, if \eref{ATT2} is not satisfied, we set $a=1$ and $n_a=0$ in the above expression of the index.

\subsubsection{'t Hooft anomalies of the one-form symmetries}
Gauging the $\SU(2)_H$ and $\SU(2)_C$ global symmetries of $T(\SU(2))$ respectively with CS levels $k_H$ and $k_C$ leads to  the $\BZ_{2, H} \times \BZ_{2, C}$ one-form symmetry arising from the centres of $SU(2)_H$ and $SU(2)_C$.  The 't Hooft anomaly of such a one-form symmetry is characterised by the following anomaly theory (see \cite[(3.62)]{Gang:2021hrd})
\bes{
\pi \int_{\CM_4} \left[ k_H \frac{\CP(w_H^{(2)}) }{2} + k_C \frac{\CP(w_C^{(2)}) }{2} + w_H^{(2)} \cup w_C^{(2)} \right]~,
}
where $w_{H/C}^{(2)}$ are the two-form background fields for the $\BZ_{2, H/C}$ one-form symmetries.  The first two terms arise as in \eref{anomth1} and the last term comes from \eref{mixedanomTSU2}.  Upon gauging with the $T_2$ theory as in \eref{singleT2withTSU2}, we have six $\SU(2)$ gauge groups but $\BZ^5_2$ one-form symmetry due to the screening effect of the matter of the $T_2$ theory.  The 't Hooft anomalies of the latter are given by
\bes{ \label{anom1formT2wS}
\pi \int_{\CM_4} \left[ \sum_{r=1}^2 \sum_{i=1}^3  k^{(r)}_i  \frac{\CP(B_i^{(r)}) }{2} + \sum_{i=1}^3 B_i^{(1)} \cup B_i^{(2)}\right] \quad \text{with}~\sum_{i=1}^3 B_i^{(1)} = 0~,
} 
where $B_i^{(r)}$ is the two-form background field associated with the $\SU(2)$ gauge group with CS level $k_i^{(r)}$, with $i=1,2,3$ and $r=1,2$.\footnote{We denote the two-form background fields for the one-form symmetries differently from the other sections in order to avoid cluttered notation.}  The last constraint comes from \eref{constrT2}.

\subsubsection{Summary of the results}
We observe the following result:
\begin{quote} \vspace{-0.7cm}
\bes{ \label{isomseifert1}
&\text{For given ratios $p_i/q_i$ (with $i=1,2,3$), the index \eref{indsingleT2withTSU2} of theory} \\
&\text{\eref{singleT2withTSU2} is independent of specific values of $k^{(a_i)}_i$ in \eref{defratios}.}
}
\end{quote}
As an immediate consequence, the following statement holds:
\begin{quote} \vspace{-0.7cm}
\bes{ \label{isomseifert}
&\text{If $p_i/q_i \in \BZ$ for all $i =1, 2,3$, the index \eref{indsingleT2withTSU2} of theory \eref{singleT2withTSU2}} \\
&\text{is equal to the index \eref{indexsingleT2} of theory \eref{singleT2} with $k_i = p_i/q_i$.}
}
\end{quote}

We remark that these statements are true, independently of whether the ATT condition \eref{ATT2} is satisfied.  This means that the aforementioned theories flow to the same interacting SCFT in the IR.  This observation may not be a surprise from the geometrical perspective, since both theories are associated with Seifert manifolds that are diffeomorphic to each other.  Note that the decoupled topological sectors in the IR may be different, since the anomalous one-form symmetries determined by \eref{Sanom1} and \eref{anom1formT2wS} may be different.  We have actually seen this phenomenon in Footnote \ref{foot:diffTQFT}.

For simplicity, we examine the indices of the following theory
\bes{
\begin{tikzpicture}[baseline, font=\footnotesize]
\node[draw=none] (T2) at (0,0) {$T_2$};
\node[draw=none] (s1) at (2,1) {$S$~};
\node[draw=none] (c1) at (4,1) {};
\node[draw=none] (c2) at (4,0) {};
\node[draw=none] (c3) at (4,-1) {};
\draw[solid, bend left] (T2) to node[above,midway] {$k^{(1)}_1$} (s1);
\draw[solid] (s1) to node[above,midway] {$k^{(2)}_1$} (c1);
\draw[solid] (T2) to node[above,midway] {$k_2$} (c2);
\draw[solid, bend right] (T2) to node[above,midway] {$k_3$} (c3);
\end{tikzpicture}
}
with various CS levels as follows.
\bes{ \label{tableoneT2oneS}
\scalebox{0.94}{
\begin{tabular}{|c|c|c|c|c|c|c|}
\hline
$k^{(1)}_1$ & $k^{(2)}_1$ & $\frac{p_1}{q_1}$ & $k_2$ & $k_3$ & ATT \eref{ATT2} & Index  \\
\hline
$-4$ & $-1$ & $-3$ & $6$ & $6$ & \ding{51} & \eref{indmk2k2k} with $k=3$ \\
\hline
$3$ & $2$ & $\frac{5}{2}$ & $-5$ & $-5$ & \ding{51} & $1+(2a^4-1)x^2+(a^6 -a^2 +a^{-2}) x^3 +$ \\
$2$ &$-2$ & & & & &  $(3a^8-a^4-2)x^4+\ldots$ \\
\hline
$3$ & $3$ & $\frac{8}{3}$ & $-4$ & $-8$ & \ding{51} & $1+(a^4-1)x^2 + a^{-2} x^3$ + \\
& & & & & &  $2(a^8-1)x^4+\ldots$ \\
\hline
$2$ & $2$ & $\frac{3}{2}$ & $-3$ & $-3$ & \ding{51} & diverges\\
\hline \hline
$1$ & $3$ & $\frac{2}{3}$ & $1$ & $1$ & \ding{55} & $1 +0x - x^2 + 2 x^3 - 2 x^4 +\ldots$ \\
\hline
$2$ & $2$ & $\frac{3}{2}$ & $1$ & $1$ & \ding{55} & $1 +0x- x^2 + 2 x^3 - 2 x^4 +\ldots$ \\
$1$ & $-2$ & & & & &  \\
\hline
$-2$ & $-2$ & $-\frac{3}{2}$ & $1$ & $1$ &  \ding{55} & $1$ \\
\hline
$1$ & $7$ & $\frac{6}{7}$ & $-2$ & $3$ &  \ding{55} & $1+0x+0x^2+x^3-x^4+\ldots$ \\
\hline
\end{tabular}}
}
For reference, we also provide the contribution of zero gauge magnetic fluxes:
\bes{ \label{indzerofluxoneT2oneS}
1+0x +(a^4-1)x^2 + a^{-2} x^3 +(a^8-2)x^4 +\ldots~.
}
The term $a^4 x^2$ indicates that there is one marginal operator coming from the zero gauge flux sector, the term $0x$ indicates the absence of the $\CN=3$ flavour symmetry current, and so the term $-x^2$ indicates that there is one $\CN=3$ extra SUSY-current.  Therefore, when the ATT condition is satisfied, the IR SCFT has enhanced $\CN=4$ supersymmetry.  However, when the index diverges, we cannot conclude the IR behaviour from it.

When the ATT condition is not satisfied, there are cases in which the IR SCFTs have {\bf enhanced $\CN=4$ supersymmetry}; these are explicitly shown in the fifth and sixth rows of Table \eref{tableoneT2oneS}.  This can be deduced by the same reasoning as above.  Note that the marginal operator discussed below \eref{indzerofluxoneT2oneS} is set to zero in the chiral ring by an $F$-term equation ({\it cf.} Section \ref{sec:nonATToneT2}) and so we do not have a positive term at order $x^2$.  However, in the final row of Table \eref{tableoneT2oneS}, we find no indication of supersymmetry enhancement from the index, assuming that there is no additional marginal operator.   If the index is unity, then the IR theory is a TQFT.  

Next, let us report some indices for theory \eref{singleT2withTSU2} with various CS levels such that the ATT condition \eref{ATT2} is satisfied:
\bes{
\begin{tabular}{|c|c|c|c|c|c|c|c|c|c|}
\hline
$k^{(1)}_1$ & $k^{(2)}_1$ & $\frac{p_1}{q_1}$ & $k^{(1)}_2$ & $k^{(2)}_2$ & $\frac{p_2}{q_2}$ & $k^{(1)}_3$ & $k^{(2)}_3$ & $\frac{p_3}{q_3}$  & Index  \\
\hline
$-2$ & $1$ & $-3$ & $7$ & $1$ & $6$ & $7$ & $1$ & $6$ & \eref{indmk2k2k} with $k=3$ \\
\hline
$-3$ & $1$ & $-4$ & $9$ & $1$ & $8$ & $9$ & $1$ & $8$ & \eref{indmk2k2k} with $k=4$ \\
\hline
$-4$ & $1$ & $-5$ & $11$ & $1$ & $10$ & $11$ & $1$ & $10$ & \eref{indmk2k2k} with $k=5$ \\
\hline
\end{tabular}
}
These results support the statement \eref{isomseifert}.  In all of these cases, the IR interacting SCFT has enhanced $\CN=4$ supersymmetry.

Finally, we observe that if a pair of Seifert fibres with parameters $q_i/p_i$ and $\tilde{q}_i/\tilde{p}_i$ are isomorphic by an orientation-preserving diffeomorphism, namely \cite[Prop. 2.1]{Hatcher:3manifolds}
\ben
\item after possibly permuting indices, $q_i/p_i = \tilde{q}_i/\tilde{p}_i$ $(\mod \, 1)$ for each $i$, and
\item $\sum_i q_i/p_i = \sum_i \tilde{q}_i/\tilde{p}_i$,
\een
then the indices of the theories associated with these Seifert fibres are equal.  In other words, the corresponding IR SCFTs are the same.

\section{Theories with two $T_2$ building blocks} \label{sec:twoT2}
Let us now couple two copies of the $T_2$ theory together by gauging a diagonal subgroup of the two $\SU(2)_i$ flavour symmetries (with $i=1,2,3$), belonging to different copies of the $T_2$ theories, with CS levels $k_i$.  We denote this diagrammatically as
\bes{ \label{twoT2}
\begin{tikzpicture}[baseline, font=\footnotesize]
\node[draw=none] (T2a) at (0,0) {$T_2$};
\node[draw=none] (T2b) at (3,0) {$T_2$};
\draw[solid, bend left] (T2a) to node[above,midway] {\scriptsize $k_1$} (T2b);
\draw[solid] (T2a) to node[above,midway] {\scriptsize $k_2$} (T2b);
\draw[solid, bend right] (T2a) to node[above,midway] {\scriptsize $k_3$} (T2b);
\end{tikzpicture}
}
We denote by $\mu_i^{(I)}$, with $i=1,2,3$ and $I=1,2$, the moment maps of the $\SU(2)_i$ flavour symmetry of the $I$-th $T_2$ theory.  Their explicit expression for each $I$ is given by \eref{momentmaps}.  We also have the analogue of \eref{rel1}, namely 
\bes{
\tr(\mu^{(I)\, \, 2}_1)= \tr(\mu^{(I)\, \, 2}_2)= \tr(\mu^{(I)\, \, 2}_3)  \equiv \tr(\mu^{(I)\, \, 2})~
}
for each $I=1,2$.  The effective superpotential after integrating out the adjoint scalar fields is (see \cite[(2.16)]{Assel:2022row})
\bes{ \label{suptwoT2}
W &=\frac{1}{2} \left(\frac{1}{k_1}+\frac{1}{k_2}+ \frac{1}{k_3} \right) \left[\tr \left(\mu^{(1) \,\, 2} \right)+ \tr \left(\mu^{(2) \,\, 2} \right) \right]+ \sum_{i=1}^3 \frac{1}{k_i} \tr \left( \mu_i^{(1)} \mu_i^{(2)} \right)~.
}

When the ATT condition \eref{ATT} is satisfied, the first term vanishes and there is a flavour symmetry that assigns charge $+1$ to every chiral field of the first $T_2$ theory and charge $-1$ to every chiral field of the second $T_2$ theory.  As a consequence, $\mu_i^{(1)}$ and $\mu_i^{(2)}$ carry charges $+2$ and $-2$, respectively.  We denote by $a$ the fugacity associated with this flavour symmetry.  We will shortly see that, if the CS levels obey the ATT condition, the flavour symmetry algebra is actually $\su(2)_a$. On the other hand, if the ATT condition \eref{ATT} is not satisfied, this flavour symmetry is explicitly broken by the first term of the superpotential. 

The index for theory \eref{twoT2} is given by
\bes{ \label{indextwoT2}
\CI_{\eref{twoT2}}(a, n_a; x) &= \left( \frac{1}{8}\prod_{i=1}^3 \oint \frac{d z_i}{2\pi i z_i} \right) \sum_{(m_1, m_2, m_3) \in \BZ^3}  \left( \prod_{i=1}^3 z_i^{2 k_i m_i}   \CZ^{\SU(2)}_{\text{vec}} (z_i;m_i;x) \right) \times  \\
&\qquad  \prod_{s_1, s_2, s_3 =\pm 1} \CZ^{1/2}_{\chi}(z_1^{s_1} z_2^{s_2} z_3^{s_3} a; s_1 m_1+s_2 m_2+s_3 m_3+n_a;x) \times \\
&\qquad  \prod_{s'_1, s'_2, s'_3 =\pm 1} \CZ^{1/2}_{\chi}(z_1^{s'_1} z_2^{s'_2} z_3^{s'_3} a^{-1}; s'_1 m_1+s'_2 m_2+s'_3 m_3-n_a;x)~.
}
As before, we will set $n_a=0$ and drop $n_a$ from the argument in $\CI_{\eref{twoT2}}(a, n_a; x)$ when we study the series expansion of the index.  For the cases that do not satisfy the ATT condition, $a$ should be set to $1$ and $n_a$ should be set to zero.

Without the CS levels, theory \eref{twoT2} can be viewed as 3d reduction of the 4d $\CN=2$ $A_1$ class $\CS$ theory associated with a Riemann surface of genus 2 with no puncture.  The one-form symmetry of the latter is $\BZ_2 \times \BZ_2$; see \cite[(3.12)]{Bhardwaj:2021pfz}. With the CS levels turned on, their 't Hooft anomalies are characterised by \eref{anomtheory} and \eref{anompmatrix}.

\subsection{Cases that satisfy the ATT condition} \label{sec:ATTtwoT2}
We consider theories \eref{twoT2} with CS levels satisfying \eref{ATT}.

\subsubsection{Special case of $(k_1, k_2, k_3) = (-k,2k,2k)$}
This theory is equivalent to the $\USp(2)_{-k} \times \Spin(4)_{2k}$ gauge theory with two copies of bifundamental half-hypermultiplets in the representation $[\mathbf{2}; \mathbf{4}]$ :
\bes{ \label{USp2Spin4ABJ}
\begin{tikzpicture}[baseline, font=\footnotesize]
\node[draw, circle, minimum size=1.25cm] (USp) at (-1.5,0) {\scalebox{0.65}{$\USp(2)_{-k}$}};
\node[draw, circle, minimum size=1.25cm] (Spin) at (1.5,0) {\scalebox{0.65}{$\Spin(4)_{2k}$}};
\draw[solid, bend left] (USp) to (Spin);
\draw[solid, bend right] (USp) to (Spin);
\end{tikzpicture}
}  
It is indeed closely related to the ABJ theory \cite{Aharony:2008gk}, described by the $\USp(2)_{-k} \times \mathrm{O}(4)_{2k}$ gauge theory with the same matter content.  We can start from the $\USp(2)_{-k} \times \mathrm{SO}(4)_{2k}$ variant of the theory: gauging the $\BZ_2$ zero-form charge conjugation symmetry associated with the $\SO(4)$ gauge group leads to the original ABJ theory, whereas gauging the $\BZ_2$ zero-form magnetic symmetry leads to theory \eref{USp2Spin4ABJ}.  This type of arguments was used to studied variants of ABJM \cite{Aharony:2008ug} and ABJ theories in \cite{Tachikawa:2019dvq, Bergman:2020ifi, Beratto:2021xmn, Mekareeya:2022spm}.  The index of the $\USp(2)_{-k} \times \mathrm{SO}(4)_{2k}$ variant is
\bes{ \label{indUSp2SO4ABJ}
&\CI_{\USp(2)_{-k} \times \mathrm{SO}(4)_{2k}}(\zeta, a;x) \\
&= \frac{1}{8} \sum_{(\fm_1, \fm_2) \in \BZ^2} \, \sum_{\fn \in \BZ} \left( \prod_{j=1}^2 \oint \frac{d v_j}{2 \pi i v_j}  \, v_j^{2k \fm_j}  \right)   \zeta^{\fm_1+\fm_2} \oint \frac{d u}{2 \pi i u}\, u^{-2k \fn} \times \\
& \quad  \CZ^{\SO(4)}_{\text{vec}}(v_1,v_2; \fm_1, \fm_2; x)  \CZ^{\USp(2)}_{\text{vec}}(u; \fn; x)  \times \\
& \quad  \prod_{i=1}^2 \,  \prod_{s_1, s_2 = \pm 1}  \CZ^{1/2}_{\chi} \left( v_i^{s_1} u^{s_2} a; s_1 \fm_i+{s_2} \fn; x \right) \CZ^{1/2}_{\chi} \left( v_i^{s_1} u^{s_2} a^{-1}; s_1 \fm_i+{s_2} \fn; x \right) ~.
}
where $\zeta$ (with $\zeta^2=1$) denotes the fugacity for the $\BZ_2$ zero-form magnetic symmetry and we have set the fugacity $\chi$ for the charge conjugation symmetry to $1$. The index of \eref{USp2Spin4ABJ} can be obtained by gauging the magnetic symmetry as follows:
\bes{
\CI_{\eref{USp2Spin4ABJ}}(a;x)  = \frac{1}{2} \left[ \CI_{\USp(2)_{-k} \times \mathrm{SO}(4)_{2k}}(\zeta=1, a;x)+\CI_{\USp(2)_{-k} \times \mathrm{SO}(4)_{2k}}(\zeta=-1, a;x) \right]~.
}

Let us first consider the index of the $\USp(2)_{-k} \times \mathrm{SO}(4)_{2k}$ variant of the ABJ theory, or equivalently the $\SU(2)_{-k} \times [\SU(2)_{2k} \times \SU(2)_{2k} ]/\BZ_2$ theory, where a $\BZ_2$ one-form symmetry of \eref{twoT2} is gauged.  We tabulate the results, up to order $x^4$, below.
\bes{ \label{indmk2k2kmodZ2}
\scalebox{0.82}{
\begin{tabular}{|c|c|}
\hline
CS levels & Index \eref{indUSp2SO4ABJ} \\
$(-k, 2k, 2k)$ &  \\
\hline
$k=1$ &  ${1}+{\left[{1}+{\zeta}{\chi_{[2]}^{\mathfrak{su}(2)}}\left({a}\right)\right]}{x}+{\left[{\left({2}+{\zeta}\right)}{{\chi_{[4]}^{\mathfrak{su}(2)}}\left({a}\right)}+{2}-{{\chi_{[2]}^{\mathfrak{su}(2)}}\left({a}\right)}\right]}{x^2}+$\\ 
&${\left\{{{\left({1}+{2}{\zeta}\right)}{{\chi_{[6]}^{\mathfrak{su}(2)}}\left({a}\right)}-{\left[{\zeta}{{\chi_{[4]}^{\mathfrak{su}(2)}}\left({a}\right)}+{\left({3}+{2}{\zeta}\right)}{{\chi_{[2]}^{\mathfrak{su}(2)}}\left({a}\right)}+{\zeta}\right]}}\right\}}{x^3}+$\\ 
&$\left\{{\left({3}+{2}{\zeta}\right)}{{\chi_{[8]}^{\mathfrak{su}(2)}}\left({a}\right)}+{3}{\left({1}+{\zeta}\right)}{{\chi_{[2]}^{\mathfrak{su}(2)}}\left({a}\right)}+{1}+{2}{\zeta}\right.-$\\ 
&$\left.{\left[{\left({2}+{\zeta}\right)}{{\chi_{[6]}^{\mathfrak{su}(2)}}\left({a}\right)}+{\left({1}+{\zeta}\right)}{{\chi_{[4]}^{\mathfrak{su}(2)}}\left({a}\right)}\right]}\right\}{x^4}+{\ldots}$ \\
\hline
$k=2$ &  ${1}+{x}+{\left[{\left({1}+{\zeta}\right)}{{\chi_{[4]}^{\mathfrak{su}(2)}}\left({a}\right)}+{2}-{{\chi_{[2]}^{\mathfrak{su}(2)}}\left({a}\right)}\right]}{x^2}+$\\
& ${\left[{\zeta}{{\chi_{[6]}^{\mathfrak{su}(2)}}\left({a}\right)}+{2}-{\left({2}+{\zeta}\right)}{{\chi_{[2]}^{\mathfrak{su}(2)}}\left({a}\right)}\right]}{x^3}+$\\ 
&$\left\{{\left({2}+{\zeta}\right)}{{\chi_{[8]}^{\mathfrak{su}(2)}}\left({a}\right)}+{2}-{\left[{\left({1}+{\zeta}\right)}{{\chi_{[6]}^{\mathfrak{su}(2)}}\left({a}\right)}+{\zeta}{{\chi_{[4]}^{\mathfrak{su}(2)}}\left({a}\right)}+{\zeta}{{\chi_{[2]}^{\mathfrak{su}(2)}}\left({a}\right)}\right]}\right\}{x^4}+{\ldots}$ \\
\hline
$k=3$ & ${1}+{x}+{\left[{{\chi_{[4]}^{\mathfrak{su}(2)}}\left({a}\right)}+{2}-{{\chi_{[2]}^{\mathfrak{su}(2)}}\left({a}\right)}\right]}{x^2}+{\left[{{\zeta}{{\chi_{[6]}^{\mathfrak{su}(2)}}\left({a}\right)}+{2}-{2}{{\chi_{[2]}^{\mathfrak{su}(2)}}\left({a}\right)}}\right]}{x^3}+$\\ 
&$\left\{{\left({1}+{\zeta}\right)}{{\chi_{[8]}^{\mathfrak{su}(2)}}\left({a}\right)}+{2}-{\left[{{\chi_{[6]}^{\mathfrak{su}(2)}}\left({a}\right)}+{\zeta}{{\chi_{[4]}^{\mathfrak{su}(2)}}\left({a}\right)}\right]}\right\}{x^4}+{\ldots}$ \\
\hline
$k=4$ & ${1}+{x}+{\left[{{\chi_{[4]}^{\mathfrak{su}(2)}}\left({a}\right)}+{2}-{{\chi_{[2]}^{\mathfrak{su}(2)}}\left({a}\right)}\right]}{x^2}+{2}{\left[{{1}-{{\chi_{[2]}^{\mathfrak{su}(2)}}\left({a}\right)}}\right]}{x^3}+$\\ 
&$\left\{{\left({1}+{\zeta}\right)}{{\chi_{[8]}^{\mathfrak{su}(2)}}\left({a}\right)}+{2}-{{\chi_{[6]}^{\mathfrak{su}(2)}}\left({a}\right)}\right\}{x^4}+{\ldots}$ \\
\hline
\end{tabular}}
}
For $k\geq 5$, the terms with fugacity $\zeta$ appear at a higher order than $x^4$.  The index of the case of $k=1$ was studied in \cite[(3.80)]{Beratto:2021xmn}, where it was pointed out that the $\USp(2)_{-1} \times \mathrm{SO}(4)_{2}$ ABJ theory is dual to another variant of the ABJ theory, namely the $[\U(3)_4 \times \U(1)_{-4}]/\BZ_2$ gauge theory with two bifundamental hypermultiplets, whose IR SCFT has {\bf enhanced $\CN=6$ supersymmetry}.  The indices for the cases of $k\geq 2$ indicate that supersymmetry gets {\bf enhanced to $\CN=5$}.  This can be seen as follows. All of such indices have the coefficient of $x$ equal to $1$ indicating that there is one $\CN=3$ flavour current, but since there is the term $-{{\chi_{[2]}^{\mathfrak{su}(2)}}\left({a}\right)} =-(a^2+1+a^{-2})$ at order $x^2$, the term $-1$ corresponds to the $\CN=3$ flavour current and the terms $-a^2$ and $-a^{-2}$ correspond to the $\CN=3$ extra SUSY-currents, rendering supersymmetry enhancement from $\CN=3$ to $\CN=5$.

Now we report the index of theory \eref{twoT2} with CS levels $(-k, 2k, 2k)$, or equivalently \eref{USp2Spin4ABJ}, given by  $\CI_{\eref{USp2Spin4ABJ}}(a;x)$ up to order $x^4$ below.
\bes{ \label{indmk2k2ktwoT2}
\scalebox{0.89}{
\begin{tabular}{|c|c|}
\hline
CS levels & Index $\CI_{\eref{USp2Spin4ABJ}}(a;x)$ \\
$(-k, 2k, 2k)$ &  \\
\hline
$k=1$ & ${1}+{x}+{\left\{{2}{\left[{\chi_{[4]}^{\mathfrak{su}(2)}}\left({a}\right)+{1}\right]}-{{\chi_{[2]}^{\mathfrak{su}(2)}}\left({a}\right)}\right\}}{x^2}+{\left[{{{\chi_{[6]}^{\mathfrak{su}(2)}}\left({a}\right)}-{3}{{\chi_{[2]}^{\mathfrak{su}(2)}}\left({a}\right)}}\right]}{x^3}+$ \\
& ${\left\{{3}{\left[{{\chi_{[8]}^{\mathfrak{su}(2)}}\left({a}\right)}+{{\chi_{[2]}^{\mathfrak{su}(2)}}\left({a}\right)}\right]}+{1}-{\left[{2}{{\chi_{[6]}^{\mathfrak{su}(2)}}\left({a}\right)}+{{\chi_{[4]}^{\mathfrak{su}(2)}}\left({a}\right)}\right]}\right\}}{x^4}+{\ldots}$ \\
\hline
$k=2$ & ${1}+{x}+{\left[{{\chi_{[4]}^{\mathfrak{su}(2)}}\left({a}\right)}+{2}-{{\chi_{[2]}^{\mathfrak{su}(2)}}\left({a}\right)}\right]}{x^2}+{2}{\left[{{1}-{{\chi_{[2]}^{\mathfrak{su}(2)}}\left({a}\right)}}\right]}{x^3}+$\\ & ${\left\{{{2}{\left[{{\chi_{[8]}^{\mathfrak{su}(2)}}\left({a}\right)}+{1}\right]}-{{\chi_{[6]}^{\mathfrak{su}(2)}}\left({a}\right)}}\right\}}{x^4}+{\ldots}$ \\
\hline
$k\geq3$ & ${1}+{x}+{\left[{{\chi_{[4]}^{\mathfrak{su}(2)}}\left({a}\right)}+{2}-{{\chi_{[2]}^{\mathfrak{su}(2)}}\left({a}\right)}\right]}{x^2}+$\\ &${2}{\left[{{1}-{{\chi_{[2]}^{\mathfrak{su}(2)}}\left({a}\right)}}\right]}{x^3}+{\left[{{{\chi_{[8]}^{\mathfrak{su}(2)}}\left({a}\right)}+{2}-{{\chi_{[6]}^{\mathfrak{su}(2)}}\left({a}\right)}}\right]}{x^4}+ \ldots$ \\
\hline
\end{tabular}}
}
For ${k}\ge{3}$ the indices differ from each other at a higher order than $x^4$.  By the same reasoning as before, the indices indicate that the IR SCFT has enhanced $\CN=5$ supersymmetry.   

The 't Hooft anomalies of the one-form symmetry of these theories are as presented in \eref{anommk2k2k}.  For $k$ odd, theory \eref{twoT2} with CS levels $(-k, 2k, 2k)$ or equivalently \eref{USp2Spin4ABJ} flows to an interacting $\CN=5$ SCFT with a non-anomalous $\BZ_2$ one-form symmetry, along with the TQFT $\CA^{2,1}$.\footnote{For $k$ odd, we can infer from \eref{condMN} that the $\USp(2)_{-k} \times \SO(4)_{2k}$ theory does not admit a $\BZ_2$ quotient.  This means that, for $k$ odd, the non-anomalous one-form symmetry of the $\USp(2)_{-k} \times \Spin(4)_{2k}$ is $\BZ_2$, which we can gauge in order to obtain the $\USp(2)_{-k} \times \SO(4)_{2k}$ theory, and there is no further $\BZ_2$ one-form symmetry that we can gauge in the latter.}  For $k$ even, the theory flows to an interacting $\CN=5$ SCFT that has a $\BZ_2^2$ one-form symmetry and there is no decoupled topological sector.

The above indices receive the contributions of the gauge invariant dressed monopole operators, whose properties are similar to that discussed around \eref{foot:chargesbare}.  Explicitly, the bare monopole operator with magnetic fluxes $(m_1, m_2, m_3)$ has dimension 
\bes{ \label{foot:chargesbaretwoT2}
\Delta(m_1, m_2, m_3)=  \frac{1}{2} \sum_{s_1, s_2, s_3 =\pm 1} \left|  \sum_{i=1}^3 s_i m_i \right| -\sum_{i=1}^3 2|m_i| ~,
}
is neutral under the flavour symmetry and it carries charge $2 k_i m_i$ under the Cartan subalgebra of each $\SU(2)_i$ gauge factor. 

We can compute the Coulomb branch and Higgs branch limits of the index as in \eref{CBHBlimits}.  We find that they are equal, as expected for SCFTs with $\CN \geq 5$ supersymmetry.  In particular, we have
\bes{
&\text{Higgs limit $\eref{twoT2}_{(-k,2k,2k)}$} = \text{Coulomb limit $\eref{twoT2}_{(-k,2k,2k)}$} \\
&= \PE \left[ t^2 + t^{2k} + t^{2k+1} - t^{4k+2} \right] ~, \quad \text{with $t = h$ or $c$}~. 
}
This is the Hilbert series of $\BC^2/\hat{D}_{2k+2}$, indicating that the Higgs and Coulomb branches are isomorphic to this singularity.  The generators of the Higgs branch are
\bes{
w= \tr (\mu^{(1)\, 2})~, \quad v= X_{(2,1,1)} \left( Q^{(1)} \right)^{4k}~, \quad u = X_{(2,1,1)}  \left( Q^{(1)} \right)^{4k-4} \mu^{(1)}_1 \mu^{(1)}_2  \mu^{(1)}_3~,
}
satisfying the relation 
\bes{
u^2 + v^2 w = w^{2k+1}~.
}
The generators of the Coulomb branch can be obtained simply by replacing the superscript $(1)$ by $(2)$.  Note that the generators $u$ and $v$ are the bare monopole operator $X_{(2,1,1)}$, whose dimension is zero, dressed by appropriate chiral fields from each copy of $T_2$ such that the combinations become gauge invariant.  Indeed, from \eref{indmk2k2ktwoT2}, we see that the dressed monopoles that are related to $v$ contribute the term $\chi^{\mathfrak{su}(2)}_{[4k]}(a) x^{2k}$ to the index.

We can also examine  the $\USp(2)_{-k} \times \SO(4)_{2k}$ version of the ABJ theory, or equivalently $\SU(2)_{-k} \times  [\SU(2)_{2k} \times \SU(2)_{2k}]/\BZ_2$, which comes from gauging a non-anomalous $\BZ_2$ subgroup of the $\BZ_2^2$ one-form symmetry of the aforementioned theory.  In this case, we have
\bes{
&\text{Higgs limit $\eref{twoT2}_{(-k,2k,2k)}/\BZ_2^{[1]}$} = \text{Coulomb limit $\eref{twoT2}_{(-k,2k,2k)}/\BZ_2^{[1]}$} \\
&= \PE \left[ t^2 + t^{k} + t^{k+1} - t^{2k+2} \right] ~, \quad \text{with $t = h$ or $c$}~. 
}
This is the Hilbert series of $\BC^2/\hat{D}_{k+2}$.  The generators of the Higgs branch are
\bes{
w= \tr (\mu^{(1)\, 2})~, \quad v= X_{\left(1,\frac{1}{2},\frac{1}{2} \right)} \left( Q ^{(1)} \right)^{2k}~, \quad u = X_{\left(1,\frac{1}{2},\frac{1}{2}\right)} \left( Q^{(1)} \right)^{2k-4} \mu^{(1)}_1 \mu^{(1)}_2  \mu^{(1)}_3~,
}
satisfying the relation 
\bes{
u^2+v^2w = w^{k+1}~.
}
Again, the Coulomb branch generators can be obtained by replacing the superscript $(1)$ by $(2)$.  From \eref{indmk2k2kmodZ2}, we see that the dressed monopoles that are related to $v$ contribute the term $\zeta \chi^{\mathfrak{su}(2)}_{[2k]}(a) x^{k}$ to the index.  This explains why, when $k \geq 5$, the contribution from the dressed monopole operators appears at a higher order than $x^4$.  In particular, for $k=1$, these operators are related to $\CN=3$ flavour currents that are necessary for the enhanced $\CN=6$ supersymmetry in the IR.

\subsubsection{General results for  $(k_1, k_2, k_3) = k(\fp \fq, -\fp \fr, -\fq \fr )$ with $\fr = \fp+\fq$} \label{sec:twoT2gen}
We consider the CS levels \eref{genCS} for the theories formed by gauging two copies of the $T_2$ theory.  The information about the one-form symmetries and their 't Hooft anomalies are as tabulated in \eref{oneformpqrt}.

Let us now consider the index \eref{indextwoT2}. The contribution from flux $(m_1, m_2, m_3) = (0,0,0)$, up to order $x^4$, reads 
\bes{ \label{index000twoT2}
&{1}+{x}+{\left[{{\chi_{[4]}^{\mathfrak{su}(2)}}\left({a}\right)}+{2}-{{\chi_{[2]}^{\mathfrak{su}(2)}}\left({a}\right)}\right]}{x^2}+ \\
&{2}{\left[{{1}-{{\chi_{[2]}^{\mathfrak{su}(2)}}\left({a}\right)}}\right]}{x^3}+{\left[{{{\chi_{[8]}^{\mathfrak{su}(2)}}\left({a}\right)}+{2}-{{\chi_{[6]}^{\mathfrak{su}(2)}}\left({a}\right)}}\right]}{x^4}+ \ldots~.
}
The term at order $x$ is the contribution from
\bes{ \label{momentmaptwoT2ATT}
\epsilon^{\alpha_1 \alpha'_1} \epsilon^{\alpha_2 \alpha'_2} \epsilon^{\alpha_3 \alpha'_3} Q^{(1)}_{\alpha_1 \alpha_2 \alpha_3} Q^{(2)}_{\alpha'_1 \alpha'_2 \alpha'_3}~.
}
This is the moment map operator associated with the $\U(1)$ $\CN=3$ flavour symmetry current. The marginal operators contributing the positive terms, namely ${{\chi_{[4]}^{\mathfrak{su}(2)}}\left({a}\right)}+{2}$,  at order $x^2$ are
\bes{ \label{marginaltwoT2ATT}
Q^{((I_1)}Q^{(I_2)}Q^{(I_3)}Q^{(I_4))_S}~, \quad \tr(\mu^{(1)}_1 \mu^{(2)}_1)~, \quad   \tr(\mu^{(1)}_2 \mu^{(2)}_2)~,
}
where $\tr(\mu^{(1)}_3 \mu^{(2)}_3)$ can be written as a linear combination of the latter two due to the $F$-terms.  Also, in the first quantities, the contractions of gauge indices, which we have suppressed, are done in such a way that $I_1, \ldots I_4$ are completely symmetric; the latter is denoted by $()_S$.  Note also that the first quantities contain $\tr(\mu^{(1)\, 2})$ and $\tr(\mu^{(2)\, 2})$.

There is also a contribution from eight gauge fluxes $(m_1, m_2, m_3) = (\pm \fr, \pm \fq, \pm \fp)$ corresponding to the gauge-invariant dressed monopole operators.    According to the discussion around \eref{foot:chargesbaretwoT2}, the bare monopoles associated with these fluxes have dimension $0$, are neutral under the flavour symmetry, and the charges under the Cartan subalgebra of each $\SU(2)_i$ are $2k \fp \fq \fr (\pm 1, \mp 1, \mp 1 )$.   They have to be dressed with $2k \fp \fq \fr $ chiral fields from each copy of $T_2$, or  to form gauge invariant quantities.  The gauge invariant dressed monopole operators therefore contribute to the index as $\chi^{\su(2)}_{[2 k \fp \fq \fr ]}(a) x^{ k \fp \fq \fr}$.  If $ k \fp \fq \fr$ is sufficiently large, the index at sufficiently low order does not get affected by these operators.  In any case, using the same argument as above, we see from \eref{index000twoT2} that the SCFT in the IR has enhanced $\CN=5$ supersymmetry, regardless of the contribution of the dressed monopole operators.

\subsubsection{Higgs and Coulomb branches} \label{sec:HBCBtwoT2}
For the theories discussed in Section \eref{sec:twoT2gen}, the Higgs and Coulomb branch limits of the index are both equal to the Hilbert series of $\BC^2/\hat{D}_{K+2}$, namely
\bes{
\PE \left[ t^2 + t^{K} + t^{K+1} - t^{2K+2} \right]~, \quad K= \fp \fq \fr k~, ~\text{$t = h$ or $c$}~.
}
The generators of the Higgs or Coulomb branch are
\bes{
w= \tr (\mu^{(I)\, 2})~, \quad v= X_{(\fr, \fq, \fp) } (Q^{(I)})^{2\fp \fq \fr k}~, \quad u = X_{(\fr, \fq, \fp) } (Q^{(I)})^{2\fp \fq \fr k-4} \mu^{(I)}_1 \mu^{(I)}_2  \mu^{(I)}_3~,
}
with $I=1$ or $2$.  They satisfy the relation 
\bes{
u^2 + v^2 w = w^{K+1}~.
}

\subsection{Cases that do not satisfy the ATT condition} \label{sec:nonATTtwoT2}
From the effective superpotential \eref{suptwoT2}, we see that the first term can be viewed as an $\CN=3$ preserving exactly marginal deformation of the $\CN=5$ theory whose superpotential contains only the second term.   The index of the latter, excluding the contribution from the dressed monopoles, is given by \eref{index000twoT2}, which indeed indicates $\CN=5$ supersymmetry.  The aforementioned exactly marginal deformation explicitly breaks the $\su(2)_a$ flavour symmetry, corresponding to the term $-{\chi_{[2]}^{\mathfrak{su}(2)}}\left({a}\right) x^2$ in the index, to its Cartan subalgebra.  The latter can be seen from the term $+x$ of the index which indicates that the $\CN=3$ flavour symmetry is $\U(1)$.\footnote{To see the action of this symmetry, we view the moment maps $\mu^{(I)}$ as $2 \times 2$ symmetric matrices $\mu^{(I)} = \begin{pmatrix} a^{(I)} & b^{(I)}  \\ b^{(I)}  & c^{(I)} \end{pmatrix}$.  Under this $\U(1)$ symmetry, the elements $a^{(I)}$ and $c^{(I)}$ carry charges $+1$ and $-1$ respectively, whereas $b^{(I)}$ carry charge $0$.  Indeed, the terms $\tr(\mu^{(I) \, 2}) = 2 a^{(I)} c^{(I)}  -  2b^{(I)\, 2}$ and $\tr(\mu^{(1)} \mu^{(2)}) =  a^{(2)} c^{(1)} + a^{(1)} c^{(2)} -  2b^{(1)}b^{(2)}$ in the superpotential are neutral under this flavour symmetry, as it should be.}. From the index \eref{index000twoT2} with $a=1$, we have
\bes{
1 + x + (5-1) x^2 - 4 x^3 + 4 x^4+\ldots~,
}
where we see that 7 marginal operators as listed in \eref{marginaltwoT2ATT} get reduced to 5 due to the two $F$-term relations coming from the non-vanishing first term in the superpotential \eref{suptwoT2}.  Indeed this is due to the terms $a^2$ and $a^{-2}$ in the term $-{\chi_{[2]}^{\mathfrak{su}(2)}}\left({a}\right) x^2$ being set to $1$.  In general, we expect that the cases that do not satisfy the ATT condition have $\CN=3$ supersymmetry.  Although we have not taken into account the contributions of the dressed monopole operators, we do not expect the latter to make supersymmetry enhanced.  

Nevertheless, the dressed monopole operators can make the {\it flavour symmetry} enhanced.  An example is the case of CS levels $(k_1,k_2,k_3) = (-1,1,1)$, whose index is
\bes{
1 + 3 x + (9-3) x^2 - 7 x^3 + 16 x^4 + \ldots~.
}
The contributions to $3x$ come from \eref{momentmaptwoT2ATT}, along with the dressed monopole operators $X_{(1,1,0)} Q^{(1)} Q^{(2)}$ and $X_{(1,0,1)} Q^{(1)} Q^{(2)}$.  We propose that these form a triplet of the {\it enhanced $\SO(3)$ flavour symmetry} which is indeed the moment map of this symmetry.  For convenience, let denote by $b$ the fugacity of this $\SO(3)$ flavour symmetry, and so the term $3x$ can be rewritten as $\chi^{\su(2)}_{[2]}(b)x$. Note that the square of such dressed monopole operators contribute to the index at order $x^2$; therefore, the term $(9-3)x^2$ can be rewritten as $\left[\chi^{\su(2)}_{[4]}(b)+4- \chi^{\su(2)}_{[2]}(b)\right] x^2$.  The term $b^0=1$ in $\chi^{\su(2)}_{[4]}(b)$, together with $+4$, accounts for the 5 marginal operators, as mentioned above; the remaining positive terms are marginal operators that are dressed monopole operators.  We do not see any contribution of the $\CN=3$ extra SUSY-current, and so we conclude that the theory has $\CN=3$ supersymmetry.

\subsection{Gluing with $T(\SU(2))$ theories} \label{sec:twoT2gluewithS}
Let us consider gauging with the $T(\SU(2))$ theories in a similar fashion to the theories discussed in Section \ref{sec:singleT2gluewithS}. In particular, we focus on the following class of theories:
\bes{ \label{twoT2withTSU2}
\begin{tikzpicture}[baseline, font=\footnotesize]
\node[draw=none] (T2a) at (0,0) {$T_2$};
\node[draw=none] (s1) at (2,1) {$S$~};
\node[draw=none] (s2) at (2,0) {$S$~};
\node[draw=none] (s3) at (2,-1) {$S$~};
\node[draw=none] (T2b) at (4,0) {$T_2$};
\draw[solid, bend left] (T2a) to node[above,midway] {$k^{(1)}_1$} (s1);
\draw[solid, bend left] (s1) to node[above,midway] {$k^{(2)}_1$} (T2b);
\draw[solid] (T2a) to node[above,midway] {$k^{(1)}_2$} (s2);
\draw[solid] (s2) to node[above,midway] {$k^{(2)}_2$} (T2b);
\draw[solid, bend right] (T2a) to node[above,midway] {$k^{(1)}_3$} (s3);
\draw[solid, bend right] (s3) to node[above,midway] {$k^{(2)}_3$} (T2b);
\end{tikzpicture}
}
where $S$ stands for the $T(\SU(2))$ theory.  The $\SU(2)_i$ global symmetry left $T_2$ theory is diagonally gauged with the $\SU(2)_C$ symmetry of the $i$-th copy of the $T(SU(2))$ theory with CS level $k_i^{(1)}$ (with $i=1,2,3$), and the $\SU(2)_i$ global symmetry right $T_2$ theory is diagonally gauged with the $\SU(2)_H$ symmetry of the $i$-th copy of the $T(SU(2))$ theory with CS level $k_i^{(2)}$.  The analogue of the ATT condition \eref{ATT2} is the following \cite{Assel:2022row}:
\bes{ \label{ATT3}
\frac{q_1}{p_1}+ \frac{q_2}{p_2} + \frac{q_3}{p_3} = 0~, \qquad \frac{q'_1}{p_1}+ \frac{q'_2}{p_2} + \frac{q'_3}{p_3} = 0~,
}
where
\bes{
\frac{p_i}{q_i} = k^{(1)}_i - \frac{1}{k^{(2)}_i} ~, \qquad \frac{p_i}{q'_i} = k^{(2)}_i - \frac{1}{k^{(1)}_i}~.
}
For convenience, we will also refer to \eref{ATT3} as the ATT conditions. The index of these theories is given by the following expression:
\bes{ \label{indtwoT2withTSU2}
&\CI_{\eref{twoT2withTSU2}}(a, n_a; x) \\
&= \left( \frac{1}{8}\prod_{i=1}^3 \oint \frac{d z_i}{2\pi i z_i} \right) \sum_{(m_1, \cdots, m_3) \in \BZ^3}  \left( \frac{1}{8}\prod_{i=1}^3 \oint \frac{d f_i}{2\pi i f_i} \right) \sum_{(\hat{m}_1, \cdots, \hat{m}_3) \in \BZ^3} \times \\
& \quad \ \left( \prod_{i=1}^3 z_i^{2 k^{(1)}_i m_i}   \CZ^{\SU(2)}_{\text{vec}} (z_i;m_i;x) \right) \left( \prod_{i=1}^3 f_i^{2 k^{(2)}_i \hat{m}_i}   \CZ^{\SU(2)}_{\text{vec}} (f_i;\hat{m}_i;x) \right) \times  \\
&\quad \ \prod_{s_1, s_2, s_3 =\pm 1} \CZ^{1/2}_{\chi}(z_1^{s_1} z_2^{s_2} z_3^{s_3} a; s_1 m_1+s_2 m_2+s_3 m_3+  n_a;x) \times \\
&\quad \ \prod_{s_1, s_2, s_3 =\pm 1} \CZ^{1/2}_{\chi}(z_1^{s_1} z_2^{s_2} z_3^{s_3} a^{-1}; s_1 m_1+s_2 m_2+s_3 m_3  -n_a;x) \times \\
& \quad \ \prod_{i=1}^3 \CI_{T(\SU(2))} (z_i, m_i| f_i, \hat{m}_i |a, n_a; x) ~.
}
As before, when both conditions in \eref{ATT3} are satisfied, there is a flavour symmetry that assigns charge $+1$ to the chiral fields of the first copy of the $T_2$ theory, charge $-2$ to the Coulomb branch moment map of $T(\SU(2))$, $+2$ to the Higgs branch moment map of $T(\SU(2))$, and charge $-1$ to the chiral fields of the second copy of the $T_2$ theory.  If one of the conditions in \eref{ATT3} is not satisfied, this symmetry is explicitly broken and we should set $a=1$ and $n_a=0$ in the above expression.

\subsubsection*{'t Hooft anomalies of the one-form symmetries}

Similarly to \eref{anom1formT2wS}, here we have six $\SU(2)$ gauge groups but $\BZ^4_2$ one-form symmetry due to the screening effect of the matter of two copies of the $T_2$ theory.  The 't Hooft anomalies are characterised by
\bes{ \label{anom1formtwoT2wS}
&\pi \int_{\CM_4} \left[ \sum_{r=1}^2 \sum_{i=1}^3  k^{(r)}_i  \frac{\CP(B_i^{(r)}) }{2} + \sum_{i=1}^3 B_i^{(1)} \cup B_i^{(2)}\right] \\&\quad \text{with}~\sum_{i=1}^3 B_i^{(1)} = \sum_{i=1}^3 B_i^{(2)} = 0~,
} 
where $B_i^{(r)}$ is the two-form background field associated with the $\SU(2)$ gauge group with CS level $k_i^{(r)}$, with $i=1,2,3$ and $r=1,2$.  The last constraint comes from \eref{constrT2}.

\subsubsection*{Summary of the results}

For simplicity, we focus on the following theories
\bes{
\begin{tikzpicture}[baseline, font=\footnotesize]
\node[draw=none] (T2a) at (0,0) {$T_2$};
\node[draw=none] (s1) at (2,1) {$S$~};
\node[draw=none] (T2b) at (4,0) {$T_2$};
\draw[solid, bend left] (T2a) to node[above,midway] {$k^{(1)}_1$} (s1);
\draw[solid, bend left] (s1) to node[above,midway] {$k^{(2)}_1$} (T2b);
\draw[solid] (T2a) to node[above,midway] {$k_2$} (T2b);
\draw[solid, bend right] (T2a) to node[above,midway] {$k_3$} (T2b);
\end{tikzpicture}
}
with various CS levels.  We find the following results:
\bi
\item When both conditions in \eref{ATT3} are satisfied, IR SCFT always has enhanced $\CN=4$ supersymmetry.
\item When either of the conditions \eref{ATT3} is not satisfied, we generally do not find an indication of supersymmetry enhancement from the index. 
\ei
We report the indices as follows.
\bes{ \label{indATT3}
\scalebox{0.74}{
\begin{tabular}{|c|c|c|c|c|c|c|}
\hline
$k^{(1)}_1$ & $k^{(2)}_1$ & $\frac{p_1}{q_1}, \frac{p_1}{q'_1}$ & $k_2$ & $k_3$ & \eref{ATT3} & Index  \\
\hline
$2$ & $2$ & $\frac{3}{2}$, $\frac{3}{2}$ & $-1$ & $3$ & \ding{51}\,\, \ding{51} & $1+(a^4+a^{-4}+2-1)x^2 - (2 a^{2}+2a^{-2}) x^3 + $ \\
& & & & & &  $(2a^8+2a^{-8}-1)x^4+\ldots$ \\
\hline
$2$ & $2$ & $\frac{3}{2}$, $\frac{3}{2}$ & $-3$ & $-3$ & \ding{51}\,\, \ding{51} & $1+(a^4+a^{-4}+2-1)x^2 +(a^6-2a^2-2a^{-2}+a^{-6}) x^3 + $ \\
& & & & & &  $(2a^8-a^4-2-a^{-4}+2a^{-8})x^4+\ldots$ \\
\hline \hline
$1$ & $-1$ & $2$, $-2$ & 1 & 1 & \ding{55} \,\, \ding{55} & $1 +0x + 3 x^2 - 4 x^3 + 3 x^4+\ldots$ \\
\hline
$-2$ & $-1$ & $-1, -\frac{1}{2}$ & $1$ & $1$ & \ding{55}\, \, \ding{51} & $1 + x + 4 x^2 - 4 x^3 + 13 x^4 +\ldots$ \\
\hline
$-2$ & $-1$ & $-1, -\frac{1}{2}$ & $2$ & $2$ & \ding{51}  \,\, \ding{55} & $1 +0x + 4 x^2 - 3 x^3 + 3 x^4 +\ldots$ \\
\hline
\end{tabular}}
}

Let us first consider the cases in which both conditions in \eref{ATT3} are satisfied. The contribution of the gauge fluxes that are all zero is
\bes{
1+0 x+(a^4+a^{-4}+2 -1) x^2 -(2a^2+2a^{-2}) x^3 +(a^8 +a^{-8})x^4 +\ldots~.
}
Comparing with the first two rows in the above table, we see that the contributions of the higher monopole fluxes generally appear at higher order of $x$.  Since the coefficient of $x$ vanishes, there is no $\CN=3$ flavour current.  The current associated with the $\U(1)_a$ flavour symmetry, contributing the term $-x^2$, acts as the $\CN=3$ extra-SUSY current.  The latter implies that the IR SCFTs have {\bf enhanced $\CN=4$ supersymmetry}.  Due to the vanishing coefficient of $x$, the index does not satisfy the sufficient conditions to have $\CN \geq 5$ supersymmetry \cite{Evtikhiev:2017heo}.  The marginal operators contributing the terms $a^4+a^{-4}+2$ at order $x^2$ are
\bes{ \label{margATT3}
a^{\pm 4}: &\quad \epsilon^{a_1 b_1} \epsilon^{c_1 d_1} \epsilon^{a_2 c_2} \epsilon^{b_2 d_2} \epsilon^{a_3 c_3} \epsilon^{b_3 d_3} Q^{(I)}_{a_1 a_2 a_3} Q^{(I)}_{b_1 b_2 b_3} Q^{(I)}_{c_1 c_2 c_3} Q^{(I)}_{d_1 d_2 d_3}~, \quad I=1,2~, \\
1: &\quad \epsilon^{a_1 b_1} \epsilon^{\hat{c}_1 \hat{d}_1} \epsilon^{a_2 c_2} \epsilon^{b_2 d_2} \epsilon^{a_3 c_3} \epsilon^{b_3 d_3} Q^{(1)}_{a_1 a_2 a_3} Q^{(1)}_{b_1 b_2 b_3} Q^{(2)}_{\hat{c}_1 c_2 c_3} Q^{(2)}_{\hat{d}_1 d_2 d_3}~, \\
1: &\quad \epsilon^{a_1 b_1} \epsilon^{\hat{c}_1 \hat{d}_1} \epsilon^{a_2 c_2} \epsilon^{b_2 d_2} \epsilon^{a_3 d_3} \epsilon^{b_3 c_3} Q^{(1)}_{a_1 a_2 a_3} Q^{(1)}_{b_1 b_2 b_3} Q^{(2)}_{\hat{c}_1 c_2 c_3} Q^{(2)}_{\hat{d}_1 d_2 d_3}~. 
}
The other gauge invariant combinations with $R$-charge $2$ are related to these combinations by the identities of the epsilon tensors or the $F$-term conditions.  

However, in the cases in which one or both of the conditions \eref{ATT3} is not satisfied, the $\U(1)_a$ flavour symmetry is explicitly broken, and so the marginal operators listed in \eref{margATT3} may be related to each other by the $F$-terms ({\it cf.} Sections \ref{sec:nonATToneT2} and \ref{sec:nonATTtwoT2}).  In these cases, we do not see clear evidence of supersymmetry enhancement from the index.


\section{Theories with $T_3$ building blocks} \label{sec:T3}
We now consider theories whose building blocks are the 3d $T_3$ theory. Let us start by summarising the important information of the $T(\SU(3))$ and $T_3$ theories.  The $T(\SU(3))$ theory has an $\SU(3)_H \times \SU(3)_C$ global symmetry with a mixed anomaly characterised by
\bes{
\frac{2 \pi}{3} \int_{\CM_4} w_2^H \cup w_2^C ~,
}
where $w_2^{H/C}$ is the second Stiefel-Whitney class which measures the obstruction to lifting the $(\SU(3)/\BZ_3)_{H/C}$ bundle to the $SU(3)_{H/C}$ bundle. The 3d $\CN=4$ $T_3$ theory can then be constructed by gauging the diagonal $\SU(3)/\BZ_3$ subgroup of the $\SU(3)^3_H$ symmetry coming from three copies of the $T(\SU(3))$ theory.  Note that the $\SU(3)^3_C$ manifest flavour symmetry of the $T_3$ theory gets enhanced to $E_6$ in the IR.  The moment map in the adjoint representation of $E_6$ can be decomposed into fields in representations of the $\SU(3)^3$ maximal subgroup as follows:
\bes{
\begin{tabular}{ccccccc}
$\mathbf{78}$ &\,$ \rightarrow$ \,&$ [\mathbf{8} ; \mathbf{1}; \mathbf{1}]$ \,\, $\oplus$  & $[\mathbf{1} ; \mathbf{8}; \mathbf{1}]$\,\, $\oplus$ &$ [\mathbf{1} ; \mathbf{1}; \mathbf{8}] $ \,\, $\oplus$ & $[\mathbf{3} ; \mathbf{3}; \mathbf{3}]$\,\,  $\oplus$ & $[\bar{\mathbf{3}} ; \bar{\mathbf{3}}; \bar{\mathbf{3}}]$ \\
& & $X^{i_1}_{j_1}$ & $Y^{i_2}_{j_2}$ & $Z^{i_3}_{j_3}$ & $\CQ^{i_1 i_2 i_3}$ & $\tilde{\CQ}_{i_1 i_2 i_3}$ \\
\end{tabular}~.
}
They satisfy the following relations (see \cite[Section 2.2]{Maruyoshi:2013hja} and \cite[Section 5.3]{Tachikawa:2015bga}):
\bes{ \label{chiralringT3}
\scalebox{0.9}{$
\begin{split}
&\mathrm{tr}_1(X^2) = \mathrm{tr}_2(Y^2) = \mathrm{tr}_3 (Z^2) \equiv \mathbb{M}_2~, \\
&\mathrm{tr}_1(X^3) = \mathrm{tr}_2(Y^3) = \mathrm{tr}_3 (Z^3) \equiv \mathbb{M}_3~, \\
& X^{i_1}_{j_1} \CQ^{j_1 i_2 i_3} =   Y^{i_2}_{j_2} \CQ^{i_1 j_2 i_3} =   Z^{i_3}_{j_3} \CQ^{i_1 i_2 j_3}~, \\
& X^{j_1}_{i_1} \tilde{\CQ}_{j_1 i_2 i_3} =   Y^{j_2}_{i_2} \tilde{\CQ}_{i_1 j_2 i_3} =   Z^{j_3}_{i_3} \tilde{\CQ}_{i_1 i_2 j_3}~, \\
& \CQ^{i_1 i_2 i_3}\tilde{\CQ}_{j_1 j_2 i_3} = \sum_{l=0}^3 v_l \sum_{m=0}^{2-l} (X^{2-l-m})^{i_1}_{j_1} (Y^m)^{i_2}_{j_2}~, \quad  v_0=1~, \,\, v_1=0~, \,\, (X^0)^i_j = (Y^0)^i_j = \delta^i_j ~, \\
& \frac{1}{2} \CQ^{i_1 i_2 i_3} \CQ^{j_1 j_2 j_3} \epsilon_{i_2 j_2 k_2} \epsilon_{i_3 j_3 k_3} = \tilde{\CQ}_{k_1k_2 k_3} \delta^{i_1}_{p_1} X^{j_1}_{q_1} \epsilon^{p_1 q_1 k_1}~, \\
& \frac{1}{2} \tilde{\CQ}_{i_1 i_2 i_3} \tilde{\CQ}_{j_1 j_2 j_3} \epsilon^{i_2 j_2 k_2} \epsilon^{i_3 j_3 k_3} = \CQ^{k_1k_2 k_3} \delta_{i_1}^{p_1} X_{j_1}^{q_1} \epsilon_{p_1 q_1 k_1} ~,
\end{split}$}
}
where $\mathrm{tr}_i$ denotes the trace over the fundamental representation of the $\SU(3)_i$ symmetry (with $i=1,2,3$) of the $T_3$ theory.

\subsection{One $T_3$ building block}

The theory of our interest is obtained by gauging each $\SU(3)_C$ factor of the $\SU(
3)^3_C$ symmetry with CS levels $k_1$, $k_2$ and $k_3$.  Similarly to \eref{singleT2}, we denote this theory by
\bes{ \label{singleT3}
\begin{tikzpicture}[baseline, font=\footnotesize]
\node[draw=none] (T3) at (0,0) {$T_3$};
\node[draw=none] (c1) at (2,1) {};
\node[draw=none] (c2) at (2,0) {};
\node[draw=none] (c3) at (2,-1) {};
\draw[solid, bend left] (T3) to node[above,midway] {$k_1$} (c1);
\draw[solid] (T3) to node[above,midway] {$k_2$} (c2);
\draw[solid, bend right] (T3) to node[above,midway] {$k_3$} (c3);
\end{tikzpicture}
}

\subsubsection*{'t Hooft anomalies of the one-form symmetry}
Since the faithful manifest flavour symmetry of the $T_3$ theory is $\SU(3)^3/(\BZ_3 \times \BZ_3)$ \cite[(4.40)]{Bhardwaj:2021ojs}, it follows that theory \eref{singleT3} has a $\BZ_3^2$ one-form symmetry.  The 't Hooft anomaly of the one-form symmetry in theory \eref{singleT3} with CS levels $(k_1, k_2, k_3)$ is characterised by the 4d anomaly theory whose action is \cite{Benini:2017dus}
\bes{ \label{anomT3}
\frac{2 \pi}{3} \int_{\CM_4}  \sum_{i=1}^3 k_i \frac{\CP(w_i^{(2)})}{2} ~, \quad \text{with  \,\, $\sum_{i=1}^3 w_i^{(2)} =0$}~,
}
where $w_i^{(2)}$ is the two-form background field for the $\BZ_3$ one-form symmetry arising from the $\SU(3)_i$ gauge group of \eref{singleT3}. 

\subsubsection*{Superconformal indices}

The index of the $T(\SU(3))$ theory is given by
\bes{
&\CI_{T(\SU(3))} (\vec w, \vec n| \vec f, \vec m |a, n_a;x) \\
&= \frac{1}{ 2!}  \sum_{h \in \BZ + \vec \epsilon( \vec m)}   \oint \frac{d u}{ 2\pi i u}  w_1^{h} u^{n_1} \CZ_{\chi}^{1}(a^{-2};-2 n_a; x) \times  \\
& \qquad \sum_{l_1, l_2 \in \BZ + \vec \epsilon( \vec m)}  \oint \left( \prod_{\alpha=1}^2 \frac{d z_\alpha}{ 2\pi i z_\alpha} z^{n_2}_{\alpha} \right)  w_2^{l_1+l_2} \prod_{\alpha, \beta=1}^2 \CZ_{\chi}^{1}(z_\alpha z^{-1}_\beta a^{-2};l_\alpha - l_\beta -2 n_a; x) \times \\
& \qquad \ \CZ^{\U(2)} _{\text{vec}} (\{z_1, z_2 \}; \{ l_1, l_2 \}; x) \times \prod_{\alpha=1}^2 \prod_{s = \pm 1} \CZ^{\frac{1}{2}}_\chi \left(a (u z^{-1}_\alpha)^s ;s(h-l_\alpha)+n_a;x \right) \times \\
& \qquad \ \prod_{i=1}^3 \prod_{\alpha=1}^2 \prod_{s = \pm 1} \CZ^{\frac{1}{2}}_\chi \left(a (z_\alpha f^{-1}_i)^s ;s(l_\alpha-m_i)+n_a;x \right)~,
}
where $(\vec w, \vec n)$, $(\vec f, \vec m)$ and $(a, n_a)$ are $\text{(fugacities, background fluxes)}$ for the topological, flavour, and axial symmetries respectively.  In the above,  $\vec \epsilon( \vec m)$ denotes the fractional part of the background fluxes $m_i$.  The $\U(N)$ vector multiplet contribution is
\bes{
\CZ^{\U(N)}_{\text{vec}} \left(\vec z; \vec n; x \right) &= x^{-\sum_{1 \leq i<j\leq N} |n_i-n_j|} \prod_{1 \leq i \neq  j\leq N}(1- (-1)^{n_i -n_j} x^{|n_i-n_j|} z_i z_j^{-1})~.
}
Note that the index of the $T(\SU(3))$ theory is invariant under the mirror symmetry in the following sense:
\bes{
&\hat{\CI}_{T(\SU(3))}(\{w_1; w_2 \} , \{n_1, n_2\} | \{f_1, f_2 \}, \{m_1,m_2\}| a, n_a ;x) \\
&= \hat{\CI}_{T(\SU(3))}(\{f_1, f_2 \} , \{m_1; m_2\} | \{w_1, w_2 \}, \{n_1,n_2\}| a^{-1}, -n_a ;x) ~,
}
where we have defined
\bes{
&\hat{\CI}_{T(\SU(3))}(\{w_1, w_2 \} , \{n_1; n_2\} | \{f_1, f_2 \}, \{m_1,m_2\}| a, n_a ;x)   \\
&:= \CI_{T(\SU(3))} (\{w_1 w_2^{-1}, w_1^{-2} w_2^{-1} \}, \{n_1-n_2, -2n_1-n_2\} | \\ 
& \qquad\qquad\qquad \{ f_1, f_2, f_1^{-1} f_2^{-1} \}, \{ m_1 ,m_2, -m_1-m_2\} |a, n_a;x)~.
}
The index of the $T_3$ theory is therefore
\bes{
& \CI_{T_3}(\vec w^{(1)} ,\vec n^{(1)}| \vec w^{(2)} ,\vec n^{(2)}| \vec w^{(3)} ,\vec n^{(3)} |a, n_a ; x) \\
&= \frac{1}{3!} \sum_{r=0}^2 \,\, \sum_{m_1, m_2 \in \BZ + \frac{r}{3}} \,\, \oint \left( \prod_{\alpha=1}^2 \frac{d f_\alpha}{ 2\pi i f_\alpha} \right) \CZ^{\SU(3)}_{\text{vec}}\left(\vec f; \vec m; x \right) \times \\
& \qquad \   \prod_{I=1}^3 \hat{\CI}_{T(\SU(3))}(\vec w^{(I)} ,\vec n^{(I)}  | \vec f, \vec m| a, n_a ;x)~,
}
where the $\SU(3)$ vector multiplet contribution is given by
\bes{
\scalebox{0.95}{$
\begin{split}
\CZ^{\SU(3)}_{\text{vec}} \left(\{z_1,z_2\}; \{ n_1, n_2\}; x \right) &= \CZ^{\U(3)}_{\text{vec}} \left( \{z_1, z_2, z_1^{-1} z_2^{-1} \} ; \{ n_1, n_2, -n_1-n_2 \}; x \right)~.
\end{split}$}
}
The index of the theory of our interest \eref{singleT3} is then
\bes{
 & \CI_{\eref{singleT3}} (a, n_a; x)\\
 & = \frac{1}{(3!)^3}\prod_{i=1}^3 \sum_{n^{(i)}_1, n^{(i)}_2 \in \BZ} \oint \frac{d w_1^{(i)}}{2 \pi w_1^{(i)}} \frac{d w_2^{(i)}}{2 \pi w_2^{(i)}}\,\, (w_1^{(i)})^{k_i (2 n^{(i)}_1+n^{(i)}_2)} (w_2^{(i)})^{k_i  (n^{(i)}_1+2n^{(i)}_2)}  \times  \\
 & \quad \ \left[ \prod_{i=1}^3 \CZ^{\SU(3)}_{\text{vec}} \left(\vec w^{(i)}; \vec n^{(i)}; x \right) \right]\times \CI_{T_3}(\vec w^{(1)} ,\vec n^{(1)}| \vec w^{(2)} ,\vec n^{(2)}| \vec w^{(3)} ,\vec n^{(3)} |a, n_a ; x)~.
}
As before, if the ATT condition \eref{ATT} is not satisfied, we set $a=1$ and $n_a=0$.

Due to the technicality of the computation, let us discuss the results only in certain cases.  We first focus on the theories that satisfy the ATT condition and we set $n_a=0$. The contribution from the fluxes $n^{(i)}_1, n^{(i)}_2=0$ for all $i=1,2,3$ is 
\bes{ \label{indexsingleT3zero}
\scalebox{0.99}{$
1 +0 x+ (a^4-1) x^2 + ( a^6 - a^2 + a^{-2} + a^{-6} ) x^3 + ( a^8 -1 -  a^{-4} + a^{-8}  ) x^4+ \ldots~.
$}}
This index holds when the CS levels are sufficiently high so that the contributions from non-zero fluxes $n^{(i)}_1, n^{(i)}_2$ appear at a higher order.\footnote{For example, the index for theory \eref{singleT3} with CS levels $(-1,2,2)$ is $1 +0x + (a^4-1) x^2 + ( a^6 - a^2 + a^{-2} -3a^{-6} ) x^3 + ( a^8 -1 - 5 a^{-4} + 2 a^{-8}   ) x^4+\ldots$.  On the other hand, for CS levels $(-3,6,6)$, the index up to order $x^4$ is given by \eref{indexsingleT3zero}.}  The term $0x$ implies that there is no $\CN=3$ flavour symmetry current.  Therefore, the term $-x^2$ indicates that there is one extra SUSY-current, implying that the IR SCFT has enhanced $\CN=4$ supersymmetry.  The term $a^4 x^2$ corresponds to the marginal operator $\mathbb{M}_2$, defined in \eref{chiralringT3}.  When the ATT condition is not satisfied, we set $a=1$ in \eref{indexsingleT3zero}.  In this case, the index does not provide any evidence for supersymmetry enhancement.

\subsection{Two $T_3$ building blocks}
Similarly to \eref{twoT2}, we can couple two copies of the $T_3$ theory together by gauging a diagonal subgroup of the two $\SU(3)_i$ flavour symmetries (with $i=1,2,3$), belonging to different copies of the $T_3$ theories, with CS levels $k_i$.  
\bes{ \label{twoT3}
\begin{tikzpicture}[baseline, font=\footnotesize]
\node[draw=none] (T3a) at (0,0) {$T_3$};
\node[draw=none] (T3b) at (3,0) {$T_3$};
\draw[solid, bend left] (T3a) to node[above,midway] {\scriptsize $k_1$} (T3b);
\draw[solid] (T3a) to node[above,midway] {\scriptsize $k_2$} (T3b);
\draw[solid, bend right] (T3a) to node[above,midway] {\scriptsize $k_3$} (T3b);
\end{tikzpicture}
}
This theory has a $\BZ_3^2$ one-form symmetry, whose 't Hooft anomaly is given by \eref{anomT3}.  The index of this theory is given by
\bes{
 & \CI_{\eref{twoT3}} (a, n_a; x)\\
 & = \frac{1}{(3!)^3}\prod_{i=1}^3 \sum_{n^{(i)}_1, n^{(i)}_2 \in \BZ} \oint \frac{d w_1^{(i)}}{2 \pi w_1^{(i)}} \frac{d w_2^{(i)}}{2 \pi w_2^{(i)}}\,\, (w_1^{(i)})^{k_i (2 n^{(i)}_1+n^{(i)}_2)} (w_2^{(i)})^{k_i  (n^{(i)}_1+2n^{(i)}_2)}  \times  \\
 & \quad \left[ \prod_{i=1}^3 \CZ^{\SU(3)}_{\text{vec}} \left(\vec w^{(i)}; \vec n^{(i)}; x \right) \right]  \prod_{s= \pm1} \CI_{T_3}(\vec w^{(1)} ,\vec n^{(1)}| \vec w^{(2)} ,\vec n^{(2)}| \vec w^{(3)} ,\vec n^{(3)} | a^s, s n_a ; x)~.
}
Once again, the fugacity $a$ and background magnetic flux $n_a$ for the flavour symmetry should be set to $1$ and $0$ respectively if the ATT condition \eref{ATT} is not satisfied.  

Due to the technicality of the computation, we set $n_a=0$ and we report only the contributions of the zero gauge fluxes $n_1^{(i)}, n_2^{(i)} =0$:
\bes{
{1}+0x+{\left({a^4}+{a^{-4}}+{4}-1\right)}{x^2}+{2}{\left({a^6}+{a^{-6}}\right)}{x^3}+{\left[{2}{\left({a^8}+{a^{-8}}\right)}+{3}\right]}{x^4}+{\ldots}~.
}
Note that this is the index of theory \eref{twoT3} with sufficiently large CS levels $k_{1,2,3}$.  As before, the term $0x$ indicates that there is no $\CN=3$ flavour symmetry current, and so the term $-1 x^2$ indicates that there is one extra SUSY-current.  The IR SCFT indeed has enhanced $\CN=4$ supersymmetry.  The positive terms at order $x^2$ correspond to the following marginal operators:
\bes{
{a^{\pm 4}} :& \quad  \mathbb{M}^{(1)}_2~,\,\, \mathbb{M}^{(2)}_2~, \\
{4} :& \quad  {\mathrm{tr}_{1}{\left({X^{(1)}}{X^{(2)}}\right)}}~, \ {\mathrm{tr}_{2}{\left({Y^{(1)}}{Y^{(2)}}\right)}}~, \ {\mathrm{tr}_{3}{\left({Z^{(1)}}{Z^{(2)}}\right)}} \ {\text{and}} \ {\CQ^{{(1)}{i_1}{i_2}{i_3}}}{\tilde{\CQ}^{(2)}_{{i_1}{i_2}{i_3}}}~,
}
where we have used the same notation as in \eref{chiralringT3} with an extra superscript $(I)$ such that $I=1,2$ to denote the $I$-th copy of the $T_3$ theory.

\acknowledgments
We express our gratitude to Matteo Sacchi and Alessandro Tomasiello for a number of useful discussions and for carefully reading through the manuscript as well as providing us with insightful comments.  N.M. thanks the visiting research fellowship of the CNRS and the LPTENS, ENS Paris, where part of this project was conducted.  We thank Dongmin Gang for a fruitful discussion as well as for comparing the result in \cite{Gang:2023rei} with ours.  In this regard, N.M. thanks the workshop ``Recent Trends in Supersymmetric Field Theories 2023'' for making this discussion happen.

\appendix
\section{Theories with four $T_2$ building blocks} \label{app:fourT2}
In this appendix, we consider theories obtained by gauging four copies of the $T_2$ theory in the way depicted below:
\bes{ \label{T24copies}
\scalebox{0.75}{
\begin{tikzpicture}[baseline]
\node[draw=none] (T21) at (-9,0) {$T_2^{\left(1\right)}$};
\node[draw=none] (T22) at (-6,-2) {$T_2^{\left(2\right)}$};
\node[draw=none] (T23) at (-3,0) {$T_2^{\left(3\right)}$};
\node[draw=none] (T24) at (-6,2) {$T_2^{\left(4\right)}$};
\draw[solid, thick, black] (T24)--(T21) node[above,midway] {$-k$};
\draw[solid, thick, black] (T21) to [bend right] node[below,midway] {${2}{k}$} (T22);
\draw[solid, thick, black] (T21) to [bend left] node[above,midway] {${2}{k}$} (T22);
\draw[solid, thick, black] (T22)--(T23) node[below,midway] {$-k$};
\draw[solid, thick, black] (T23) to [bend right] node[above,midway] {${2}{k}$} (T24);
\draw[solid, thick, black] (T23) to [bend left] node[below,midway] {${2}{k}$} (T24);
\end{tikzpicture}}
}
where each line with label $k$ denotes the gauging with CS level $k$ of the diagonal $\SU(2)$ subgroup of the $\SU(2) \times \SU(2)$ flavour symmetry belonging to a pair of $T_2$ theories. Theory \eref{T24copies} admits an {\it equivalent} quiver description in terms of the ${\USp({2})_{-k}}\times{\Spin({4})_{{2}{k}}}\times{\USp({2})_{-k}}\times{\Spin({4})_{{2}{k}}}$ circular quiver with a bifundamental half-hypermultiplet corresponding to each line between adjacent gauge nodes.
\bes{ \label{T24copiesa}
\scalebox{0.75}{
\begin{tikzpicture}[baseline]
\draw (5,0) circle (2);
\node[draw, circle, fill=white, minimum size=1.5cm] (c1) at (5,-2) {\scalebox{0.65}{$\Spin({4})_{{2}{k}}$}};
\node[draw, circle, fill=white, minimum size=1.5cm] (c2) at (7,0) {\scalebox{0.65}{$\USp({2})_{-k}$}};
\node[draw, circle, fill=white, minimum size=1.5cm] (c3) at (5,2) {\scalebox{0.65}{$\Spin({4})_{{2}{k}}$}};
\node[draw, circle, fill=white, minimum size=1.5cm] (c4) at (3,0) {\scalebox{0.65}{$\USp({2})_{-k}$}};
\end{tikzpicture}}
}
 
Let us compute the index of our theory, which is given by
\bes{ \label{IT24copies}
\scalebox{0.83}{$
\begin{split}
&\mathcal{I}_{\eref{T24copies}} \left({a},{n_a};{x}\right)\\
&= {\frac{1}{64}}\,\,  \sum_{(m_1, \ldots, m_6) \in {\mathbb{Z}^{6}}} \,\,\oint {\left(\prod_{{b}={1}}^{6}{\frac{dz_b}{{2}{\pi}{i}{z_b}}}\right)}\times \\ & \qquad {z_1^{-{2}{k}{m_1}}}{z_2^{{4}{k}{m_2}}}{z_3^{{4}{k}{m_3}}}{z_4^{-{2}{k}{m_4}}}{z_5^{{4}{k}{m_5}}}{z_6^{{4}{k}{m_6}}} \prod_{{b}={1}}^{6} {{\mathcal{Z}_{\text{vec}}^{SU(2)}}\left({z_b};{m_b};x\right)} \prod_{{s_1},{s_2},{s_3}={\pm{1}}} \times \\ &\left[{{\mathcal{Z}^{\frac{1}{2}}_\chi}\left({z_1^{s_1}}{z_2^{s_2}}{z_3^{s_3}}{a};{{s_1}{m_1}}+{{s_2}{m_2}}+{{s_3}{m_3}}+{n_a};x\right)} {{\mathcal{Z}^{\frac{1}{2}}_\chi}\left({z_2^{s_1}}{z_3^{s_2}}{z_4^{s_3}}{a^{-1}};{{s_1}{m_2}}+{{s_2}{m_3}}+{{s_3}{m_4}}-{n_a};x\right)}\times\right.\\ 
&\left.{{\mathcal{Z}^{\frac{1}{2}}_\chi}\left({z_4^{s_1}}{z_5^{s_2}}{z_6^{s_3}}{a};{{s_1}{m_4}}+{{s_2}{m_5}}+{{s_3}{m_6}}+{n_a};x\right)}{{\mathcal{Z}^{\frac{1}{2}}_\chi}\left({z_5^{s_1}}{z_6^{s_2}}{z_1^{s_3}}{a^{-1}};{{s_1}{m_5}}+{{s_2}{m_6}}+{{s_3}{m_1}}-{n_a};x\right)}\right]~, 
\end{split}$}
}
where $a$ and $n_a$ are the fugacity and background magnetic flux for the flavour symmetry that assigns charge $+1$ to the chiral fields of theories $T_2^{(1)}$ and $T_2^{(3)}$ and $-1$ to those of theories $T_2^{(2)}$ and $T_2^{(4)}$. For simplicity, we will set ${n_a}={0}$ upon computing the series expansion of the index.

On the other hand, the index of theory \eref{T24copiesa} can be obtained starting from the one of the ${\USp({2})_{-k}}\times{\SO({4})_{{2}{k}}}\times{\USp({2})_{-k}}\times{\SO({4})_{{2}{k}}}$ circular quiver theory.
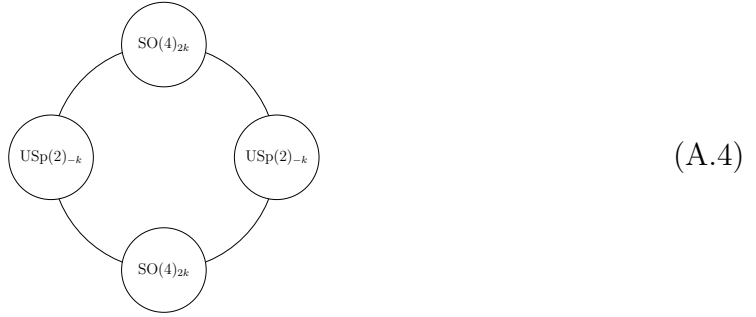
\bes{ \label{T24copiesb}
\scalebox{0.75}{
\begin{tikzpicture}[baseline]
\draw (5,0) circle (2);
\node[draw, circle, fill=white, minimum size=1.5cm] (c1) at (5,-2) {\scalebox{0.65}{$\SO({4})_{{2}{k}}$}};
\node[draw, circle, fill=white, minimum size=1.5cm] (c2) at (7,0) {\scalebox{0.65}{$\USp({2})_{-k}$}};
\node[draw, circle, fill=white, minimum size=1.5cm] (c3) at (5,2) {\scalebox{0.65}{$\SO({4})_{{2}{k}}$}};
\node[draw, circle, fill=white, minimum size=1.5cm] (c4) at (3,0) {\scalebox{0.65}{$\USp({2})_{-k}$}};
\end{tikzpicture}}
}
\bes{\label{ISO4USp2SO4USp2}
\scalebox{0.9}{$
\begin{split}
&\mathcal{I}_{\eref{T24copiesb}}{\left({\zeta_1},{\zeta_2},{a};x\right)}\\ 
&= {\frac{1}{64}}\,\,  \sum_{\fm_1, \ldots, \fm_4, \fn_1,\fn_2 \in \BZ} \,\,\oint {\left(\prod_{{b}={1}}^{4}{\frac{dv_b}{{2}{\pi}{i}{v_b}}}{v_b^{{2}{k}{\mathfrak{m}_b}}}\right)}{\zeta_{1}^{{\mathfrak{m}_1}+{\mathfrak{m}_2}}}{\zeta_{2}^{{\mathfrak{m}_3}+{\mathfrak{m}_4}}} \oint {\left(\prod_{{b}={1}}^{2}{\frac{du_b}{{2}{\pi}{i}{u_b}}}{u_b^{-{2}{k}{\mathfrak{n}_b}}}\right)}{\times}\\ 
&\qquad  {{\mathcal{Z}_{\text{vec}}^{\SO(4)}}\left({v_1},{v_2};{\mathfrak{m}_1},{\mathfrak{m}_2};x\right)}{{\mathcal{Z}_{\text{vec}}^{\USp(2)}}\left({u_1};{\mathfrak{n_1}};x\right)}{\times}\\ 
& \qquad {{\mathcal{Z}_{\text{vec}}^{\SO(4)}}\left({v_3},{v_4};{\mathfrak{m}_3},{\mathfrak{m}_4};x\right)}{{\mathcal{Z}_{\text{vec}}^{\USp(2)}}\left({u_2};{\mathfrak{n_2}};x\right)}{\times}\\ &{\prod_{{b}={1}}^{2} \ \prod_{{s_1},{s_2}={\pm{1}}}{{\mathcal{Z}^{\frac{1}{2}}_\chi}\left({v_b^{s_1}}{u_{1}^{s_2}}{a};{{s_1}{\mathfrak{m}_b}}+{{s_2}{\mathfrak{n}_{1}}};x\right)}{{\mathcal{Z}^{\frac{1}{2}}_\chi}\left({v_b^{s_1}}{u_{2}^{s_2}}{a^{-1}};{{s_1}{\mathfrak{m}_b}}+{{s_2}{\mathfrak{n}_{2}}};x\right)}}{\times}\\ &{\prod_{{b}={3}}^{4} \ \prod_{{s_1},{s_2}={\pm{1}}}{{\mathcal{Z}^{\frac{1}{2}}_\chi}\left({v_b^{s_1}}{u_{1}^{s_2}}{a^{-1}};{{s_1}{\mathfrak{m}_b}}+{{s_2}{\mathfrak{n}_{1}}};x\right)}{{\mathcal{Z}^{\frac{1}{2}}_\chi}\left({v_b^{s_1}}{u_{2}^{s_2}}{a};{{s_1}{\mathfrak{m}_b}}+{{s_2}{\mathfrak{n}_{2}}};x\right)}}~,
\end{split}$}
}
where $\zeta_{1}$ and $\zeta_{2}$ are the fugacities associated with the $\mathbb{Z}_2$ zero-form magnetic symmetries of the two $\SO(4)_{{2}{k}}$ gauge factors, and the fugacities for the zero-form charge conjugation symmetries are set to unity. In order to obtain the index of the circular quiver \eref{T24copiesa} with $\Spin$ gauge groups, we have to sum over both ${\zeta_{1}}=\pm{1}$ and ${\zeta_{2}}=\pm{1}$  and divide by four:
\begin{equation}\label{ISpin4USp2Spin4USp2}
\begin{split}
\mathcal{I}_{\eref{T24copiesa}}\left({a};{x}\right) ={\frac{1}{4}} \sum_{\zeta_1, \zeta_2 = \pm 1} \mathcal{I}_{\eref{T24copiesb}}\left({\zeta_1},{\zeta_2},{a};{x}\right) .
\end{split}
\end{equation}
It is easy to check that the indices \eqref{IT24copies} and \eqref{ISpin4USp2Spin4USp2} are equal if we perform the following map between the gauge fugacities and magnetic fluxes of the two theories:
\begin{equation}\label{map4}
\begin{array}{lll}
{z_1}={u_1}~, &\qquad {z_2^{2}}={v_1}{v_2}~, & \qquad {z_3^{2}}={v_1}{v_2^{-1}}~, \\
{z_4}={u_2}~, & \qquad {z_5^{2}}={v_3}{v_4}~, & \qquad {z_6^{2}}={v_3}{v_4^{-1}}~, \\
{m_1}={\mathfrak{n}_1}~, &  \qquad {2}{m_2}={\mathfrak{m}_1}+{\mathfrak{m}_2}~, & \qquad  {2}{m_3}={\mathfrak{m}_1}-{\mathfrak{m}_2}~, \\
{m_4}={\mathfrak{n}_2}~, &  \qquad {2}{m_5}={\mathfrak{m}_3}+{\mathfrak{m}_4}~, &\qquad  {2}{m_6}={\mathfrak{m}_3}-{\mathfrak{m}_4}~.
\end{array}
\end{equation}

It can now be checked that the IR SCFT in question exhibits ${\mathcal{N}}={4}$ supersymmetry enhancement, as expected from the general prescription described in the main text, by computing the index as a series expansion in the variable $x$.  We report the results up to order $x^4$ for various values of $k$:
\bes{
&{\mathcal{I}_{\eref{T24copies}}}{\left({a};{x}\right)}={\mathcal{I}_{\eref{T24copiesa}}}{\left({a};{x}\right)}\\
k=1: \quad & = {1}+{\left({2}{a^4}+{2}{a^{-4}}+{3}\right)}{x^2}-{4}{\left({a^2}+{a^{-2}}\right)}{x^3}+\\
&\qquad {\left({6}{a^8}+{3}{a^4}+{3}{a^{-4}}+{6}{a^{-8}}+{7}\right)}{x^4}+\ldots~,\\
k=2: \quad & ={1}+{\left({2}{a^4}+{2}{a^{-4}}+{3}\right)}{x^2}-{4}{\left({a^2}+{a^{-2}}\right)}{x^3}+\\
&\qquad{\left({3}{a^8}+{2}{a^4}+{2}{a^{-4}}+{3}{a^{-8}}+{4}\right)}{x^4} +\ldots~.
}
Note that for ${k}\ge{2}$ the indices differ at a higher order than $x^4$ in the expansion.  We notice that the coefficient of $x$ vanishes, meaning that there is no ${\mathcal{N}}={3}$ flavour current.  The coefficient of $x^2$ should be written as
\begin{equation}
{2}{a^4}+{2}{a^{-4}}+{4}-{1}~,
\end{equation}
where the term $-1$ is the contribution of the ${\mathcal{N}}={3}$ extra SUSY-current that leads to ${\mathcal{N}}={4}$ supersymmetry enhancement. The marginal operators, contributing the positive terms ${2}{a^4}+{2}{a^{-4}}+{4}$, can be listed as follows:
\bes{
{2}{a^4} :& \ \tr(\mu^{(1)\, 2}) \ \text{and} \ \tr(\mu^{(3)\, 2})~, \\
{2}{a^{-4}} :& \ \tr(\mu^{(2)\, 2}) \ \text{and} \ \tr(\mu^{(4)\, 2})~, \\
{4} :& \ \text{gauge invariant combinations of two chiral fields ${Q^{(I)}_{{i_I}{j_I}{k_I}}}$ of $T_2^{(I)}$} \\ 
& \ \text{and two chiral fields ${Q^{(J)}_{{i_J}{j_J}{k_J}}}$ of $T_2^{(J)}$, with ${I}={1},{3}$ and ${J}={2},{4}$}~.
}

We can gauge the one-form symmetry of the theory \eref{T24copies} to obtain the ${SU(2)_{-k}}\times{\left[{SU(2)_{{2}{k}}}\times{SU(2)_{{2}{k}}}\right]/{\mathbb{Z}_{2}}}\times{SU(2)_{-k}}\times{\left[{SU(2)_{{2}{k}}}\times{SU(2)_{{2}{k}}}\right]/{\mathbb{Z}_{2}}}$ gauge theory, as depicted below.
\bes{ \label{T24copiesmodZ2}
\scalebox{0.75}{
\begin{tikzpicture}[baseline]
\node[draw=none] (T21) at (-9,0) {$T_2^{\left(1\right)}$};
\node[draw=none] (T22) at (-6,-2) {$T_2^{\left(2\right)}$};
\node[draw=none] (T23) at (-3,0) {$T_2^{\left(3\right)}$};
\node[draw=none] (T24) at (-6,2) {$T_2^{\left(4\right)}$};
\draw[solid, thick, black] (T24)--(T21) node[above,midway] {$-k$};
\draw[solid, thick, black] (T21) to [bend right] node[below,midway] {${2}{k}$} (T22);
\draw[solid, thick, black] (T21) to [bend left] node[above,midway] {${2}{k}$} (T22);
\draw[solid, thick, black] (T22)--(T23) node[below,midway] {$-k$};
\draw[solid, thick, black] (T23) to [bend right] node[above,midway] {${2}{k}$} (T24);
\draw[solid, thick, black] (T23) to [bend left] node[below,midway] {${2}{k}$} (T24);
\draw[solid, thick, black] (-8.5,-1.7) to node[right, midway] {\scalebox{0.75}{${\mathbb{Z}_2^{[1]}}$}} (-6.5,-0.3);
\draw[solid, thick, black] (-5.5,0.3) to node[right, midway] {\scalebox{0.75}{${\mathbb{Z}_2^{[1]}}$}} (-3.5,1.7);
\end{tikzpicture}
}}
It turns out that this theory is equivalent to the circular quiver theory \eref{T24copiesb}. The index of theory \eqref{T24copiesmodZ2} can be obtained from \eref{IT24copies} by replacing the summation as
\bes{ \label{replacesumT24copies}
\sum_{(m_1, \ldots, m_6) \in {\mathbb{Z}^{6}}} \quad \longrightarrow \quad {\sum_{{p}={0}}^{1}}{{\zeta_1}^{p}} \,\, {\sum_{{p'}={0}}^{1}}{{\zeta_2}^{p'}}{\sum_{{{m_1},{m_4}} \in {\mathbb{Z}^{2}}}} \ \ {\sum_{{{m_2},{m_3}} \in {\left(\mathbb{Z}+{\frac{p}{2}}\right)^{2}}}} \,\,\, {\sum_{{{m_5},{m_6}} \in {\left(\mathbb{Z}+{\frac{p'}{2}}\right)^{2}}}}~,   
}
where $\zeta_1$ and $\zeta_2$ are the fugacities of the zero-form symmetry of \eqref{T24copiesmodZ2}. The result of this procedure is equal to the index \eqref{ISO4USp2SO4USp2}.\footnote{Indeed, by taking into account the map \eqref{map4}, $\zeta_1$ and $\zeta_2$ are such that
\begin{equation*}
{{\zeta_1}^{{\mathfrak{m}_1}\pm{\mathfrak{m}_2}}}=\begin{cases} {1} & {{\mathfrak{m}_1}\pm{\mathfrak{m}_2}}\in{\mathbb{Z}_{\text{even}}}\longleftrightarrow{{m_2},{m_3}\in{\mathbb{Z}}}\longleftrightarrow{{p}={0}}\\
{\zeta_1} & {{\mathfrak{m}_1}\pm{\mathfrak{m}_2}}\in{\mathbb{Z}_{\text{odd}}}\longleftrightarrow{{m_2},{m_3}\in{\frac{\mathbb{Z}_{\text{odd}}}{2}}}\longleftrightarrow{{p}={1}}\end{cases}~,
\end{equation*}
\begin{equation*}
{{\zeta_2}^{{\mathfrak{m}_3}\pm{\mathfrak{m}_4}}}=\begin{cases} {1} & {{\mathfrak{m}_3}\pm{\mathfrak{m}_4}}\in{\mathbb{Z}_{\text{even}}}\longleftrightarrow{{m_5},{m_6}\in{\mathbb{Z}}}\longleftrightarrow{{p'}={0}}\\
{\zeta_2} & {{\mathfrak{m}_3}\pm{\mathfrak{m}_4}}\in{\mathbb{Z}_{\text{odd}}}\longleftrightarrow{{m_5},{m_6}\in{\frac{\mathbb{Z}_{\text{odd}}}{2}}}\longleftrightarrow{{p'}={1}}\end{cases}~.
\end{equation*}} 
Let us report the results for various values of $k$ up to order $x^4$.
For ${k}={1}$, the expansion of the index up to order $x^4$ reads
\bes{
&\mathcal{I}_{\eref{T24copiesmodZ2}}{\left({\zeta_1},{\zeta_2},a;{x}\right)}=\mathcal{I}_{\eref{T24copiesb}}{\left({\zeta_1},{\zeta_2},{a};{x}\right)}\\ k=1: \quad & = {1}+{\left[{\left({2}+{\zeta_1}+{\zeta_2}+{\zeta_1\zeta_2}\right)}{\left({a^4}+{a^{-4}}\right)}+{3}+{\zeta_1}+{\zeta_2}+{\zeta_1\zeta_2}\right]}{x^2}+\\ 
&\quad \left[{\left({\zeta_1}+{\zeta_2}+{2}{\zeta_1\zeta_2}\right)}{a^6}+{\left(-{4}-{\zeta_1}-{\zeta_2}+{2}{\zeta_1\zeta_2}\right)}{\left({a^2}+{a^{-2}}\right)}+\right. \\ & \quad\left.{\left({\zeta_1}+{\zeta_2}+{2}{\zeta_1\zeta_2}\right)}{a^{-6}}\right]{x^3}+\\ 
&\quad \left[{\left({6}+{3}{\zeta_1}+{3}{\zeta_2}+{3}{\zeta_1\zeta_2}\right)}{a^8}+{\left({3}-{\zeta_1}-{\zeta_2}-{3}{\zeta_1\zeta_2}\right)}{\left({a^4}+{a^{-4}}\right)}\right.+\\ 
&\quad \left.{\left({6}+{3}{\zeta_1}+{3}{\zeta_2}+{3}{\zeta_1\zeta_2}\right)}{a^{-8}}+{7}-{3}{\zeta_1\zeta_2}\right]{x^4}+\ldots~,
}
For ${k}={2},{3}$ we get
\bes{
&\mathcal{I}_{\eref{T24copiesmodZ2}}{\left({\zeta_1},{\zeta_2},a;{x}\right)}=\mathcal{I}_{\eref{T24copiesb}}{\left({\zeta_1},{\zeta_2},{a};{x}\right)}\\
k=2: \quad & = {1}+{\left({2}{a^4}+{2}{a^{-4}}+{3}\right)}{x^2}+\\ &\quad  \left[{\left({\zeta_1}+{\zeta_2}\right)}{a^6}-{4}{\left({a^2}+{a^{-2}}\right)}+{\left({\zeta_1}+{\zeta_2}\right)}{a^{-6}}\right]{x^3}+\\ &\quad \left[{\left({3}+{\zeta_1}+{\zeta_2}+{\zeta_1\zeta_2}\right)}{a^8}+{\left({2}+{\zeta_1\zeta_2}\right)}{\left({a^4}+{a^{-4}}\right)}+\right.\\ &\quad \left. {\left({3}+{\zeta_1}+{\zeta_2}+{\zeta_1\zeta_2}\right)}{a^{-8}}+{4}+{\zeta_1\zeta_2}\right]{x^4}+\ldots~, \\
k=3: \quad & = {1}+{\left({2}{a^4}+{2}{a^{-4}}+{3}\right)}{x^2}-{4}{\left({a^2}+{a^{-2}}\right)}{x^3}+\\ &\quad \left[{\left({3}+{\zeta_1}+{\zeta_2}\right)}{a^8}+{2}{\left({a^4}+{a^{-4}}\right)}+{\left({3}+{\zeta_1}+{\zeta_2}\right)}{a^{-8}}+{4}\right]{x^4}+\ldots~.
}
For ${k}\ge4$ the fugacities ${\zeta_1}$ and ${\zeta_2}$ appear at a higher order than $x^4$.

From the expansion of the indices, it is clear that we have two independent $\BZ_2$ fugacities $\zeta_1$ and $\zeta_2$ satisfying ${{\zeta_1}^2}={\zeta_2^{2}}={1}$. Hence, the elements of $\left\{{1},{\zeta_1},{\zeta_2},{\zeta_1\zeta_2}\right\}$ corresponding to the possible choices of ${p}=\left\{{0},{1}\right\}$ and ${p'}=\left\{{0},{1}\right\}$ in \eqref{replacesumT24copies} indicate the presence of the $\BZ_2 \times \BZ_2$ zero-form symmetry in theory \eref{T24copiesb} = \eref{T24copiesmodZ2}, and hence the ${\mathbb{Z}_{2}}\times{\mathbb{Z}_{2}}$ one-form symmetry of theory \eref{T24copies} = \eref{T24copiesa}.  

From the point of view of the ${\USp({2})_{-k}}\times{\SO({4})_{{2}{k}}}\times{\USp({2})_{-k}}\times{\SO({4})_{{2}{k}}}$ circular quiver theory \eref{T24copiesb}, it is expected by a similar argument that leads to \eqref{oneformUSp2SO4} that there is a non-anomalous $\BZ_2$ one-form symmetry for $k$ even. In principle, one can further gauge this one-form symmetry, generalising what we did in \eref{gaugefurtherZ2}.

\section{Mixed gauge/zero-form monopole operators} \label{sec:mixedflvoneT2}
In this appendix, we examine the mixed gauge/zero-form monopole operators with {\it fractional} magnetic flux for both the Cartan subalgebra of the gauge group and the $\U(1)_a$ flavour symmetry group \cite{Bhardwaj:2022dyt, Bhardwaj:2023zix} for theories \eref{singleT2} with one $T_2$ building block such that the ATT condition \eref{ATT} is satisfied.  We adopt the same method as described in \cite{Mekareeya:2022spm} (see also \cite{Sacchi:2023omn}) which relies on the supersymmetric index.  In particular, we take $n_a=1/2$ in \eref{indexsingleT2} and examine the contribution of the gauge fluxes $m_{1,2,3}$ satisfying the Dirac quantisation condition, namely $\pm m_1 \pm m_2 \pm m_3+n_a \in \BZ$, \ie~ 
\bes{ \label{diracnahalf}
\sum_i m_i \in \BZ+\frac{1}{2}~.
}  

Let us report the result for various theories with one $T_2$ building block below.
\bes{
\scalebox{0.75}{
\begin{tabular}{|c|c|c|}
\hline
CS levels & Gauge fluxes  & Contribution to the index \eref{indexsingleT2} \\
 & $(m_1,m_2,m_3)$ & \\
\hline
$(-1,2,2)$ & $\left( \frac{1}{2} ,0,0\right)$& 0 \\
                & $\left(0 , \frac{1}{2},0\right)$& $X_2 \equiv \left(\frac{1}{2}+ \frac{1}{2 a^2}\right) x+\left(\frac{a^4}{2}-1-\frac{1}{a^2}-\frac{1}{2 a^4}\right) x^5 +O(x^7)$ \\
                & $\left(0 , 0, \frac{1}{2}\right)$& $X_2$ \\
                & $\left( \frac{1}{2} ,\frac{1}{2},\frac{1}{2}\right)$& 0 \\              
\hline \hline
$(-2,4,4)$ & $\left( \frac{1}{2} ,0,0\right)$& $X_{-2} \equiv \frac{x}{2 a^2}+\left(\frac{a^2}{2}-\frac{1}{a^2}-\frac{1}{2 a^4}\right) x^5+O(x^7)$ \\
                & $\left(0 , \frac{1}{2},0\right)$& $X_4 \equiv \left(\frac{a^2}{2}+\frac{1}{2}\right) x^2+\left(\frac{1}{2}+\frac{1}{2 a^2}\right) x^4+\left(\frac{a^6}{2}-a^2-\frac{1}{2}\right) x^6+O(x^7)$ \\
                & $\left(0 , 0, \frac{1}{2}\right)$& $X_4$ \\
                & $\left( \frac{1}{2} ,\frac{1}{2},\frac{1}{2}\right)$& $\left(\frac{a^3}{8}+\frac{a}{4}+\frac{1}{8 a}\right) x^{3/2}+\left(\frac{1}{8 a} +\frac{1}{8 a^3} \right) x^{7/2}+ O(x^{11/2})$ \\
&&                $=\frac{1}{4} \left[ \{ x^{1/2} (a^5+a^3)+\ldots \}X_{-2} + \{ x^{-1/2} a^{-1} +\ldots \} X_4 \right]$ \\       
                \hline \hline
$(-2,3,6)$ & $\left( \frac{1}{2} ,0,0\right)$& $X_{-2}$ \\
                & $\left(0 , \frac{1}{2},0\right)$& $0$ \\
                & $\left(0 , 0, \frac{1}{2}\right)$& $X_{6} \equiv \left(\frac{a^4}{2}+\frac{a^2}{2}\right) x^3 + \left(\frac{a^2}{2}+\frac{1}{2}\right) x^5+\left(\frac{a^8}{2}-a^4+\frac{1}{2}\right) x^7 + O(x^{9})$ \\
                & $\left( \frac{1}{2} ,\frac{1}{2},\frac{1}{2}\right)$& $0$ \\     
                \hline \hline
$(-4,6,12)$ & $\left( \frac{1}{2} ,0,0\right)$ & $X_{-4} \equiv \left(\frac{1}{2 a^2}+\frac{1}{2}\right) x^4+\left(\frac{1}{2 a^4}+\frac{1}{2 a^6}\right) x^6+\left(\frac{1}{2 a^6}+\frac{a^4}{2}-\frac{3}{2 a^2}-\frac{3}{2}\right) x^8 +O(x^{10})$\\
& $\left( 0, \frac{1}{2} ,0\right)$ & $X_6$  \\     
 & $\left( 0, 0, \frac{1}{2} \right)$  & $X_{12} \equiv \left(\frac{a^{10}}{2}+\frac{a^8}{2}\right) x^6+\left(\frac{a^8}{2}+\frac{a^6}{2}\right) x^8+\left(\frac{a^{14}}{2}-a^{10}+\frac{a^6}{2}\right) x^{10}+O(x^{12})$\\   
 & $\left( \frac{1}{2} , \frac{1}{2} , \frac{1}{2} \right)$  &  $\left(\frac{a^{11}}{8}+\frac{a^9}{4}+\frac{a^7}{8}\right) x^{11/2}+\left(\frac{a^9}{8}+\frac{a^7}{4}+\frac{a^5}{8}\right) x^{15/2} +O(x^{19/2})$ \\                
\hline                
\end{tabular}}
}
From the above results, we observe that whenever all CS levels are even, the contribution of the gauge fluxes $(\frac{1}{2}, \frac{1}{2}, \frac{1}{2})$ contains odd powers of $a$. This corresponds to a different quantisation in comparison to, for example \eref{indmk2k2k}, where only even powers of $a$ appear.  Note, however, that  this is not the case if there exists an odd CS level.  The presence of such mixed gauge/zero-form monopole operators, when all CS levels are even, potentially implies a mixed anomaly between a $\BZ_2$ one-form symmetry and the $\U(1)_a$ symmetry \cite{Bhardwaj:2022dyt, Bhardwaj:2023zix} (see also \cite{Mekareeya:2022spm, Sacchi:2023omn}).  In our case, such a mixed anomaly is characterised by the 4d anomaly action $\pi \int_{\CM_4} w^{(2)} \cup c_1^a$, where $w^{(2)}$ is the two-form background field for a $\BZ_2$ subgroup of the $\BZ_2 \times \BZ_2$ non-anomalous one-form symmetry and $c_1^a$ is the first Chern class of a background $\U(1)_a$ flavour symmetry bundle.\footnote{Let us compare the above interpretation with respect to \eref{gaugefurtherZ2}.  Recall that in the latter we gauge the $\BZ_2 \times \BZ_2$ one-form symmetry of theory \eref{singleT2} with CS levels $(k_1,k_2,k_3)=(-4,8,8)$ and obtain the dual $\BZ_2 \times \BZ_2$ zero-form symmetry associated with the fugacities $\zeta$ and $s$.  Observe that only even powers of $a$ appear in \eref{gaugefurtherZ2}, and we do not see any indication of the mixed anomaly between a $\BZ_2$ subgroup the $\BZ_2 \times \BZ_2$ zero-form symmetry and the $\U(1)_a$ zero-form symmetry.  A natural question that arises is whether this is in contradiction with the above interpretation regarding the mixed anomaly between the $\BZ_2$ one-form symmetry and the $\U(1)_a$ flavour symmetry. (Note also that the quantisation condition \eref{diracnahalf} with $n_a=1/2$ is different from \eref{replacesumT2} with $n_a=0$.) We leave this for a future investigation.}


A similar analysis can be performed for theories \eref{twoT2} with two $T_2$ building blocks such that the ATT condition \eref{ATT} is satisfied.  The only difference is that in the present case the faithful flavour symmetry group is $\SO(3)_a$.  This can be seen from the discussion in the previous subsection, where only representations of $\su(2)_a$ with even Dynkin labels appear in the index.   To determine whether there is a mixed anomaly involving the flavour symmetry, we fix $n_a=1/2$ in \eref{indextwoT2}; this amounts to turning on the second Stiefel-Whitney class $w_2^a$ that obstructs the lift from the $\SO(3)_a$ bundle to the $\SU(2)_a$ bundle.  The Dirac quantisation condition requires that the gauge fluxes $m_{1,2,3}$ satisfy $\pm m_1 \pm m_2 \pm m_3 \pm n_a \in \BZ$, \ie~$\sum_i m_i \in \BZ+\frac{1}{2}$.  We tabulate the contributions of some of such gauge fluxes in certain examples below.
\bes{
\scalebox{0.75}{
\begin{tabular}{|c|c|c|}
\hline
CS levels & Gauge fluxes  & Contribution to the index \eref{indextwoT2} \\
 & $(m_1,m_2,m_3)$ & \\
\hline
$(-2,3,6)$ & $\left( \frac{1}{2} ,0,0\right)$& $X_{-2}\equiv \left(\frac{1}{2}+\frac{1}{2 a^2}\right) x^2+\left(\frac{a^2}{2}+\frac{3}{2}+\frac{3}{2 a^2}+\frac{1}{2 a^4}\right) x^5+O(x^7)$ \\
                & $\left(0 , \frac{1}{2},0\right)$& $0$ \\
                & $\left(0 , 0, \frac{1}{2}\right)$& $X_{6} \equiv \left(\frac{a^6}{2}+\frac{a^4}{2}\right) x^4+\left(-\frac{a^4}{2}-\frac{a^2}{2}\right) x^5 +\left(\frac{a^4}{2}+\frac{3 a^2}{2}+1\right) x^6+O(x^7)$ \\
                & $\left( \frac{1}{2} ,\frac{1}{2},\frac{1}{2}\right)$& $0$ \\     
                \hline \hline
$(-4,6,12)$ & $\left( \frac{1}{2} ,0,0\right)$& $X_{-4}\equiv \left(\frac{1}{2 a^2}+\frac{1}{2 a^4}\right) x^3+\left(-\frac{1}{2}-\frac{1}{2 a^2}\right) x^4+\left(\frac{a^2}{2}+1+\frac{1}{2 a^2}\right) x^5+O(x^6)$ \\
                & $\left(0 , \frac{1}{2},0\right)$& $X_6$ \\
                & $\left(0 , 0, \frac{1}{2}\right)$& $X_{12} \equiv \left(\frac{a^{12}}{2}+\frac{a^{10}}{2}\right) x^7+\left(-\frac{a^{10}}{2}-\frac{a^8}{2}\right) x^8+O(x^9)$ \\
                & $\left( \frac{1}{2} ,\frac{1}{2},\frac{1}{2}\right)$& $\left(\frac{a^{14}}{8}+\frac{a^{12}}{4}+\frac{a^{10}}{8}\right) x^7+\left(-\frac{a^{12}}{8}-\frac{a^{10}}{4}-\frac{a^8}{8}\right) x^8+O(x^9)$ \\     
                \hline \hline                
\end{tabular}}                
}
In contrast to the precedent example of one $T_2$ building block, we see that only even powers of $a$ appear in the above.  We therefore do not observe a mixed anomaly between a $\BZ_2$ one-form symmetry and the $\SO(3)_a$ flavour symmetry. 

\bibliographystyle{ytphys}
\bibliography{ref}

\end{document}